\documentclass{aa}  

\usepackage{epsfig}
\usepackage{graphicx}
\usepackage{txfonts}
\usepackage{amssymb}
\usepackage{natbib}
\bibpunct{(}{)}{;}{a}{}{,}

\newcommand{\be}{\begin{equation}}
\newcommand{\ee}{\end{equation}}
\newcommand{\bea}{\begin{eqnarray}}
\newcommand{\eea}{\end{eqnarray}}

\newcommand{\epse}{\epsilon_{\rm e}}
\newcommand{\epsB}{\epsilon_{\rm B}}
\newcommand{\Rd}{R_{\rm dec}}
\newcommand{\RB}{R_{\rm B}}
\newcommand{\mpr}{m_{\rm p}}
\newcommand{\mel}{m_{\rm e}}
\newcommand{\Ne}{N_{\rm e}}
\newcommand{\tad}{t_{\rm ad}}
\newcommand{\tB}{t_{\rm B}}
\newcommand{\gc}{\gamma_{\rm c}}
\newcommand{\sT}{\sigma_{\rm T}}
\newcommand{\nph}{n_{\rm ph}}
\newcommand{\Epk}{E_{\rm pk}}
\newcommand{\gp}{\gamma_{\rm pk}}
\newcommand{\Qi}{Q_{\rm inj}}
\newcommand{\tesc}{t_{\rm esc}}
\newcommand{\nsyn}{\dot{N}_{\rm syn}}
\newcommand{\ncs}{\dot{N}_{\rm cs}}
\newcommand{\npp}{\dot{N}_{\rm pp}}
\newcommand{\nad}{\dot{N}_{\rm ad}}
\newcommand{\gad}{\dot{\gamma}_{\rm ad}}
\newcommand{\nel}{\dot{n}_{\rm el}}
\newcommand{\Epc}{E_{\rm pk,IC}}
\newcommand{\tex}{t_{\rm exp}}
\newcommand{\tco}{t_{\rm cool}}
\newcommand{\Gsh}{\Gamma_{\rm sh}}
\newcommand{\Aw}{A_{\rm w}}
\newcommand{\bel}{\beta_{\rm e}}
\newcommand{\xt}{x_{\rm thr}}
\newcommand{\xhe}{x_{\rm HE}}
\newcommand{\xpk}{x_{\rm pk}}
\newcommand{\Ae}{A_{\rm e}}
\newcommand{\Be}{B_{\rm e}}
\newcommand{\tstar}{\tau_*}
\newcommand{\asyn}{\alpha_{\rm syn}}
\newcommand{\app}{\alpha_{\rm pp}}
\newcommand{\scs}{\alpha_{\rm cs}}
\newcommand{\Awn}{A_{\rm w,35}}
\newcommand{\xs}{x_{\rm syn}}
\newcommand{\gt}{\gamma_{\rm thr}}
\newcommand{\Net}{N_{\rm e,thr}}
\newcommand{\lB}{l_{\rm B}}
\newcommand{\UB}{U_{\rm B}}
\newcommand{\Ei}{E_{\rm iso}}
\newcommand{\dnph}{\dot{n}_{\rm ph}}

\begin{document}

   \title{Simulations of gamma-ray burst afterglows with a relativistic kinetic code}

   \author{T.~Pennanen\inst{1}
          \and
          I.~Vurm\inst{2,3}
          \and
          J.~Poutanen\inst{1,4}}

   \institute{Astronomy Division, Department of Physics, P.O. Box 3000, FI-90014 University of Oulu, Finland\\
              \email{tuulia.pennanen@oulu.fi,juri.poutanen@gmail.com}
         \and
             Physics Department and Columbia Astrophysics Laboratory, Columbia University, 538 West 120th Street, New York, NY 10027, USA
	 \and
	     Tartu Observatory, 61602 T\~{o}ravere, Tartumaa, Estonia\\
	     \email{indrek.vurm@gmail.com}
	 \and
	     Tuorla Observatory, University of Turku, V\"ais\"al\"antie 20, FI-21500 Piikki\"o, Finland
             }

   \date{Received 21 August 2013 / Accepted 16 February 2014}

   \authorrunning{Pennanen et al.}
   \titlerunning{Simulations of GRB afterglows}

% \abstract{}{}{}{}{} 
% 5 {} token are mandatory
 
  \abstract
  % context heading (optional)
  % {} leave it empty if necessary  
   {}
  % aims heading (mandatory)
   {This paper introduces a kinetic code that simulates
   gamma-ray burst (GRB) afterglow emission from the external forward shock
   and presents examples of some of its applications.
   One interesting research topic discussed in the paper is the high-energy radiation
   produced by Compton scattering of the prompt GRB photons against
   the shock-accelerated electrons. The difference between
   the forward shock emission in a wind-type and a constant-density
   medium is also studied, and the emission due to Maxwellian electron
   injection is compared to the case with pure power-law electrons.}
  % methods heading (mandatory)
   {The code calculates the time-evolving
   photon and electron distributions in the emission region
   by solving the relativistic kinetic equations for each particle
   species. For the first time, the full relativistic equations for
   synchrotron emission/absorption,
   Compton scattering, and pair production/annihilation were applied
   to model the forward shock emission.
   The synchrotron self-absorption thermalization mechanism, which
   shapes the low-energy end of the electron distribution, was also
   included in the electron equation.}
  % results heading (mandatory)
   {The simulation results indicate that inverse Compton scattering of the prompt GRB photons
   can produce a luminous $\gtrsim \textrm{TeV}$ emission
   component, even when pair production in the emission region
   is taken into account. This very high-energy radiation may be
   observable in low-redshift GRBs.
   The test simulations also show that the low-energy end of a pure power-law
   distribution of electrons can thermalize owing to
   synchrotron self-absorption in a wind-type environment,
   but without an observable impact on the radiation spectrum.
   Moreover, a flattening in the forward shock X-ray light curve may be
   expected when the electron injection function is assumed
   to be purely Maxwellian instead of a power law.
   The flux during such a flattening is likely
   to be lower than the {\it Swift}/XRT sensitivity in the case of a constant-density external medium,
   but a wind environment may result in a higher flux during the shallow decay.}
  % conclusions heading (optional), leave it empty if necessary 
   {}


   \maketitle
%
%________________________________________________________________

\section{Introduction}

Gamma-ray burst (GRB) afterglows are produced
by relativistic electrons radiating mainly via the synchrotron and inverse
Compton mechanisms. According to the standard afterglow model,
the electrons are accelerated to highly relativistic
energies at two shock fronts, the forward shock and the reverse shock, which are the result of
the interaction between the
relativistic jet from the GRB central engine and the surrounding
medium \citep[for reviews, see, e.g.,][]{2004RvMP...76.1143P,2006RPPh...69.2259M}.

The earliest afterglow models invoke pure synchrotron radiation from the forward shock
in a constant-density interstellar medium (ISM) or a wind-type environment
and yield analytic time-evolving synchrotron spectra of the decelerating blast wave
\citep{1998ApJ...497L..17S,2000ApJ...536..195C,2002ApJ...568..820G}.
The role of inverse Compton scattering of the synchrotron photons has also been
investigated, typically relying on an
approximate treatment of the scattering process because no analytic solution for the inverse
Compton spectrum is available \citep{1998ApJ...501..772P,1999ApJ...512..699C,
2000ApJ...543...66P,2001ApJ...548..787S}.

The GRB observations by the {\it Swift} and {\it Fermi} satellites have revealed
some surprising features in the afterglow and prompt light curves, resulting in a
need to improve the models for GRB emission. For example,
the {\it Fermi}/LAT telescope has observed $> 100\:\textrm{MeV}$ emission from several
GRBs. In the literature, the high-energy radiation has been attributed to the
prompt emission \citep[e.g.,][]{2009ApJ...706L.138A},
to pure synchrotron radiation from the external shock
\citep{2009ApJ...706L..33G,2010MNRAS.403..926G,2010MNRAS.409..226K},
to a combination of external synchrotron photons and synchrotron self-Compton emission
\citep{2013ApJ...771L..13T,2013ApJ...771L..33W,2013ApJ...773L..20L,2013ApJ...776...95F},
and to a superposition of
the prompt and afterglow emission \citep{2011MNRAS.415...77M}.
Another possibility is that some of the high-energy emission
stems from prompt photons being Compton scattered to higher
energies by the afterglow-emitting electrons
\citep{2005ApJ...618L..13B,2012ApJ...753..178H,2013arXiv1307.2663B,2013ApJ...776...95F}.

Owing to the {\it Swift} observations, it has been discovered that a typical X-ray light curve
begins with a phase of steeply decaying flux,
which is often followed by a shallow decay segment. The late-time
afterglow, on the other hand, can often be explained by the standard synchrotron model
\citep{2006MNRAS.366L..13G,2006MNRAS.369..197F}.
Energy injection to the blast wave
is currently the most popular explanation for the shallow decay phase observed
both in X-ray and optical afterglows \citep{2006MNRAS.366L..13G,2006MNRAS.369..197F,
2006ApJ...642..389N,2006ApJ...642..354Z,2011MNRAS.414.3537P,2012ApJ...758...27L}.
Other models introduced to explain the shallow decay phase include the evolution of
microphysical parameters \citep{2006MNRAS.369.2059P,2006MNRAS.370.1946G},
emission due to an outflow ejected before the prompt GRB \citep{2009ApJ...690L.118Y,
2012MNRAS.422..393B}, an off-axis viewing angle of the jet \citep{2006ApJ...641L...5E},
dust scattering of X-rays \citep{2007ApJ...660.1319S}, late prompt emission
\citep{2007ApJ...658L..75G,2011ApJ...732...77M}, and an adiabatic evolution of the shock
following a radiative phase \citep{2007ApJ...664..384D}.
It has also been suggested that the main contribution to the afterglow could come from
a long-lived reverse shock, which may also explain the shallow decay phase in
the X-ray afterglows \citep{2007ApJ...665L..93U,2007MNRAS.381..732G}.

Models aiming to explain all the different slopes seen in the light curves
have also been presented, including
accretion of different layers of the progenitor star \citep{2008Sci...321..376K} and
the curvature effect that is usually only invoked to explain the early steep X-ray decay
\citep{2008ApJ...683..900Q}.

The evolution of the GRB blast wave is described well by the self-similar
solution by \citet{1976PhFl...19.1130B}, which is valid in the deceleration phase while
the blast is still highly relativistic.
The evolution in the late non-relativistic phase is given by the Sedov-Taylor solution
\citep{1959sdmm.book.....S,1950RSPSA.201..159T}.
A mechanical model of the relativistic blast
ensuring mass, energy, and momentum conservation has been presented by
\citet{2006ApJ...651L...1B}, and it is nearly identical to the Blandford-McKee
solution at late times after the shock has started to decelerate. However, the mechanical
model gives an accurate description of the blast also before
the deceleration time, while earlier models unphysically assume an
equal pressure at the forward and reverse shock.
The evolution of the shell in the mildly relativistic phase
can be found by means of hydrodynamic simulations, which can then be
coupled to a radiation code to find the radiation spectrum from the shock.
Such simulations can also be applied to calculate the afterglow emission for
an observer with an off-axis viewing angle \citep{2010ApJ...722..235V}.

Results of one- and two-dimensional hydrodynamic simulations of the blast wave have been presented
by \citet{1999ApJ...513..669K}, \citet{2007MNRAS.376.1189M}, \citet{2009A&A...494..879M},
and \citet{2010ApJ...716.1028R}
but without discussing the radiation mechanism of the afterglow. A calculation of the
synchrotron radiation from the blast has been coupled to the hydrodynamic simulations
of \citet{2002MNRAS.332..144D}, \citet{2009ApJ...698.1261Z}, 
\citet{2010MNRAS.403..300V,2011MNRAS.410.2016V}, and \citet{2011ApJ...738L..23W}.
However, none of these works calculate the afterglow component due to Compton
scattering, which is expected to appear at high energies.

Simulations including an accurate treatment of both synchrotron and
Compton processes, as well as pair production,
have been presented by \citet{2009A&A...507..599P} (PM09),
who use the solution of \citet{1976PhFl...19.1130B} to evaluate
the evolution of the emitting shell.
The code developed by PM09 is similar to the one presented in this paper,
but it does not account for the electron heating due to synchrotron self-absorption.

For the first time, we present simulations of afterglow emission
from the forward shock with a relativistic kinetic code that treats
synchrotron emission and absorption, Compton scattering, and
electron-positron pair production/annihilation
in a self-consistent way. The kinetic
equations determining the time evolution of the electron
and photon distributions are solved simultaneously
at each timestep. We
also consider electron heating due to synchrotron self-absorption,
which shapes the electron distribution at low energies.

Our treatment accounts for the fact that electrons injected
at different times also have different cooling histories.
For example, the magnetic field that determines the synchrotron
cooling rate evolves while the electrons are cooling. It follows
that there are no sharp cooling breaks in the electron distribution
\citep{2014ApJ...780...82U}.

The current version of the code applies a one-zone model of the
emission region. It does not account for the different
locations of the particles behind the shock and assumes a
constant magnetic field throughout the shell.
A more accurate treatment
of synchrotron emission would require a model of the magnetic
field structure behind the shock. Also, knowledge of the spatial photon and
electron distributions is required for an exact calculation of
Compton scattering.

As an example of the applications of the code,
we report the results of a simulation where the
afterglow-emitting electrons interact with an external source
of photons roughly corresponding to prompt GRB emission.
The shocked electrons are expected to upscatter a small fraction of the prompt photons to
GeV$-$TeV energies as long as the prompt emission overlaps with the
shocked electrons
\citep{2005ApJ...618L..13B,2005ApJ...629..334F}. Some of the high-energy
photons then produce pairs with the prompt MeV photons, which
in turn are able to scatter radiation to higher energies.

In addition, we compare the forward shock emission in a wind environment with
the emission in a constant-density ISM.
The results indicate that a power-law electron distribution
can thermalize at low energies thanks to synchrotron self-absorption heating
in a wind medium with a typical density structure expected from
the surroundings of a Wolf-Rayet star. Along with the ambient density, the
importance of thermalization is mainly determined by the fraction of
shock-generated energy given to the magnetic field. Our simulations imply
that the thermalized electrons are unlikely to produce an observable
signature in the afterglow spectrum.

In our final example, we study the difference between the forward shock
radiation due to Maxwellian and power-law electron injection. The standard
afterglow model assumes that the injection function is a pure power law,
even though a large fraction of the shock-generated energy goes to
a thermal population of electrons. We find that
pure Maxwellian injection can lead to a flattening in the X-ray light curve.
The flux during this phase is found to be very low
compared to {\it Swift}/XRT detections for a constant-density ISM,
but detectable flux levels during the shallow decay may be achieved
in a wind-type environment.

%__________________________________________________________________

\section{Physical model of the afterglow}

\subsection{Hydrodynamic evolution}

The relativistic shell ejected by the GRB central engine initially
propagates into the surrounding medium with a constant Lorentz factor
$\Gamma_0$. After sweeping an external mass $M_0/\Gamma_0$, where
$M_0$ is the initial mass of the ejecta, the shell starts to
decelerate according to a self-similar solution found by
\citet{1976PhFl...19.1130B}.
The self-similar solution is no longer valid if the
reverse shock is long-lived \citep{2012ApJ...761..147U} or if there
is significant lateral spreading of the shell after the jet break time
\citep{1999ApJ...525..737R}. In the rest of the paper,
except for Sections \ref{ch:kin_eq} and \ref{ch:ad},
the quantities
measured in the fluid comoving frame are indicated by primes and the
unprimed quantities are given in the observer frame. However,
the electron Lorentz factor $\gamma$ and dimensionless momentum
\be \label{eq:p_def}
p \equiv \sqrt{\gamma^2-1},
\ee
along with
the electron number density per unit energy ($N(\gamma)$) or momentum ($N(p)$) interval,
are always expressed in the fluid frame but
left unprimed to simplify the notation. In the following,
we assume an adiabatic hydrodynamic evolution of the blast wave,
which means negligible radiation losses.
This assumption is valid even if the electrons cool rapidly
as long as the main fraction of the shock-generated energy is given
to the protons, which do not radiate efficiently.

The bulk Lorentz factor of the shocked fluid evolves with radius $r$
approximately as
\be \label{eq:gamma_r}
\Gamma(r) = \Gamma_0 (r/\RB)^{1-g},
\ee
where the value of the index $g$ depends on the structure of the
ambient medium.
If the medium has a constant density,
\bea \label{eq:g}
g = \left\{ \begin{array}{ll}
1 & \textrm{if} \: r < \RB
\\
5/2 & \textrm{if} \: r \geq \RB,
\end{array} \right.
\eea
where $\RB \equiv \Rd/2^{2/3}$ is the approximate radius at which
the Lorentz factor starts to decrease from its initial value $\Gamma_0$, and
$\Rd$ is the so-called deceleration radius,
i.e., the radius where the
shell has lost half of its initial kinetic energy
and $\Gamma = \Gamma_0/2$ \citep{1992MNRAS.258P..41R}.
According to this definition,
\be \label{eq:Rdec}
\Rd = \left(\frac{3E_0}{4\pi n_0 \mpr c^2 \Gamma_0^2}\right)^{1/3},
\ee
where $E_0$ is the isotropic equivalent energy of the blast wave
after the prompt GRB emission, and $n_0$ the
electron number density of the external medium.

In the case of a wind-type medium with a density profile $n(r)=\Aw r^{-2}$,
where $\Aw$ is a constant giving the number of particles per unit length,
\bea \label{eq:g_wind}
g = \left\{ \begin{array}{ll}
1 & \textrm{if} \: r < \RB
\\
3/2 & \textrm{if} \: r \geq \RB,
\end{array} \right.
\eea
where $\RB \equiv \Rd/4$, and
\be \label{eq:Rdec_wind}
\Rd = \frac{E_0}{4\pi \Aw \mpr c^2 \Gamma_0^2}.
\ee

The shock radius and the Lorentz factor of the shocked ejecta
are related to the fluid comoving time $t'$ according to
\be \label{eq:time}
dt' = \frac{dr}{c\Gamma(r)},
\ee
and the observed time can be obtained from
\be \label{eq:time_obs}
dt = \frac{(1+z)dr}{2c\Gamma^2(r)},
\ee
where $z$ is the redshift of the GRB.
Before the deceleration time, i.e., $r < \Rd$,
\be \label{eq:r_early}
r = \frac{2c\Gamma_0^2 t}{(1+z)}.
\ee
For $r \gg \Rd$,
the radius evolves in time as
\bea \label{eq:r_time}
r = \left\{ \begin{array}{ll}
\left[3 E_0 t/(2\pi n_0 \mpr c (1+z))\right]^{1/4} & \textrm{(ISM)}
\\
\left[E_0 t/(4\pi \Aw \mpr c (1+z))\right]^{1/2} & \textrm{(wind)}
\end{array} \right.
\eea
and the bulk Lorentz factor evolves as
\bea \label{eq:g_time}
\Gamma = \left\{ \begin{array}{ll}
\left[3 E_0 (1+z)^3/(2^{13}\pi n_0 \mpr c^5 t^3)\right]^{1/8} & \textrm{(ISM)}
\\
\left[E_0 (1+z)/(64\pi \Aw \mpr c^3 t)\right]^{1/4} & \textrm{(wind)}.
\end{array} \right.
\eea

One of the main uncertainties of the afterglow model is the strength of the
magnetic field in the emission region.
We adopt the typical approach to the problem and assume that a fraction $\epsB$ of
the shock-generated energy goes into the energy of the
shock-compressed interstellar magnetic field.
The comoving magnetic field $B'$ then evolves in the emission region with $\Gamma$ and $n$ as
\be \label{Bfield}
B' = c\Gamma\sqrt{32\pi \mpr \epsB n},
\ee
determined from the shock jump conditions \citep{1976PhFl...19.1130B}.

\subsection{Distribution of the accelerated electrons}

The forward and reverse shocks are thought to be capable of
accelerating electrons to relativistic
energies according to a power-law distribution \citep[e.g.,][]{2001MNRAS.328..393A}
of the form
\be \label{eq:pl}
\frac{dn}{d\gamma} \equiv N(\gamma) \propto \gamma^{-s}
\ee
for $\gamma_{\min} \leq \gamma \leq \gamma_{\max}$,
where $s$ is the power-law index, and $\gamma$ the random Lorentz factor of an electron
measured in the fluid frame.
However, simulations by \citet{2008ApJ...682L...5S} and \citet{2009ApJ...695L.189M}
show that shock acceleration actually leads
to a hybrid distribution with low-energy Maxwellian electrons connected
to a high-energy power-law tail. An afterglow model that assumes such
a distribution is discussed by \citet{2009MNRAS.400..330G}.

The number of electrons injected at a shock per unit time is obtained by multiplying
the number density of electrons in the downstream frame, $\Gamma n$,
by the volume swept by the shock per unit time,
$4\pi r^2 \beta c$. Dividing this quantity by the volume
of the shell, $4\pi r^2 \Delta R'$, one finds that
the number of injected electrons per unit time and volume is
\be \label{eq:n_el}
\nel' = \frac{\Gamma n \beta c}{\Delta R'}.
\ee

Assuming a pure power-law distribution,
the minimum Lorentz factor determined from the shock jump conditions
\citep{1996ApJ...473..204S} is
\be \label{eq:gmin}
\gamma_{\min} = \epse\frac{s-2}{s-1}\frac{\mpr}{\mel}\Gamma,
\ee
where $\epse$ is the fraction of the shock-generated
energy given to the electrons.
This relation holds for $\gamma_{\min} \ll \gamma_{\max}$.

The maximum Lorentz factor $\gamma_{\max}$ is determined either
by the radiation losses of the electrons, the age of the flow, or
the saturation limit discussed by \citet{2013ApJ...771...54S}, among others.
Typically it has been assumed that the value of $\gamma_{\max}$
and the corresponding synchrotron frequency
are too high to affect the observed afterglow properties.
After the detection of high-energy
($> 100\:\textrm{MeV}$) emission from several GRBs, which may be part
of the synchrotron or inverse Compton component of the afterglow, it has
become more important to determine $\gamma_{\max}$ accurately
to study whether there are electrons at high enough
energies to produce $> 100\:\textrm{MeV}$ synchrotron radiation.

\subsection{Radiation processes}

\subsubsection{Synchrotron emission and self-absorption}

The cooling of the shocked electrons is mainly determined by synchrotron
losses, with a significant contribution from adiabatic cooling
to the distribution of the low-energy electrons.
It is expected that the cooled electrons are distributed
according to $N(\gamma) \propto \gamma^{-s-1}$
above the injection energy $\gamma_{\min}$
\citep{1962SvA.....6..317K}.
If the electrons are cooling to energies below the injection
energy (the fast cooling case),
$N(\gamma) \propto \gamma^{-2}$ for $\gamma < \gamma_{\min}$
as long as the electrons do not thermalize due to self-absorption heating.

During the late stages of a typical afterglow, the magnetic field is
low, and only the most energetic electrons are cooling
(slow cooling). The uncooled
electrons below a critical energy $\gc$ are distributed
as $N(\gamma) \propto \gamma^{-s}$ similarly to the injection function.
The different parts of the electron distribution correspond to
the power-law segments in the emergent radiation spectrum
as described by \citet{2002ApJ...568..820G}.
The characteristic observed synchrotron photon frequency of a relativistic
electron is
\be \label{eq:syn_freq}
\nu(\gamma) = \frac{\Gamma \gamma^2 eB'}{(1+z)2\pi \mel c}.
\ee

The spectral slope of the electron distribution
gradually changes around the cooling
energy $\gc$, which is defined by \citet{1998ApJ...497L..17S}
according to the relation
\be \label{eq:g_crit}
\gc \mel c^2 = P'(\gc)t',
\ee
where
\be \label{eq:synpo}
P'(\gc) = \frac{4}{3}\sT c \gamma^2 \frac{(B')^2}{8\pi}
\ee
is the
synchrotron power of an electron with $\gamma \gg 1$ in the
comoving frame. The relation is based
on the definition that a cooling electron loses an energy
comparable to its own initial energy in the lifetime of the flow.
Correspondingly, the comoving cooling time of an electron with
a Lorentz factor $\gamma$ is
\be \label{eq:t_cool}
\tco' = \frac{6 \pi \mel c}{\sT (B')^2 \gamma}.
\ee

In reality, a sharp break at the energy $\gc$ does not appear
because
the actual electron distribution consists of different electron
populations with their own cooling histories. The radiation
power $P'(\gc)$ evolves in time as the magnetic field $B'$ decreases,
and an integration over $P'(\gc)dt'$ is required to find the exact
energy lost by each electron population.
Both our code and the one by \citet{2009A&A...507..599P}
calculate the shape of the electron distribution according to
the kinetic equation, and the results show that the transition
between the cooled and uncooled segments in the distribution
is very gradual.
This effect is taken into account
by, say, \citet{2014ApJ...780...82U}, who also point out that
the different locations of the electron
populations behind the shock shape the outgoing radiation spectrum.
In our current one-zone treatment, the structure of the electron
distribution and the magnetic field behind the shock are not
resolved.
A multi-zone approach is expected to contribute further to the
curvature of the particle distributions, and
the shape of the resulting spectrum will be a result of the emissions
from varying distances behind the shock by electron populations
with different cooling histories. Additional smoothing of the cooling break
would also be provided by emission from large angles, but the code assumes
at this stage that all the emission comes from the line of sight to the observer.

In addition to being cooled by synchrotron emission, low-energy electrons can be
heated due to self-absorption of the synchrotron
photons \citep[e.g.,][]{1988ApJ...334L...5G}, leading to some thermalization of the electron
distribution. Full thermalization of the low-energy electrons
takes roughly one synchrotron cooling time.
The cooling time for a given electron energy
depends on the magnetic field as $\propto (B')^{-2}
\propto n^{-1}\Gamma^{-2}$
(see Eq. (\ref{Bfield})). Because
the density in a wind-type medium at small radii is considerably
larger than in a typical constant-density ISM,
the magnetic field is also higher and correspondingly the
cooling time is shorter. This implies that electron thermalization
due to self-absorption is more likely to occur in a wind
environment.

\subsubsection{Inverse Compton scattering}

Some of the synchrotron afterglow photons are Compton scattered to higher
energies by the same electrons that emit the synchrotron
radiation. If the prompt GRB photons overlap with the
afterglow-emitting region, some of these photons are also upscattered
by the relativistic electrons up to $>$ TeV energies.
Compton scattering depends on the photon density and consequently
the geometry of the emission region, which is discussed in section
\ref{ch:em_size}.

In the Thomson regime, the location of the inverse Compton peak
is determined by the scattering of the peak photons
of the $\nu F_{\nu}$ Band spectrum against
the peak electrons of the $\gamma^2 N(\gamma)$ distribution.
However, if the scattering of the peak photons
against the peak electrons is suppressed due to the Klein-Nishina (K-N)
effect, the peak of the Compton scattered photons is located
at a lower energy. Noting that the location of the peak should
correspond to the highest possible energy that can be transferred to
a photon from a peak electron with a Lorentz factor $\gamma = \gp$,
the observed peak energy becomes
\be \label{eq:ic_peak}
\Epc \sim \frac{2\Gamma \gp \mel c^2}{(1+z)}.
\ee

\subsubsection{Pair production}\label{chap:pp}

Electron-positron pair production and annihilation may have
some impact on the observed afterglow spectra, especially
if a fraction of the prompt photons
is Compton scattered to high energies, after which they are able
to produce pairs with the unscattered GRB photons.
Defining the dimensionless photon energy as
\be
x \equiv \frac{h\nu}{\mel c^2},
\ee
high-energy photons of energy $\xhe$ produce pairs with target
photons exceeding the threshold energy
\be \label{eq:x_thr}
\xt \sim \frac{4\Gamma^2}{\xhe(1+z)^2}.
\ee
This value of $\xt$ is obtained from the threshold condition
$\xt \xhe (1+z)^2 (1-\cos \theta) > 2$, where $\theta$ is the
half-angle inside which the upscattered photons propagate.
The condition states that the invariant four-product of the
photon momenta should be greater than 2 to be able to produce
two leptons with zero kinetic energy in the center of momentum
frame. With a typical angle $\theta \sim 1/\Gamma$, one
obtains Eq. (\ref{eq:x_thr}).
For target photons just above $\xt$, the optical
depth for pair production reaches its maximum value
\citep[e.g.,][]{1988ApJ...335..786Z}
\be \label{eq:tau_gg}
\tau_{\gamma\gamma} \approx \frac{\sT}{5}\nph r(1-\cos \theta)
\approx \frac{\sT}{5}\nph\frac{r}{2 \Gamma^2},
\ee
where $\nph$ is the number density of the target photons
above the threshold energy.

For hard bursts, the pair production opacity is maximal for high-energy
photons for which the threshold energy is $\xt \sim \xpk$, where $\xpk$
is the peak energy of the $\nu F_{\nu}$ prompt GRB spectrum. The distribution
of prompt photons is typically described by the so-called Band function
\citep{1993ApJ...413..281B}
\bea \label{eq:Band}
\frac{dn}{dx} \propto \left\{ \begin{array}{ll}
x^{\alpha}\exp(-x/\xpk) & \textrm{if} \: x < \xpk
\\
x^{\beta} & \textrm{if} \: x \geq \xpk,
\end{array} \right.
\eea
where $dn/dx$ is the photon number density per dimensionless energy
interval. Because the number density of the photons of energy $x$ is
$\nph \sim x\:dn/dx \propto x^{\alpha+1}$ or $\nph \propto x^{\beta+1}$
and the opacity is proportional to $\nph$ according to Eq. (\ref{eq:tau_gg}),
one finds that $\tau_{\gamma\gamma} \propto \xt^{\alpha+1} \propto \xhe^{-\alpha-1}$  for
$\xhe > 4\Gamma^2\xpk^{-1}(1+z)^{-2}$ and $\tau_{\gamma\gamma} \propto \xt^{\beta+1}
\propto \xhe^{-\beta-1}$
for $\xhe \leq 4\Gamma^2\xpk^{-1}(1+z)^{-2}$. For a typical GRB with $\alpha \sim -1$,
the opacity remains relatively constant at $\xhe > 4\Gamma^2\xpk^{-1}(1+z)^{-2}$. One also
notes that the opacity increases for softer and decreases for harder bursts.

The number density $\nph$ is dominated by the prompt photons
while they are going through the shell of ejecta and can be
estimated as
\be \label{eq:nph}
\nph \sim \frac{L}{4\pi r^2 c \Epk},
\ee
which is obtained by dividing the prompt GRB luminosity $L$
by the volume covered by the photons per unit time, $4\pi r^2 c$, and
by the average energy of a photon, which we now assume to be the
prompt peak energy $\Epk$.

Using the expression for $\nph$ together with Eq. (\ref{eq:tau_gg}),
one finds that $\tau_{\gamma\gamma} \propto r^{-1}\Gamma^{-2}$
for high-energy photons for which the threshold energy is $\Epk$.
The prompt photons mainly overlap with the ejected shell during the coasting
phase when $\Gamma = \Gamma_0$ and $r \propto t$. In this case, the optical depth
evolves as $\tau_{\gamma\gamma} \propto r^{-1} \propto t^{-1}$. After the
deceleration time, $r \propto t^{1/4}$ and $\Gamma \propto r^{-3/2}$
in the ISM case,
from which it follows that $\tau_{\gamma\gamma} \propto r^{2}
\propto t^{1/2}$. The probability for pair production is thus expected
to decrease during the coasting phase but increase after $\RB$
if the prompt photons are still passing through the emission region.
The rising opacity is due to the increasing angular spread of the emitted
high-energy photons, which offsets the decrease due to the declining
target photon density. Taking into account the beaming of the prompt
radiation, the suppression of pair production is not exponential even at
$\tau_{\gamma\gamma} > 1$. In this case there always exists an escape
cone at sufficiently small angles, within which high-energy photons can
escape.
In a wind medium, where $\Gamma \propto r^{-1/2}$,
$\tau_{\gamma\gamma} \propto r^0$ after the deceleration time, that is to say,
the optical depth remains constant.

The above discussion on the pair production opacity
applies to the high-energy photons for which the threshold energy of target
photons is $\Epk$, and the energy of the former is generally changing with time.
However, if one assumes the canonical photon index $\alpha \sim -1$,
the photon number density $\nph \propto x^{\alpha+1} = x^0$ below the peak energy.
In this case, the density of target photons below $\Epk$ is still given by
Eq. (\ref{eq:nph}) and the pair production opacities obtained above apply to any
high-energy photon for which the threshold energy is below $\Epk$.

Once the prompt photons have crossed the emission region, the
synchrotron emission provides the target photons for pair production
with upscattered high-energy photons. The number of synchrotron photons
providing the pair production opacity is proportional to the number
of electrons emitting at $\xt$. The energy of these electrons can be
estimated from $\xt = 4\Gamma^2\xhe^{-1}(1+z)^{-2} = \xs
\propto \Gamma^2 \gamma^2$ (Eq. (\ref{eq:syn_freq})), where $\xs$ is the dimensionless synchrotron
photon energy. This yields that the electron energy corresponding to
the threshold photon energy is $\gt$ = constant for a fixed $\xhe$.
In the slow cooling regime the number of these electrons is
$\Net \propto r^3 (\gt/\gamma_{\min})^{-s+1} \propto r^3 \Gamma^{s-1}$ (Eq. (\ref{eq:gmin})).
For example, in the ISM case with $s=2$ one gets an optical depth
$\tau_{\gamma\gamma} \propto r^2 \Gamma \propto t^{1/8}$ (Eqs. (\ref{eq:r_time}) and (\ref{eq:g_time})); i.e.,
the pair production opacity is increasing in time, although very slowly.

%__________________________________________________________________

\section{Numerical treatment}

\subsection{Kinetic equations}\label{ch:kin_eq}

The numerical code we use to simulate GRB afterglows is based on the code developed by
\citet{2009ApJ...698..293V}, which
has successfully been applied to, say, modeling prompt GRB
\citep{2011ApJ...738...77V} and black hole emission
\citep{2011MNRAS.414.3330V,2013MNRAS.430.3196V}.
The code calculates the time-evolving particle distributions
in an astrophysical plasma by solving the full relativistic kinetic
equations for each particle species without any energy limitations.
This means that the equations include both differential and integral
terms, depending on the nature of each radiation process. If the energy
losses of a particle are approximately continuous, differential terms
are used to describe the process. An integral term
is necessary if the particle loses a considerable fraction of its energy due
to a single interaction. In this section,
all quantities except for $r$ and $\Gamma$ are expressed in
the fluid comoving frame and are left unprimed for ease of notation.

The equations for photons
and electrons in the fluid comoving frame both
have the same general form
\citep[for a derivation, see, e.g.,][]{1970RvMP...42..237B}
\be \label{eq:kin_gen}
\frac{\partial N}{\partial t} = 
\nsyn+\ncs+\npp+\nad+\Qi-\frac{N}{\tesc}-\frac{N}{\tex},
\ee
where $N \equiv N(x)$ (for photons) or $N \equiv N(\gamma)$ (for electrons)
is the particle number density distribution in the shocked region per dimensionless
photon or electron energy interval.
The subscripts syn, cs, pp and ad correspond to synchrotron emission and absorption,
Compton scattering, pair production and adiabatic cooling, respectively.
Each process produces a term on the right-hand side of the equation
\citep[for a detailed description of the numerical treatment of the processes, see][]
{2009ApJ...698..293V}.
The source term $\Qi$ gives the contribution of newly injected particles.
The electron injection function is assumed to be a pure power law (Eq. (\ref{eq:pl}))
in most examples presented in this paper. However, the injected electrons
have a Maxwellian distribution in some of the examples in section \ref{ex3}.
The term $N/\tesc$ accounts for the escape of particles from the emission
region and $N/\tex$ gives the dilution of the particle densities due to the
expansion of the emission region, $\tesc$ and $\tex$ being the
characteristic time scales for these processes.
The adiabatic cooling and density dilution terms are discussed
in section \ref{ch:ad}.
The photon escape time $\tesc$ is evaluated by solving the plane-parallel
radiative diffusion equation and is equal to
\be
\tesc = \frac{3 \Delta R}{4 c}\left[1 +
\frac{1-{\rm e}^{-\tstar}}{\sqrt{\epsilon}(1+{\rm e}^{-\tstar})}\right],
\ee
where $\Delta R$ is the comoving radial dimension of the emission region,
$\tstar \equiv 2\Delta R \sqrt{(\asyn+\app)(\asyn+\app+\scs)}$ gives the
effective optical depth and $\epsilon \equiv (\asyn+\app)/(\asyn+\app+\scs)$
is the probability for photon absorption, $\asyn$, $\app$ and $\scs$ being the
extinction coefficients due to synchrotron (syn) and pair production (pp)
absorption and Compton scattering (cs).

The code recalculates the radius $r$ and bulk Lorentz factor $\Gamma$
at each timestep according to Eqs. (\ref{eq:gamma_r}) and (\ref{eq:time})
and evaluates the observed time from Eq. (\ref{eq:time_obs}).
The other time-dependent quantities such as $B$, $\nel$ and $\gamma_{\min}$
(Eqs. (\ref{Bfield}), (\ref{eq:n_el}) and (\ref{eq:gmin})) can then
be evaluated to obtain, for instance, the updated synchrotron emissivities.

The main difference between our code and the similar one described in PM09
is that we include the second-order differential
terms that correspond to electron heating and diffusion due to synchrotron
and Compton processes. The
differential terms in the
kinetic equation for electrons take the form
\be
\dot{N}(\gamma) = -\frac{\partial}{\partial\gamma}\left[\Ae(\gamma)N(\gamma)-\Be(\gamma)\frac{\partial N(\gamma)}{\partial\gamma}\right]
\ee
for all continuous processes, such as synchrotron processes, Compton scattering in
the Thomson regime and adiabatic cooling, and the coefficients $\Ae$ and $\Be$
depend on the process of interest.
Synchrotron self-absorption also contributes a first- and second-order
differential term in the electron equation, where the coefficients $\Ae$ and $\Be$
depend on the number density of photons and the synchrotron emissivity of an
electron.

The kinetic equations are discretized
and solved on finite grids of photon and electron/positron energies.
Both the photon and electron energy grids consist of 200 points, with the
photon grid ranging from $x=10^{-11}$ to $x=10^8$ ($E = 5 \times 10^{-6}\:\textrm{eV}$
to $E = 50\:\textrm{TeV}$) and the electron grid ranging
from $p=10^{-4}$ to $p=10^8$.

Because of the high-energy boundary of the electron grid,
the maximum energy of the electrons is evaluated as
\be \label{eq:gmax_ev}
\gamma_{\max} = \min\left(10^8, \sqrt{\frac{3\pi e}{\sT B'}}\right).
\ee
The maximum Lorentz factor $\gamma_{\max} = \sqrt{3\pi e/(\sT B')}$
is obtained by comparing the acceleration time to the synchrotron
cooling time. This is the value used by our code as long as it does not
exceed $\gamma = 10^8$, the highest electron energy of the grid.

\subsection{Size of the emission region} \label{ch:em_size}

In the simulations presented here, we consider
only the radiation from the forward shock.
The current code applies a one-zone approximation by
assuming that the particles behind the shock front are
homogeneously distributed and that the magnetic field
has a uniform value in the region of interest.

The emission region is the spherical shell between
the forward shock and the contact discontinuity that separates the shocked
external medium and the shocked GRB ejecta. The Lorentz factor of
the forward shock is $\Gsh = \sqrt{2}\Gamma$ \citep{1976PhFl...19.1130B},
which means that the contact discontinuity is moving away from the
shock at a velocity $c/3$. This leads us to define the radial extent
of the shell as
\be \label{eq:dR}
\Delta R' = ct'/3
\ee
and the comoving volume of the emission region becomes
\be \label{eq:vol}
V' = 4\pi r^2 ct'/3.
\ee

Here it is not taken into account that the shell is most likely not spherical
but a fraction of a narrow jet. However, the luminosity per unit solid angle
is the same in both of these cases before the so-called jet break time, when
the beaming angle of the radiation exceeds the opening angle of the jet.

\subsection{Adiabatic cooling and density dilution}\label{ch:ad}

In addition to the radiation processes discussed in this paper, the code accounts for
adiabatic particle cooling due to the spreading emission region.
In this section, all quantities except for $r$ are given in the
fluid frame and are left unprimed.
The adiabatic cooling term for electrons is
\be \label{eq:ad}
\nad \equiv \frac{\partial N(\gamma)}{\partial t} = -\frac{\partial}{\partial \gamma}\left[\gad N(\gamma)\right],
\ee
where $\gad$ is the cooling rate due to adiabatic expansion.
The cooling rate is obtained from the pressure $P$ and
internal energy $E$ of the electrons.
For a monoenergetic population of $\Ne$ electrons
of energy $\gamma$ in a volume $V$,
\be
P = \frac{1}{3}\frac{\Ne}{V}\gamma\bel^2,
\ee
where $\bel$ is the random velocity of an electron in units of $c$,
and
\be
E = (\gamma-1)\Ne.
\ee
From the first law of thermodynamics, $dE=-P\:dV$, one then obtains the cooling rate
\be \label{eq:thermo}
\gad = -\frac{\gamma\bel^2}{3} \frac{d\:\ln\:V}{dt}.
\ee

Evaluating the time derivative of the comoving volume (Eq. (\ref{eq:vol})), the cooling rate
for a constant value of $g$ 
(see Eqs. (\ref{eq:g})$-$(\ref{eq:Rdec_wind})) becomes
\be \label{eq:adcool}
\gad = -\frac{\gamma\bel^2}{3t} \left(\frac{2}{g}+1\right),
\ee
where $t$ is the comoving lifetime of the shock. 
Because the simulation extends from
the coasting phase of the relativistic shell to the deceleration phase, the
value changes from $g=1$ to $g=5/2$ (constant density) or $g=3/2$ (wind)
according to Eqs. (\ref{eq:g}) and (\ref{eq:g_wind})
during the deceleration. This change is gradual in reality, but our approximation
of the Blandford-McKee solution has the side effect that the time derivative of the
volume and the cooling term of Eq. (\ref{eq:adcool}) are discontinuous at $r=\RB$.
In order to avoid the discontinuity,
the cooling rate used in the simulations is
\be
\gad = -\frac{\gamma\bel^2}{3} \left(\frac{2}{\tad}+\frac{1}{t}\right),
\ee
where
\bea \label{eq:tad}
\tad \equiv \left\{ \begin{array}{ll}
t & \textrm{if} \: r \leq \RB
\\
\tB+g(t-\tB) & \textrm{if} \: r > \RB,
\end{array} \right.
\eea
$\tB$ being the comoving time corresponding to the
transition radius $\RB$. This approach guarantees that the adiabatic
cooling term is continuous and that the solution is accurate both at
early and late times.

The term giving the dilution of the electron density
(the last term on the right-hand side of Eq. (\ref{eq:kin_gen}))
is obtained by considering a constant number of particles with
density $n = \int N(\gamma)d\gamma$ in an expanding volume $V$:
\be
\frac{d}{dt}\left(nV\right)=0.
\ee
This relation is equivalent to
\be
\frac{dN(\gamma)}{dt}=-N(\gamma)\frac{d\:\ln\:V}{dt}=-\frac{N(\gamma)}{\tex},
\ee
where
\be
\tex=\left(\frac{2}{\tad}+\frac{1}{t}\right)^{-1}
\ee
and $\tad$ is given by Eq. (\ref{eq:tad}).

\subsection{Test simulations}

\begin{figure}[!]
\begin{center}
\includegraphics[width= \columnwidth]{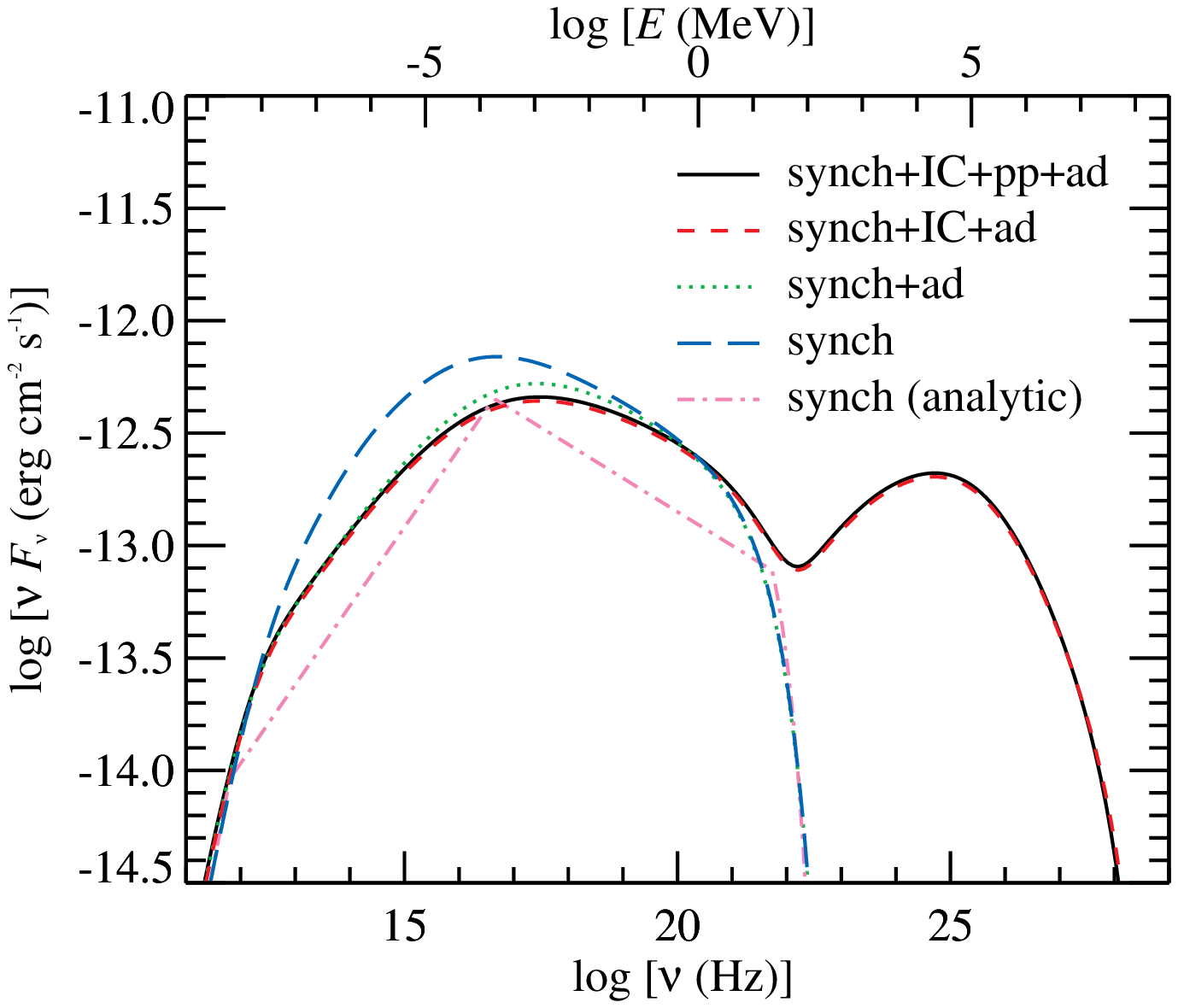}
\includegraphics[width= \columnwidth]{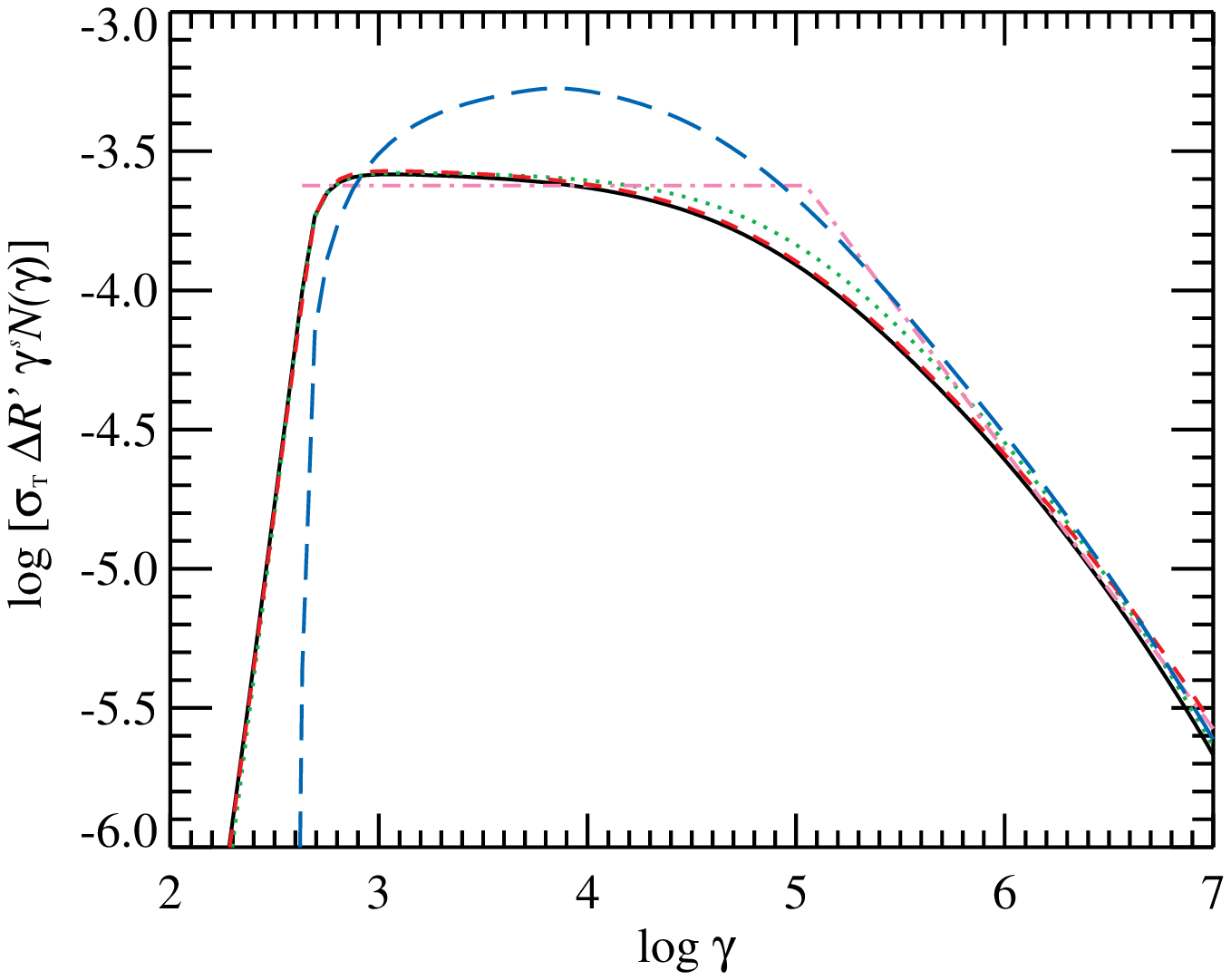}
\caption{\label{fig:1}
Simulated afterglow spectrum in the observer frame ({\it top panel})
and the electron distribution in the fluid frame ({\it bottom panel})
at radius $r = 3.4 \times 10^{17}\:\textrm{cm}$. The particle distributions
have been obtained with four different combinations of radiative
and adiabatic processes. Synchrotron emission/absorption, Compton
scattering, pair production/annihilation and adiabatic energy
losses are indicated in the figure by the abbreviations syn, IC, pp
and ad, respectively. In the simulation where only synchrotron processes
are included, self-absorption heating of the electrons is not accounted
for. The analytic synchrotron solution by
\citet{1998ApJ...497L..17S} is also shown in the figures.
The simulation parameters are $E_0 = 10^{53}\:\textrm{erg}$, $n_0 = 1\:\textrm{cm}^{-3}$,
$\epse = 0.1$, $\epsB = 10^{-3}$, $\Gamma_0 = 400$ and
$s = 2.3$. The high-energy cutoff of the electron distribution has
a constant value $\gamma_{\max} = 4 \times 10^7$, and the GRB
is located at a redshift $z = 1$.
In the {\it bottom panel},
$N(\gamma) \equiv dn/d\gamma$
is the number density of electrons per dimensionless energy interval.
The distribution function has been multiplied by
$\sT \Delta R'$ to find the Thomson optical depth and by $\gamma^s$
in order to visualize which electrons are cooling: a flat segment
in the distribution means that the slope is the same as that
of the injection function, demonstrating that the electrons
are uncooled.
The figure may be compared with Figs. 4 and 5 of PM09.
} 
\end{center} 
\end{figure}

In order to test the validity of our code,
we have compared our simulation results with those obtained by PM09
who have developed a similar numerical code. The photon spectra
and electron distributions at $r = 3.4 \times 10^{17}\:\textrm{cm}$ (Fig. \ref{fig:1})
calculated by our code should be
compared with Figs. 4 and 5 of PM09, which present the particle
distributions obtained with the same set of parameters.
In these simulations, a constant value of the maximum electron
energy, $\gamma_{\max} = 4 \times 10^7$, is assumed.
Our Fig. \ref{fig:1} shows the
results of several simulations with different combinations
of radiation processes. All the photon spectra in this section
and the rest of the paper are presented in the observer frame,
with energies
boosted by a factor of $2\Gamma/(1+z)$ from the flow frame.

The two codes produce electron distributions with
very similar shapes: the cooled electrons populate the high-energy
end of the distribution going as $N(\gamma) \propto \gamma^{-s-1}$,
with the slope of the distribution gradually approaching the slope
of the injection function at lower energies. The electrons below
$\gamma_{\min} \sim 600$
have a nearly flat $N(\gamma)$ distribution in PM09, whereas we
obtain a much steeper decline of the electron density. To study
whether this difference could be due to the inclusion of synchrotron
self-absorption heating, we turned off this process
for a test simulation, but this had no visible impact
on the electron distribution. Because the shell is in the deceleration phase and $\gamma_{\min}$
is decreasing, such a steep cutoff may appear if $\gamma_{\min}$
declines faster than the electrons are cooled adiabatically.
However,
the electrons below $\gamma_{\min}$ do not carry a large fraction of the total
electron energy and have no observable impact on the afterglow
emission.

The normalization of our electron distribution
is lower by about half an order of magnitude than that obtained
by PM09, but the origin of this difference is unclear so far.
In our radiation spectrum, the spectral slopes and
the positions of the peak and break frequencies appear identical
to those in PM09, but the normalization of the flux is again
slightly different. It is notable that the relative magnitudes
of the synchrotron and inverse Compton components are clearly
different in the two simulations. A possible cause for this
discrepancy is the slight difference in the geometries assumed
in the simulations.

The electron distribution resulting from synchrotron cooling
without adiabatic cooling or self-absorption heating is also
presented in Fig. \ref{fig:1}, showing that especially the
low-energy electrons are strongly affected by
adiabatic cooling. The distribution around the cooling break is highly
curved without this process, and there is a clear cutoff below the injection energy
$\gamma_{\min}$ since the electrons no longer have a way to cool
to lower energies.

The radiation spectrum from this simulation
is also presented in Fig. \ref{fig:1}. The curved
part of the electron distribution produces a corresponding curved segment in the
radiation spectrum. The low-energy end of the spectrum, which basically consists of
radiation from monoenergetic electrons at $\gamma_{\min}$, has the
same shape in all cases. All in all, adiabatic cooling has a clear
observable effect on the emergent spectrum but the contribution
of synchrotron self-absorption heating seems negligible.

The analytic synchrotron solution by \citet{1998ApJ...497L..17S}
is included in Fig. \ref{fig:1} for comparison with the numerical
solution. It is clear that a spectrum consisting of pure
power-law segments with sharp breaks deviates from
the actual curved synchrotron radiation component, and fitting
the simple analytic model to observed afterglow spectra may
lead to an inaccurate determination of the forward shock parameters.

%______________________________________________________________

\section{Examples}

\subsection{High-energy emission due to Compton scattering of MeV photons}\label{ch:hes}

\begin{figure*}
\begin{center}
\includegraphics[width= 0.73\columnwidth]{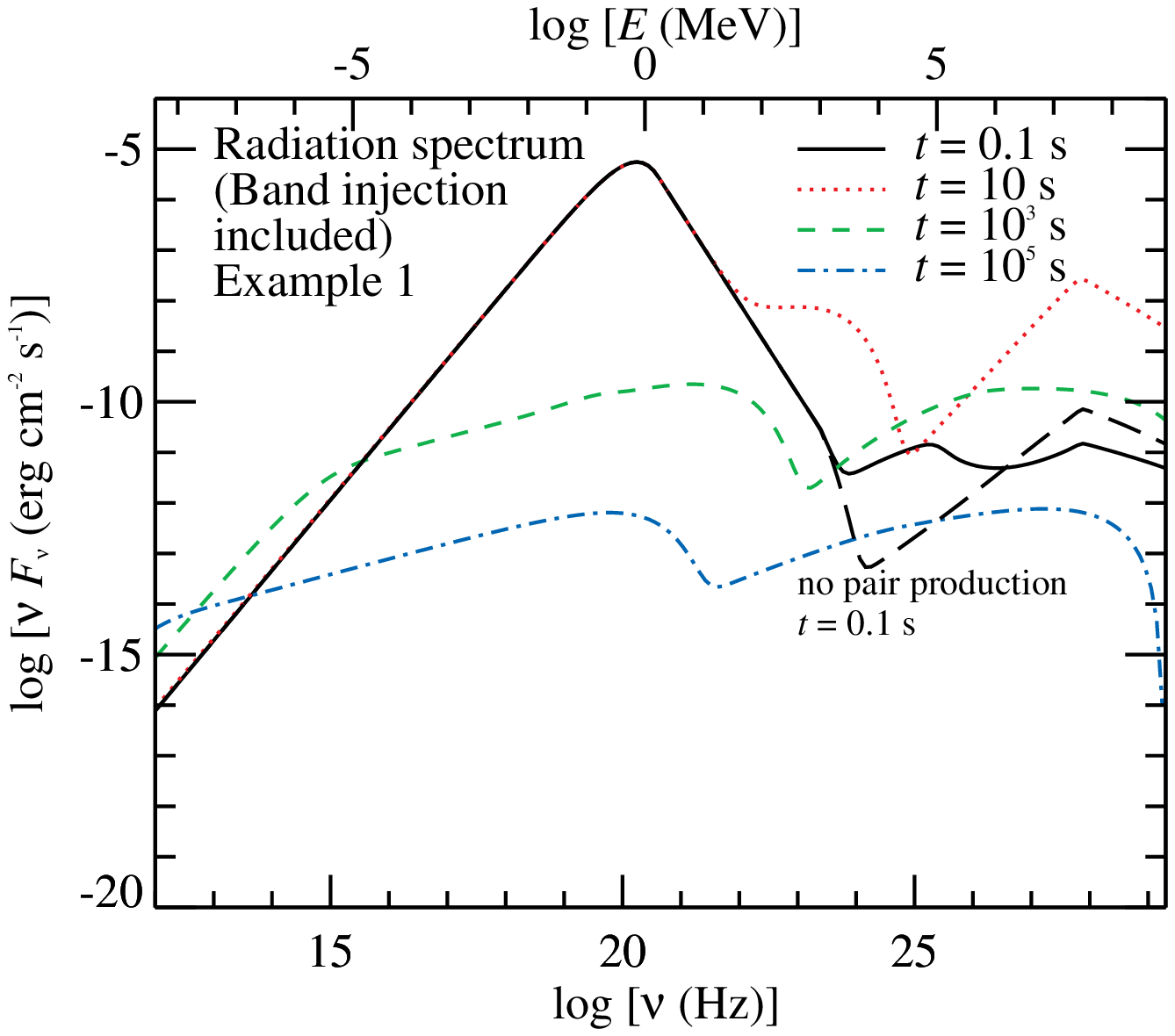}
\includegraphics[width= 0.73\columnwidth]{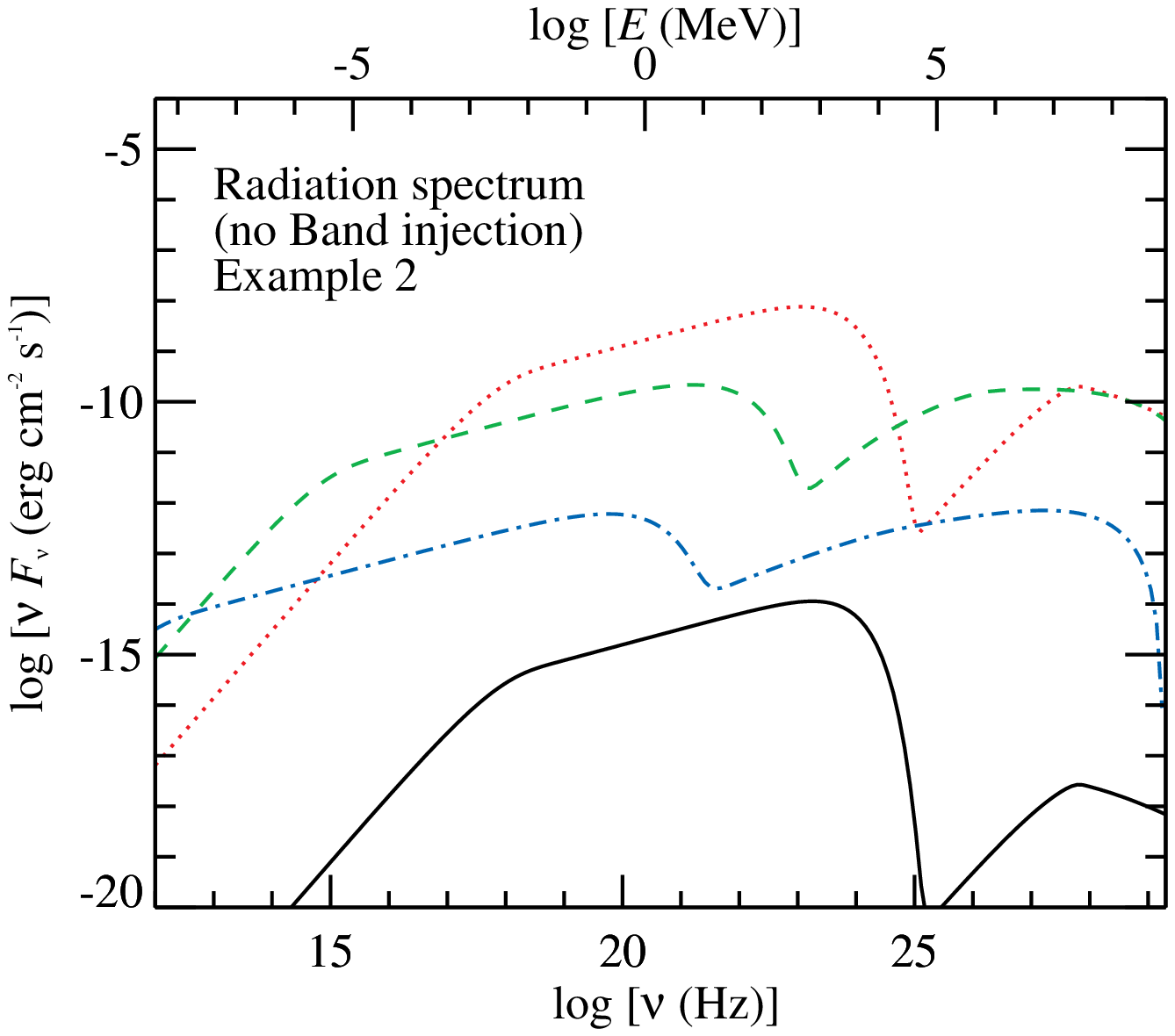}
\includegraphics[width= 0.73\columnwidth]{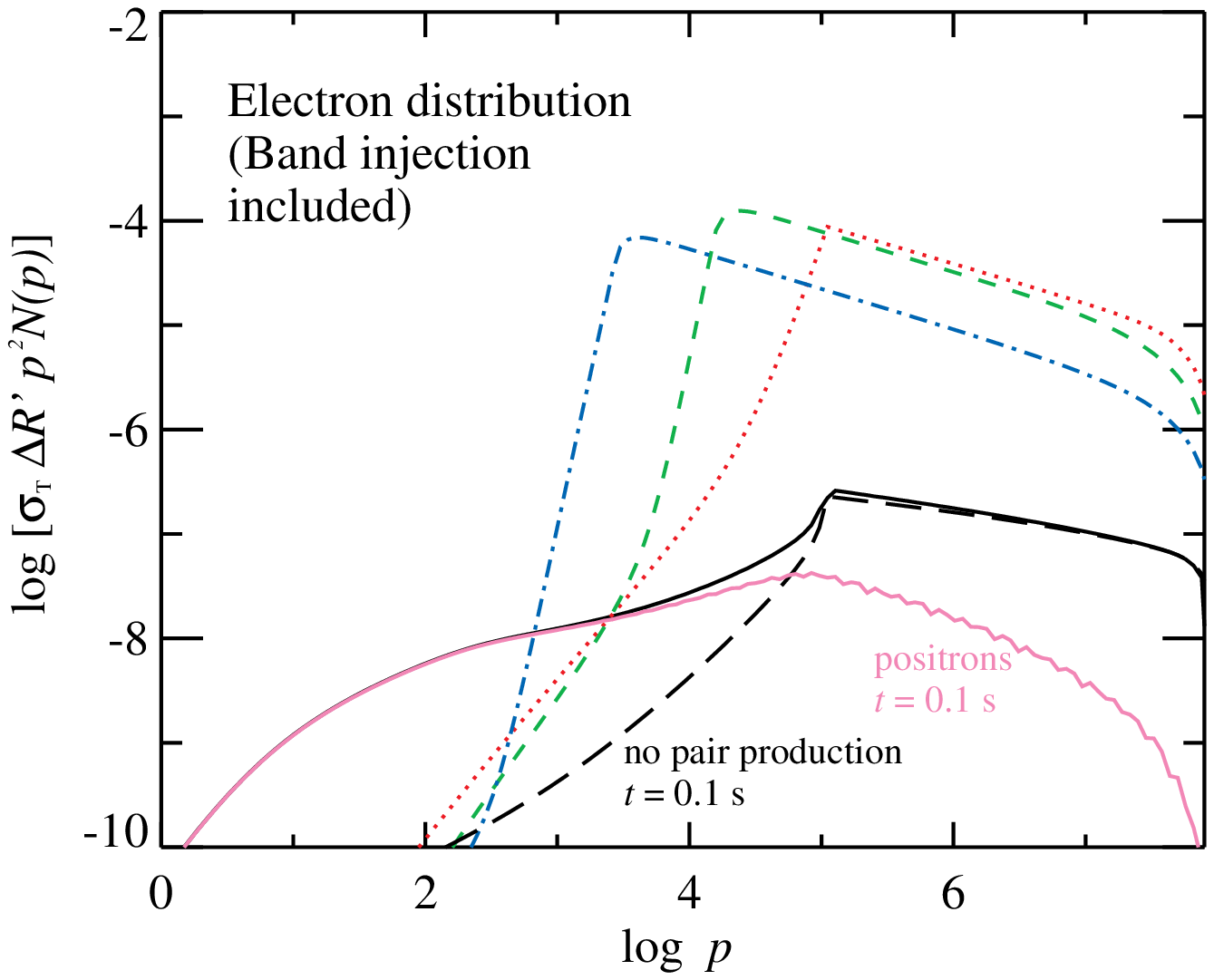}
\includegraphics[width= 0.73\columnwidth]{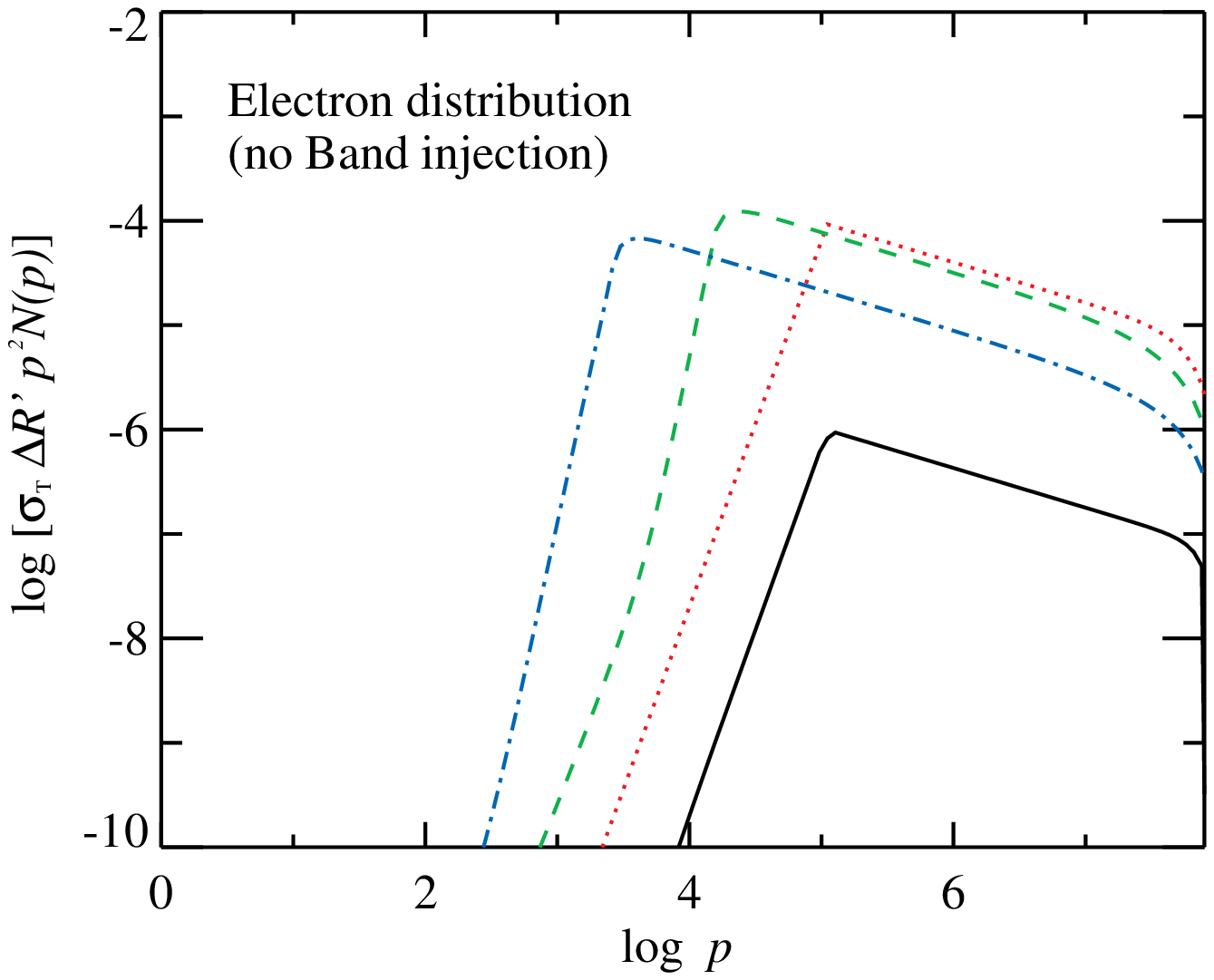}
\caption{\label{fig_he}
{\it Left panels:} the time-evolving observer frame photon spectrum
without absorption on extragalactic background light ({\it top panel})
and electron distribution ({\it bottom panel})
from the forward shock together with an additional
photon injection term that roughly represents prompt GRB emission.
The electron distributions are plotted as a function of
the dimensionless electron momentum $p$ (Eq. (\ref{eq:p_def})) instead of $\gamma$
(note that $p \rightarrow \gamma$ for $\gamma \gg 1$).
The number of electrons per
unit dimensionless momentum is $N(p) \equiv dn/dp$.
The parameters of the forward shock emission are
$E_0 = 1.4 \times 10^{55}\:\textrm{erg}$, $n_0 = 3 \times 10^{-2}\:\textrm{cm}^{-3}$,
$\epse = 0.2$, $\epsB = 10^{-6}$, $\Gamma_0 = 800$
and $s = 2.4$.
The injected photons are distributed according to a Band function,
with the parameters
$\alpha=-0.61$, $\beta=-3.8$ and $\Epk = 730\:\textrm{keV}$
(as defined in Eq. (\ref{eq:Band})). The Band function cuts off
at $E = 1\:\textrm{GeV}$ and the injection lasts for $t = 22\:\textrm{s}$,
after which the photon luminosity decreases exponentially with time.
The GRB takes place at $z = 1.8$. The black long-dashed lines
show the photon and electron distributions at $t = 0.1\:\textrm{s}$ without pair production,
and the solid magenta line in the {\it bottom left panel} represents
the positron distribution at $t = 0.1\:\textrm{s}$.
{\it Right panels}: the evolution of the observer frame photon spectrum
({\it top panel}) and electron
distribution ({\it bottom panel}) without photon injection. The forward
shock parameters are same as in the simulation presented in the {\it left panels}.
}
\end{center} 
\end{figure*}

As an example of an important application of the code, we
present the results of two simulations where the injected
electrons interact with an external source of photons
distributed according to a Band function (Eq. (\ref{eq:Band})),
which represents a typical prompt GRB spectrum.
Some of the GRB photons are scattered to higher energies by
the shock-accelerated electrons, producing an additional
spectral component extending up to TeV energies in the
observer frame. The high-energy end of the spectrum can then
be further modified as some of the high-energy photons
produce electron-positron pairs with the MeV
photons.
The $\gtrsim \:\textrm{TeV}$ component may be observable in the case
of a low-redshift GRB; otherwise the very high-energy gamma-rays
are absorbed by the extragalactic background light.
The simulations presented here demonstrate that our code can be applied
to study whether the $> 100\:\textrm{MeV}$ GRB emission is due
to Compton scattering of the prompt photons at the external shocks.
However, further modifications of the code are needed to improve
especially the treatment of the early high-energy emission.

\begin{figure}[!]
\begin{center}
\includegraphics[width= \columnwidth]{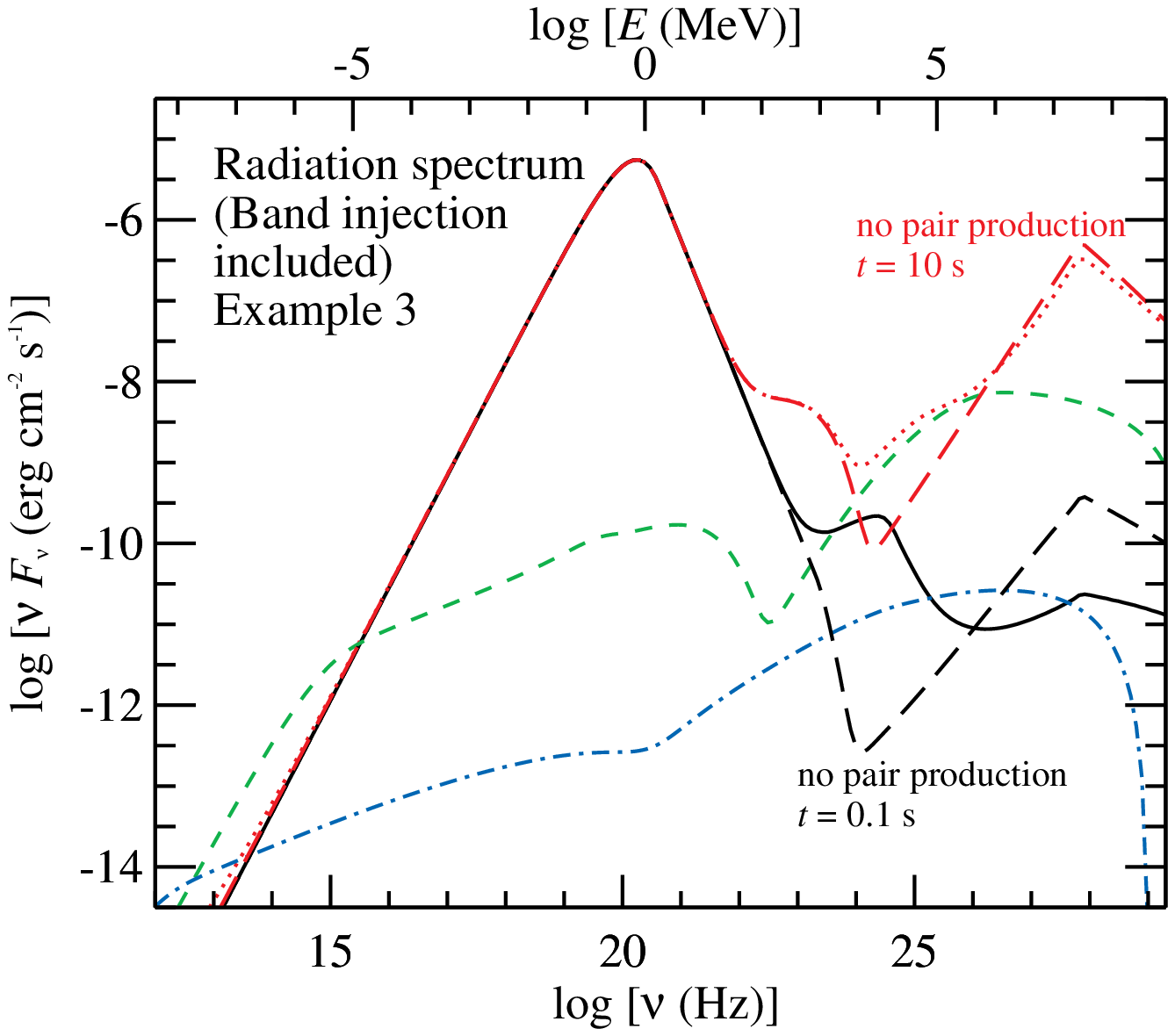}
\includegraphics[width= \columnwidth]{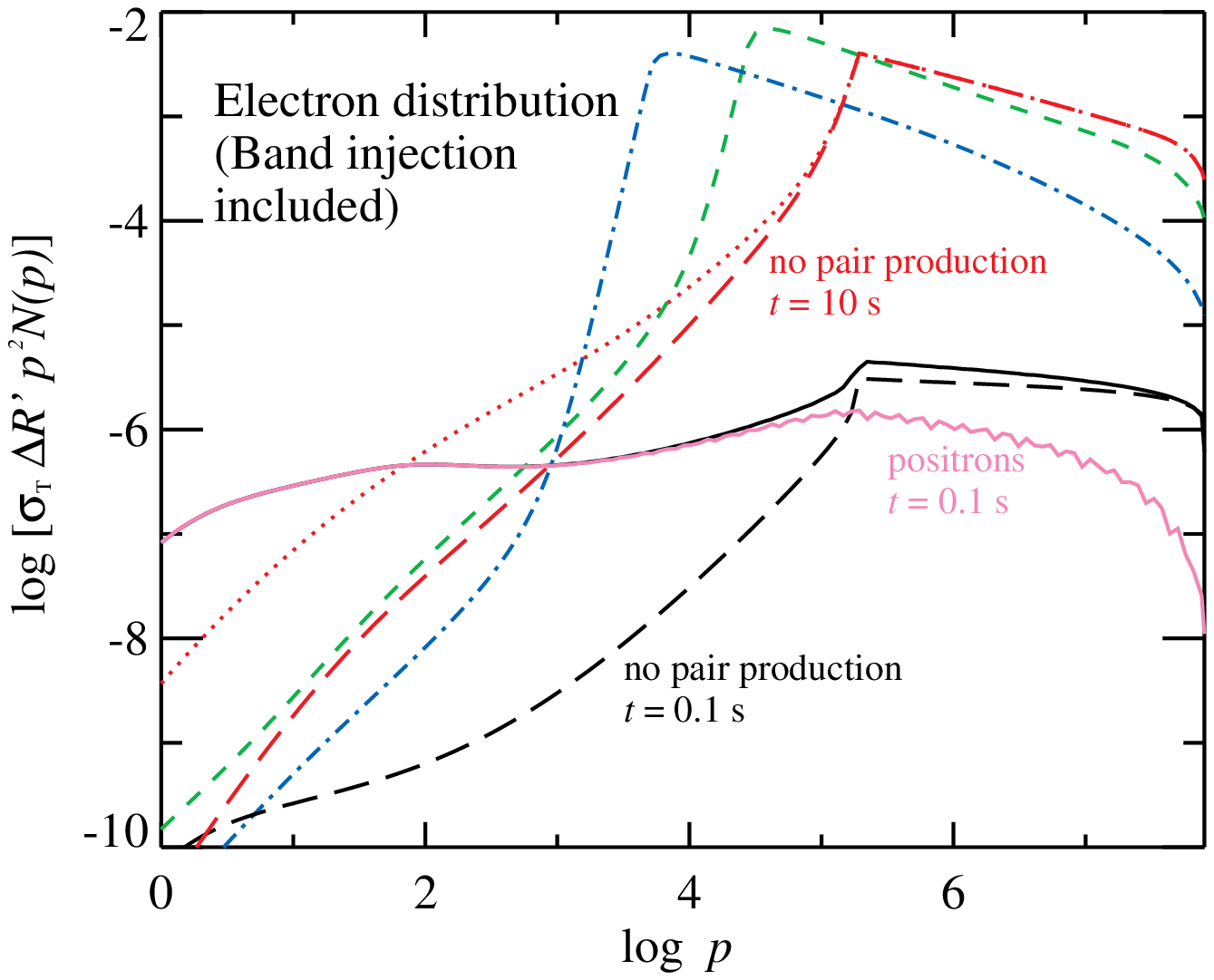}
\caption{\label{fig_ex3}
The time-evolving observer frame photon spectrum
without absorption on extragalactic background light ({\it top panel})
and electron distribution ({\it bottom panel})
from the forward shock with the same photon injection term
as in Fig. \ref{fig_he}.
The other simulation parameters are
$E_0 = 1.4 \times 10^{55}\:\textrm{erg}$, $n_0 = 3 \:\textrm{cm}^{-3}$,
$\epse = 0.8$, $\epsB = 10^{-8}$, $\Gamma_0 = 450$ and
$s = 2.4$. The redshift is $z = 1.8$. The black and red long-dashed lines
show the photon and electrons distributions at $t = 0.1\:\textrm{s}$ and $t = 10\:\textrm{s}$
without pair production,
and the positron distribution at $t = 0.1\:\textrm{s}$ is presented by a solid magenta line
in the {\it bottom panel}.
}
\end{center}
\end{figure}

The current version of the code calculates the emission from the
forward shock only, although the reverse shock emission can be
significant in the case of a long burst.
In addition, electron-positron pair production in the external
medium due to the interaction with the prompt gamma-rays
is likely to affect the afterglow emission
\citep{2005ApJ...627..346B}. This effect is not accounted for
in the code at the moment. However, both the reverse shock emission
and the pair loading of the external medium are expected to shape
the afterglow at relatively early times, and thus the late-time
afterglow emission is likely to be fairly consistent with the results of the
forward shock simulation.

Fig. \ref{fig_he} shows
the simulated time-evolving forward shock
spectrum and the corresponding electron distribution
both with and without an additional Band injection term,
and Fig. \ref{fig_ex3} presents the results of a simulation
with the same Band injection function but different
external shock parameters. The positron distribution
at $t = 0.1\:\textrm{s}$ is also presented in the cases where
pair production is important, meaning in the simulations
with a photon injection term. At the moment, the
positron distribution has a lower numerical accuracy
than that of the electrons.

The simulation parameters have been chosen to
reproduce the {\it Swift}/XRT
afterglow light curve of GRB 090902B,
which was a bright burst with a long-lasting high-energy component
\citep{2009ApJ...706L.138A}.
The Band function used in the simulations is kept constant in time
for simplicity and corresponds to the time-integrated spectral fit
presented in Table 1 of \citet{2009ApJ...706L.138A}. The Band photon
density $\nph$ in the observer frame is evaluated from Eq. (\ref{eq:nph}),
where the luminosity is obtained from dividing the isotropic equivalent
energy of the prompt GRB, $\Ei = 3.6 \times 10^{54}\:\textrm{erg}$, by the burst
duration, $t = 22\:\textrm{s}$. The number of injected photons per unit time and area
is then $\sim c\nph$ in the observer frame, and the injection rate per unit volume
in the fluid frame is
\be
\dnph' \sim \frac{c\nph(1+z)}{2\Gamma \Delta R'},
\ee
where the shell thickness $\Delta R'$ is defined in Eq. (\ref{eq:dR}).
We do not inject the additional power-law component
that is observed in the prompt emission below $\sim 50\:\textrm{keV}$ and above $100\:\textrm{MeV}$.

The allowed forward shock parameter space
based on the late-time afterglow data has been calculated by
\citet{2010MNRAS.409..226K}, who claim that the $> 100\:\textrm{MeV}$ emission
can be explained as pure synchrotron radiation.
The simulations presented in
this section correspond to two points in the allowed
$n - \epsB$ space given in Figure 3 of their paper,
with $E_0$ and $\epse$ being evaluated from their Eqs.
(12) and (13).
Our Fig. \ref{fig_he} shows the results in the case of a lower density and a
higher magnetic energy fraction than in Fig. \ref{fig_ex3}.
We assume that the
ejecta begin to decelerate at the end of the prompt burst. This assumption is used to
evaluate the initial bulk Lorentz factor of the emitting shell, $\Gamma_0$.
The extragalactic background light absorption of the $\gtrsim \textrm{TeV}$ photons is not accounted
for in the example presented here, and these photons
would in fact be unobservable from the redshift of GRB 090902B, $z = 1.8$.
However, a very high-energy component could be observed in the case of
a low-redshift burst with similar parameters.

The {\it left-hand panels} in Fig. \ref{fig_he}
show that the Band emission
at energies $E \lesssim 1\:\textrm{GeV}$
dominates over the underlying forward shock synchrotron
radiation at $t = 0.1\:\textrm{s}$,
while a double-peaked inverse Compton component appears at higher energies.
The break at $\sim 100\:\textrm{GeV}$ is due to pair production with the peak Band photons,
which are located at the threshold energy for pair production with the
photons at $\sim 100\:\textrm{GeV}$
according to Eq. (\ref{eq:x_thr}).
For target photons with a Band distribution,
the optical depth peaks at $\sim 10 \xt$, where the lowest point between the two
VHE spectral peaks is located. The maximum optical depth can be approximated using
Eqs. (\ref{eq:tau_gg}) and (\ref{eq:nph}) to obtain $\tau_{\gamma \gamma} \sim$ a few.
The importance of pair production can be confirmed by looking at the
electron and positron distributions, which are nearly identical below
$\gamma_{\min} \sim 10^5$.
The VHE peak in the radiation spectrum
is due to the scattering of Band photons against the peak electrons
with $\gp = \gamma_{\min}$. According to Eq. (\ref{eq:ic_peak}),
the peak energy is $E \sim 30\:\textrm{TeV}$, which agrees with the figure.

The {\it right-hand panels}, showing the results of the simulation
without the Band injection term, confirm that the synchrotron
spectral component at $t = 0.1\:\textrm{s}$ is clearly below the Band emission,
and the contribution from the Compton scattered synchrotron photons is also negligible.
In the simulation with the additional photon injection term, a low-energy
tail forms in the electron distribution. This is mainly caused
by pair production with a contribution from
scatterings in the K-N regime, which can take a large fraction
of the energy of an electron in one scattering.

At $t = 10\:\textrm{s}$, close to the deceleration time,
the IC component has only one peak left
in both simulations. The optical depth for
pair production is expected to decrease until the
deceleration time according to $\tau_{\gamma \gamma} \propto t^{-1}$
and thus should drop by a factor of 100 between $t = 0.1\:\textrm{s}$
(where $\tau_{\gamma \gamma} \sim 2$) and $t = 10\:\textrm{s}$,
which is in agreement with the simulation results.
In the {\it left panels},
the prompt emission dominates as the source of seed photons
for inverse Compton, as the resulting high-energy emission
is $\sim 2$ orders of magnitude more luminous than with
synchrotron self-Compton (SSC) emission alone.
The high-energy part of the synchrotron bump has become
visible at $t = 10\:\textrm{s}$ and dominates the emission at
$E \sim 100\:\textrm{MeV} - 10\:\textrm{GeV}$. The inverse Compton peak
is still located at $E \sim 30\:\textrm{TeV}$, and it seems that the
prediction for the location of the peak is very robust
during the prompt emission due to the hard spectrum of
the soft target photons. In the slow cooling regime and
in the absence of significant pair opacity, the spectral slope
below the $\sim 30\:\textrm{TeV}$ peak mimics the low-energy slope of the
prompt spectrum. Because the peak energy is proportional to
$\Gamma \gamma_{\min}$ and $\gamma_{\min} \propto \epse \Gamma$,
a measurement of the peak energy would provide us with the combination
$\epse \Gamma^2$.

After the end of the photon injection, the spectrum quickly
becomes identical to the pure forward shock emission, as can be
seen by comparing the spectra at $t = 10^3\:\textrm{s}$ and $t = 10^5\:\textrm{s}$
in the {\it left} and {\it right panels}.
The electrons are unable to cool by emitting synchrotron radiation
because such a small fraction
of the shock energy is given to the magnetic field.
The SSC cooling of the electrons on synchrotron radiation
is not effective either, which can be seen from the electrons
being distributed as
$N(\gamma) \propto \gamma^{-s}$ between $\gamma_{\min}$
and $\gamma_{\max}$. The synchrotron
spectral component produced by these electrons now corresponds
to a slow cooling spectrum as described by \citet{1998ApJ...497L..17S}.

The simulation results presented in Fig. \ref{fig_ex3} correspond
to a higher density and smaller magnetic energy density than those in
Fig. \ref{fig_he}. The VHE spectrum at $t = 0.1\:\textrm{s}$ is similar to the
one in Fig. \ref{fig_he}. In this case, the spectrum at $t = 10\:\textrm{s}$ is also
shaped by the inclusion of pair production, which leads to an increased
energy content in electrons at $\gamma < \gamma_{\min}$ and correspondingly
more energy in the upscattered photons between $\sim 3\:\textrm{GeV}$ and $\sim 1\:\textrm{TeV}$.
After the prompt emission is gone, the SSC component
is very prominent compared to the synchrotron bump because of the
high value of $\epse/\epsB$ used in the simulation. Even though a large
fraction of the shock-generated energy, $\epse = 0.8$, is given to the
electrons now, it is still valid to assume an adiabatic evolution of
the shock because the radiative cooling is inefficient.

It is notable that a power-law component similar to the one
observed during the prompt phase of GRB 090902B \citep{2009ApJ...706L.138A}
is not seen in our simulations. According to the simple model presented
here, the early power-law component cannot be of external origin.
The low-energy part of the component might be detectable
above the Band emission in the case of a low $\gamma_{\min}$ (Eq. (\ref{eq:gmin})).
Because we assume that the deceleration time is approximately
equal to the duration of the prompt GRB, we can use Eqs. (\ref{eq:Rdec}) and (\ref{eq:r_early})
in our paper, together with Eqs. (12) and (13) in \citet{2010MNRAS.409..226K} to find that
$\gamma_{\min} \cong 6 \times 10^3 (n \epsB^3)^{-1/16}$; i.e., $\gamma_{\min}$ depends
very weakly on $n$ and $\epsB$. From the allowed $n - \epsB$ space presented in Figure 3 of
\citet{2010MNRAS.409..226K}, one finds that $\gamma_{\min}$ cannot be much lower than
$\sim 10^5$.

\begin{figure} 
\begin{center}
\includegraphics[width= \columnwidth]{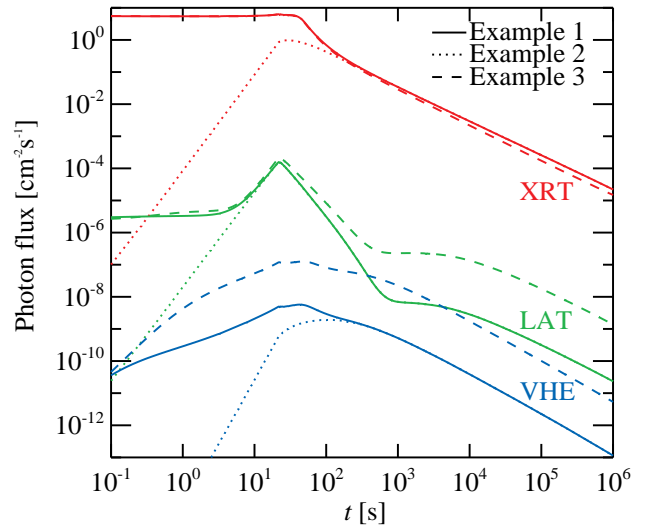}
\caption{\label{fig:lc}
Observer frame light curves in the {\it Swift}/XRT band ($0.3-10\:\textrm{keV}$,
red curves at the top of the figure), the {\it Fermi}/LAT band ($0.1-300\:\textrm{GeV}$,
green curves in the middle), and a very high energy (VHE) $0.1 - 100\:\textrm{TeV}$ band
(blue curves at the bottom). The light curves correspond to the simulations
presented in Figs. \ref{fig_he} and \ref{fig_ex3}.
} 
\end{center} 
\end{figure}

Fig. \ref{fig:lc} shows the light curves
corresponding to the simulations presented in Figs. \ref{fig_he}
and \ref{fig_ex3} (Band injection is included
in Examples 1 and 3, but not in Example 2)
in three different energy bands: the XRT and LAT bands
and a very high-energy (VHE) band
ranging from $0.1\:\textrm{TeV}$ to $100\:\textrm{TeV}$,
where the only contribution is from the inverse
Compton emission due to the upscattering of either synchrotron or
external photons. Because the Band function is kept constant in time,
the flux in both the XRT and LAT bands is also constant in the beginning
of the simulations with the photon injection term.
The inverse Compton flux in the VHE band rises at early
times as the energy of the shocked electrons increases and the
suppression of the flux due to pair production becomes less
important.

The LAT light curves are
dominated by synchrotron emission between $t\sim 10\:\textrm{s}$ and
$10^3\:\textrm{s}$.
At $10^3\:\textrm{s}$ the synchrotron component falls below
the LAT band and there is a break in the light curves, followed by a
more slowly decaying SSC emission.
The X-ray emission from the forward
shock is buried under the prompt emission and emerges only after the
latter has ended. In the TeV band the emission is dominated by the
upscattered prompt photons as long as they are available, resulting in
substantially higher luminosity at early times than would be obtained
by SSC alone.

The simulated XRT band light curves are
consistent with the observed XRT data of GRB 090902B \citep{2010ApJ...714..799P}
as intended,
but the observed LAT light curve \citep{2009ApJ...706L.138A}
is not reproduced in the current simulations.
This is simply because we assume a lower value of $\gamma_{\max}$
(Eq. (\ref{eq:gmax_ev})) than
\citet{2010MNRAS.409..226K}, who claim that the high-energy emission
consists of synchrotron photons. The maximum photon energy produced by
the synchrotron process is $\sim 2\Gamma \times 50$ MeV $/(1+z)$ in the observer
frame \citep{1983MNRAS.205..593G},
determined from the balance between the cooling and acceleration times
for electrons radiating in the background magnetic field $B'$
\citep[for synchrotron losses taking place in a magnetic
field generated by the Weibel instability, see][]{2013ApJ...771...54S}.
The most energetic photon observed from GRB 090902B, with $E = 33\:\textrm{GeV}$
in the observer frame, requires a bulk Lorentz factor $\Gamma \sim 900$
at $t = 82\:\textrm{s}$ to be consistent with the synchrotron model.
The Lorentz factor at this time can be estimated
from Eq. (\ref{eq:g_time}). For
$E_0 \sim 10^{55}\:\textrm{erg}$ and $n_0 = 10^{-3}\:\textrm{cm}^{-3}$,
the lowest possible external
density according to \citet{2010MNRAS.409..226K}, we find that $\Gamma \sim 400$
at $t = 82\:\textrm{s}$, not supporting a synchrotron origin for this photon.

The examples in this section show that a luminous
$\gtrsim \:\textrm{TeV}$ component can arise due to inverse Compton
scattering in the forward shock,
although this radiation is observable only in low-redshift cases.
The Cherenkov telescope MAGIC is the most likely candidate to
catch the upscattered prompt radiation in the VHE range due to its fast slewing
capability, but MAGIC has so far been unable to detect any emission from
GRBs \citep{2007ApJ...667..358A,2010A&A...517A...5A,2014MNRAS.437.3103A}.
However, the lack of detections does not rule out the existence of a
$\sim 10\:\textrm{TeV}$ spectral component because of the high redshifts of the observed
bursts.

\begin{figure*}
\begin{center}
\includegraphics[width= 0.73\columnwidth]{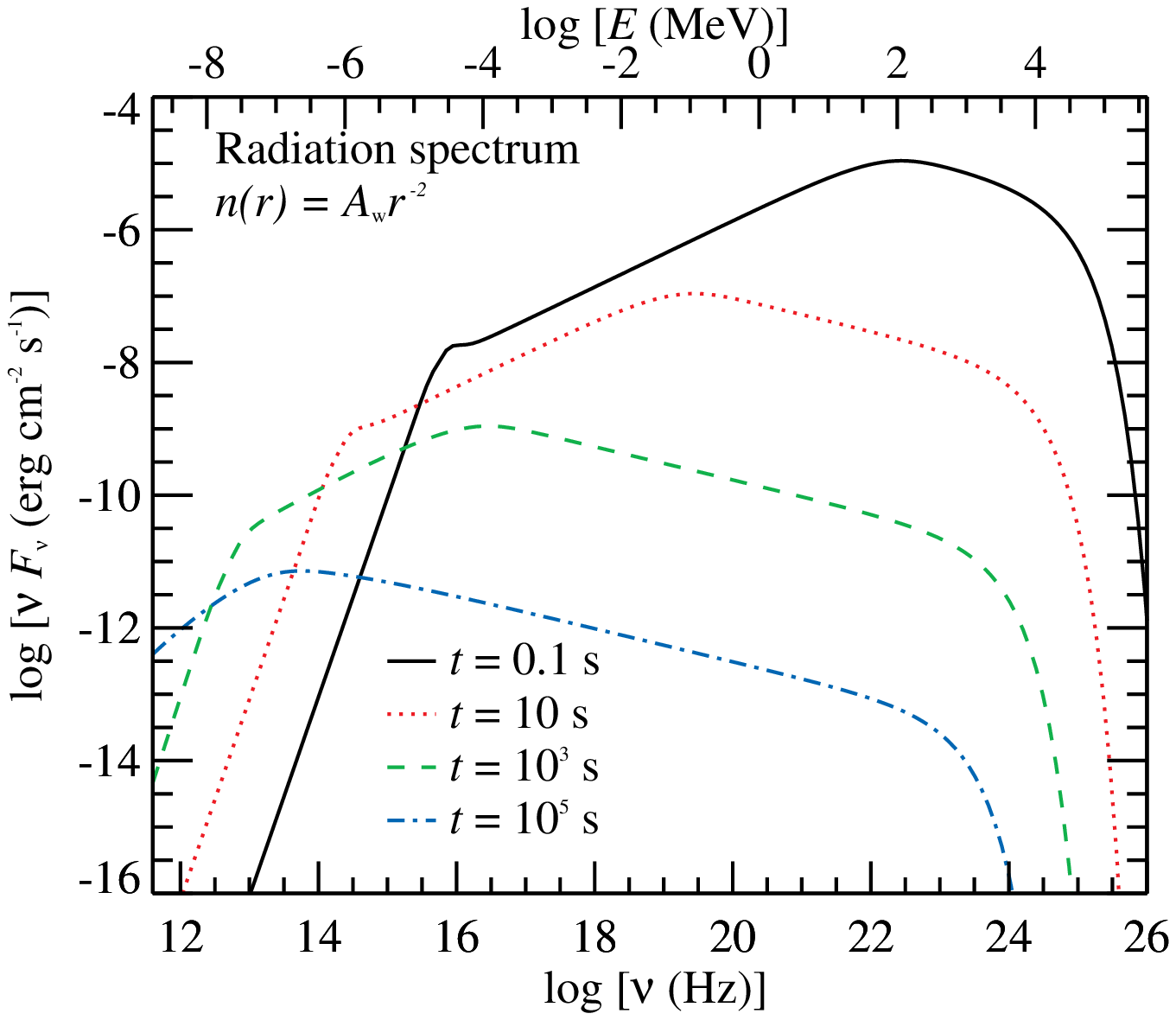}
\includegraphics[width= 0.73\columnwidth]{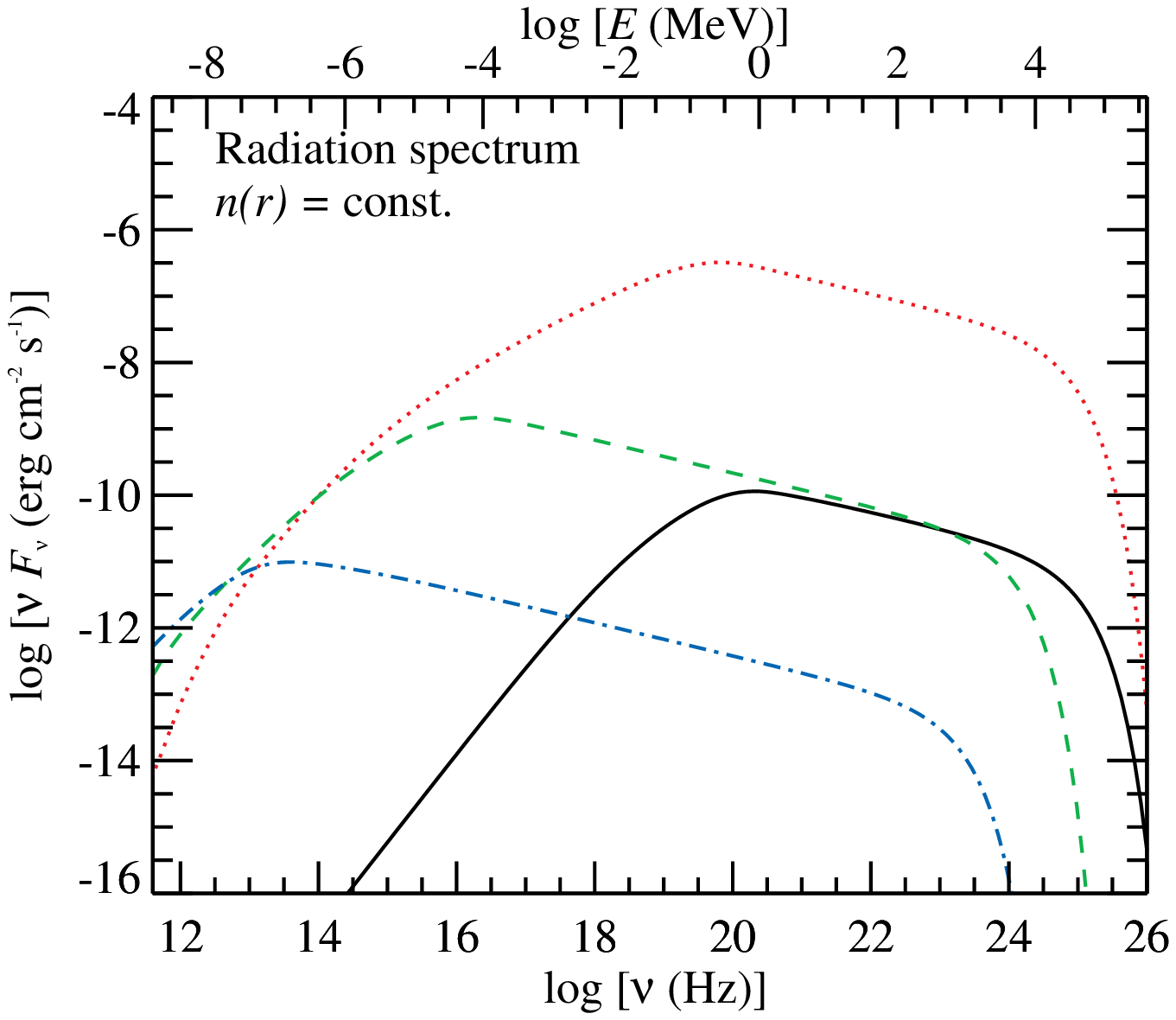}
\includegraphics[width= 0.73\columnwidth]{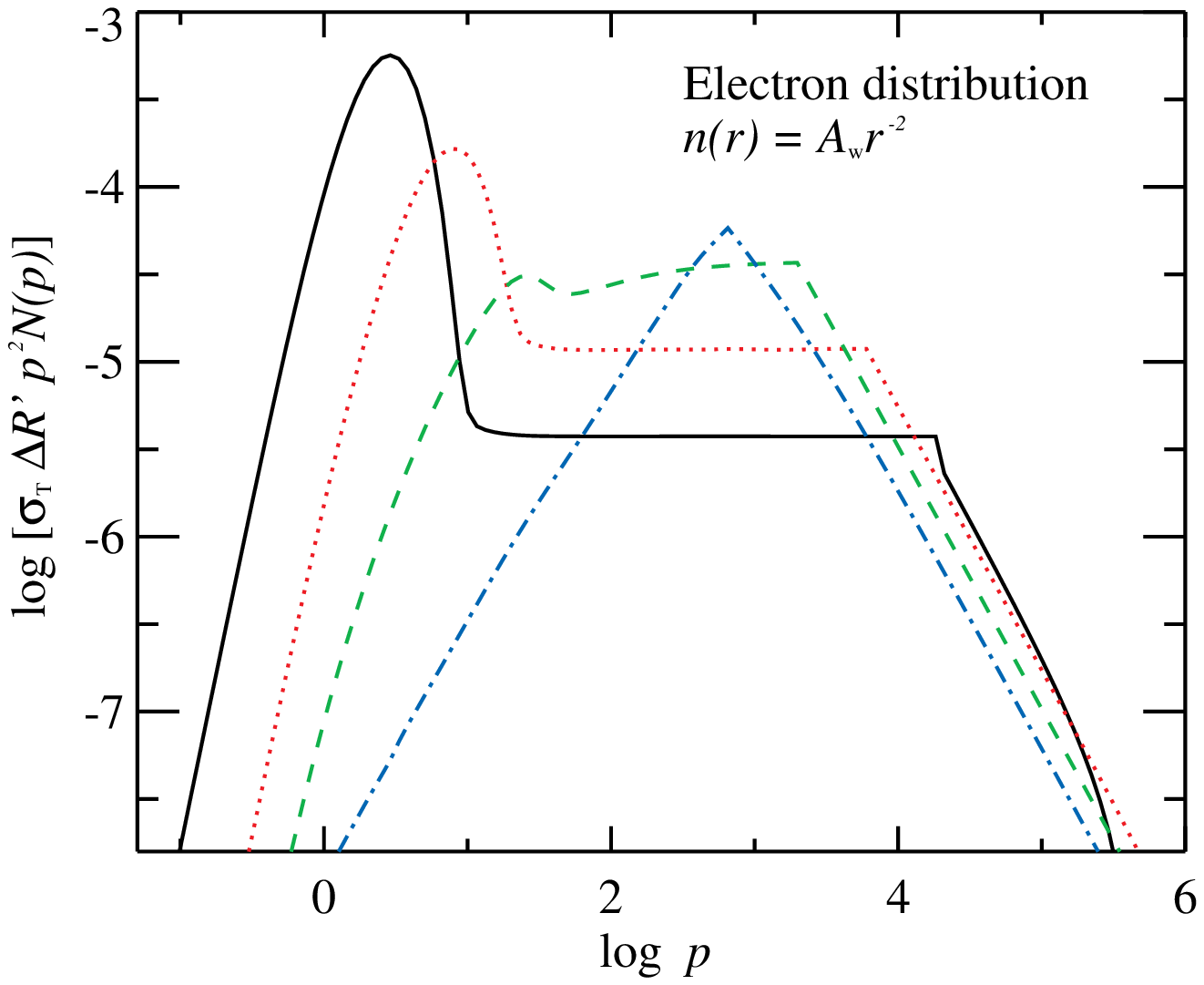}
\includegraphics[width= 0.73\columnwidth]{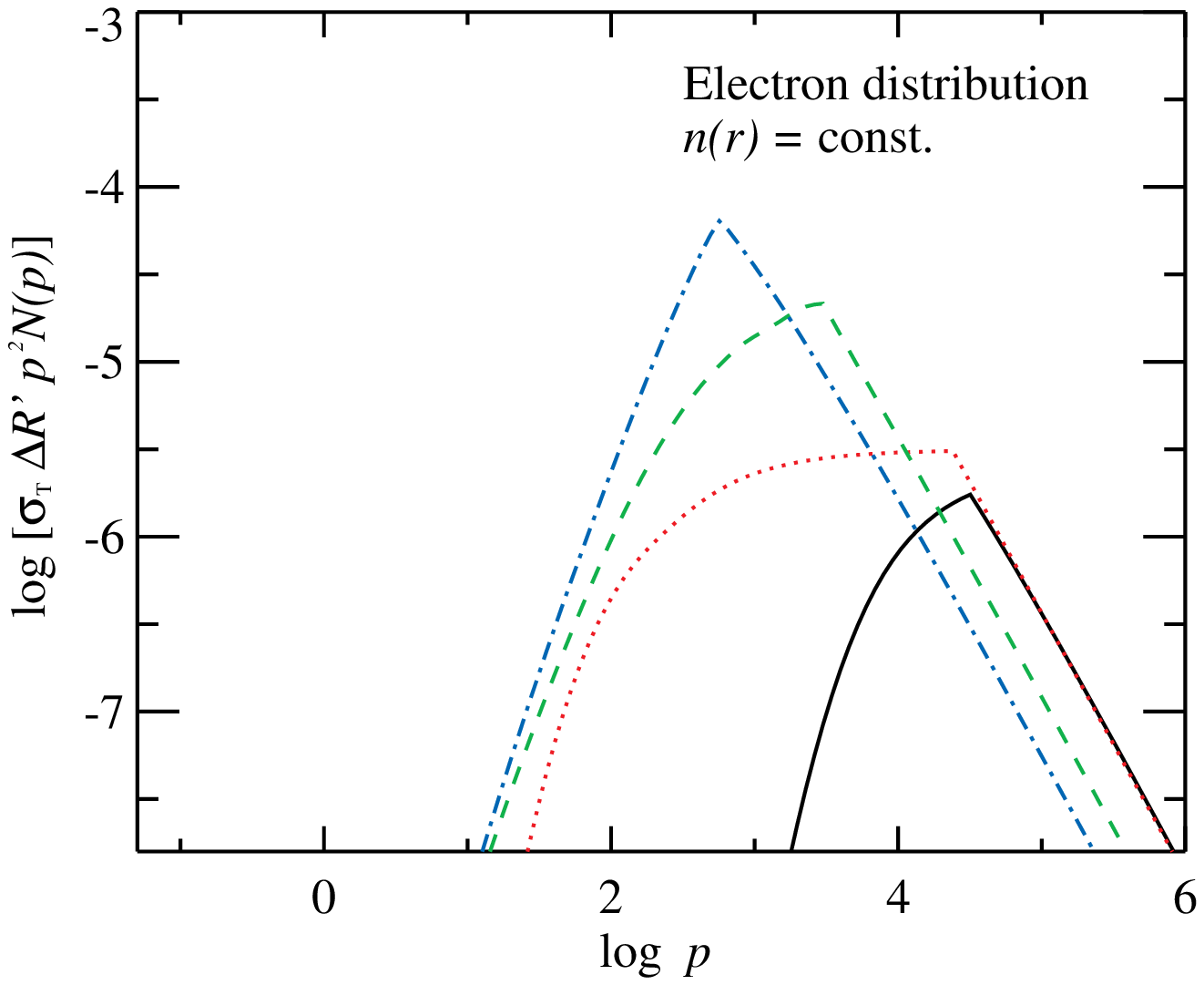}
\caption{\label{fig_wind}
{\it Left panels} show the time-evolving observer frame photon spectrum ({\it top panel})
and electron distribution ({\it bottom panel}) resulting from synchrotron
emission and absorption together with adiabatic cooling at the forward shock in
a wind-type medium. 
The simulation parameters are $E_0 = 10^{54}\:\textrm{erg}$,
$\Awn \equiv \Aw/(10^{35}\:\textrm{cm}^{-1}) = 3$, $\epse = 0.1$,
$\epsB = 0.1$, $\Gamma_0 = 500$ and
$s = 2.5$. The redshift is $z = 1$.
{\it Right panels} show the observed photon spectra ({\it top panel}) and electron
distributions ({\it bottom panel}) in the case of a constant-density medium
with a typical density $n_0 = 1$ cm$^{-3}$. The other simulation parameters
are the same as in the simulation presented in the {\it left panels}.
}
\end{center} 
\end{figure*}

If the external density were higher than in our examples,
such as in the case of a wind-type medium, the inverse Compton flux
would also increase and possibly be able to reproduce the magnitude of the
GeV photon flux in the LAT band. This idea is supported by the recent
simulations of \citet{2013arXiv1307.2663B}, who show that the GeV emission
of GRB 080916C is consistent with inverse Compton scattering of the prompt
emission. The work of \citet{2013arXiv1307.2663B} is mainly
concerned with the early stage of
the outflow and does not include SSC emission, which is important once
the prompt radiation is gone.

Our current model
cannot explain the extended duration of the observed LAT emission from GRB 090902B,
which is however reproduced by \citet{2013arXiv1307.2663B}.
The delay of the arrival times of the high-energy photons
compared to the MeV emission may result from the different propagation angle of
the Compton scattered prompt photons compared to that of the unscattered photons,
which is not accounted for in the code at the moment.
It must be noted that a full treatment of the
afterglow requires a more complete model of the prompt
emission, the inclusion of the reverse shock emission
and the effect of the pair loading of the external medium.

\subsection{Constant-density ISM vs. wind medium}\label{ex2}

\begin{figure}
\begin{center}
\includegraphics[width= \columnwidth]{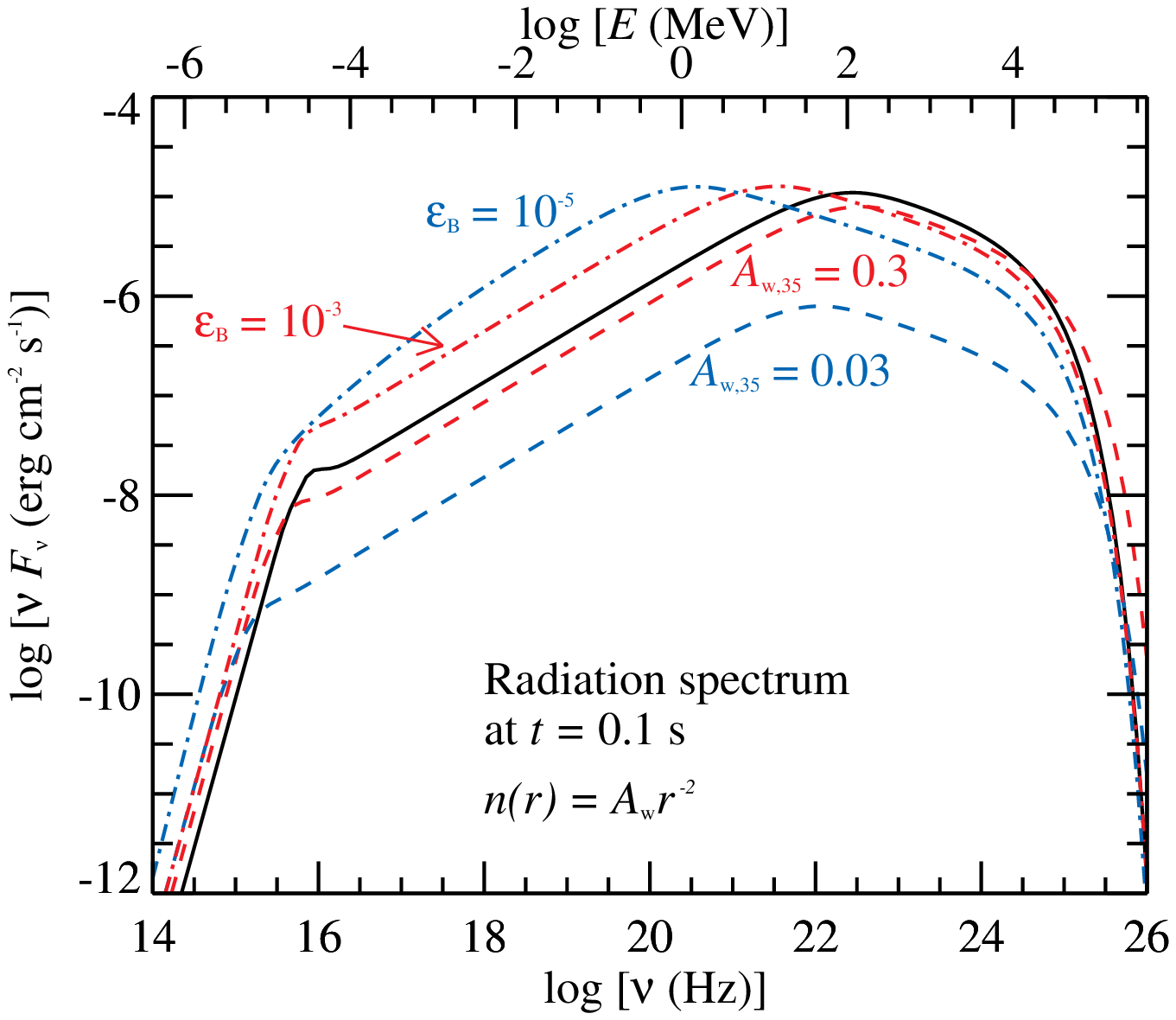}
\includegraphics[width= \columnwidth]{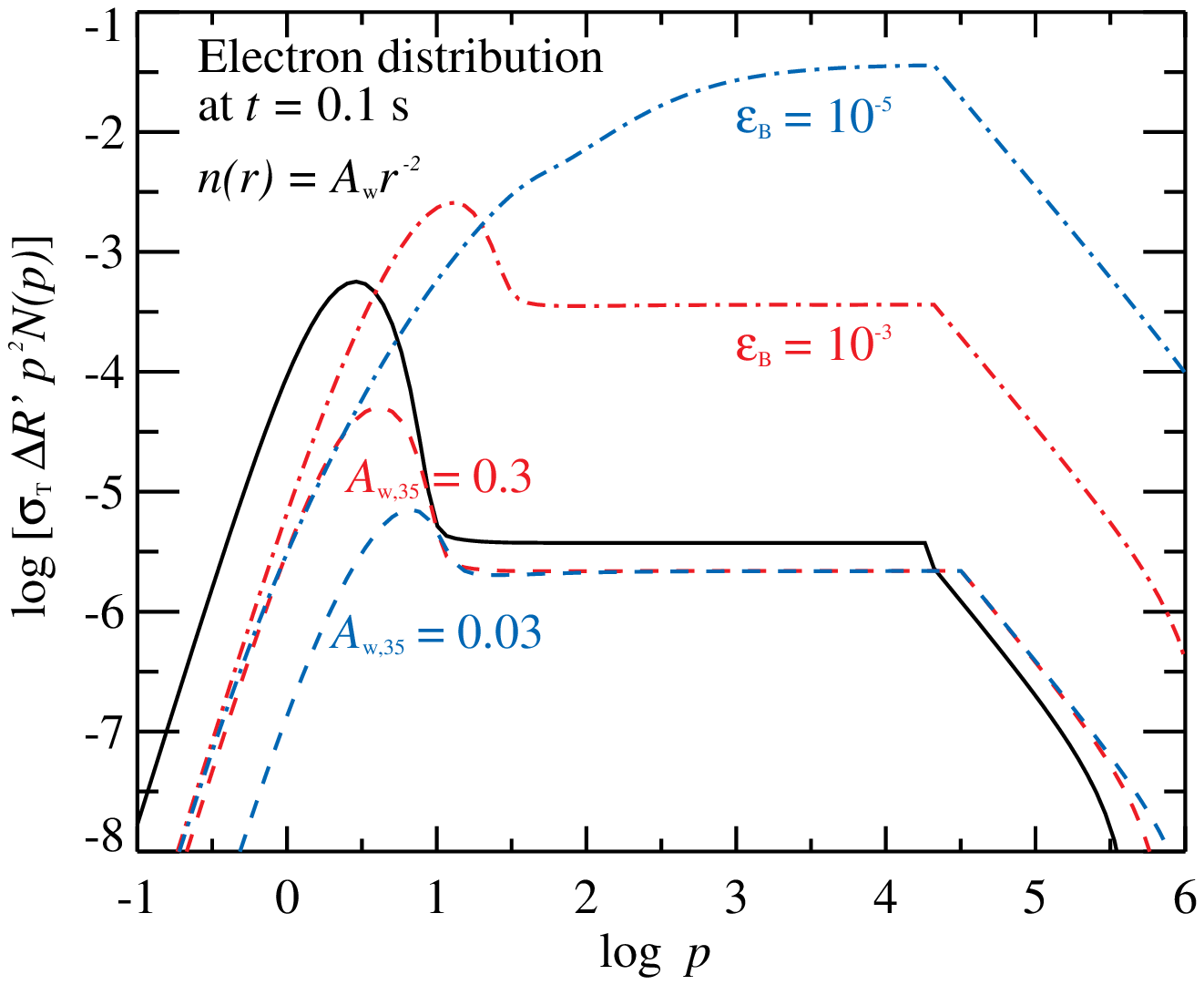}
\caption{\label{fig_wind_var}
Comparison of the radiation spectrum ({\it top panel}) and electron
distribution ({\it bottom panel})
at $t = 0.1\:\textrm{s}$ from Fig. \ref{fig_wind} (black solid line)
to four cases where the value of one parameter is changed at a time.
In two simulations, the wind density is decreased
by a factor of 10 (red dashed line)
and 100 (blue dashed line) from the fiducial value. In the other two cases,
the magnetic energy fraction is decreased by a factor of $10^2$
(red dash-dot line) and $10^4$ (blue dash-dot line).
}
\end{center} 
\end{figure}

\begin{figure}
\begin{center}
\includegraphics[width= \columnwidth]{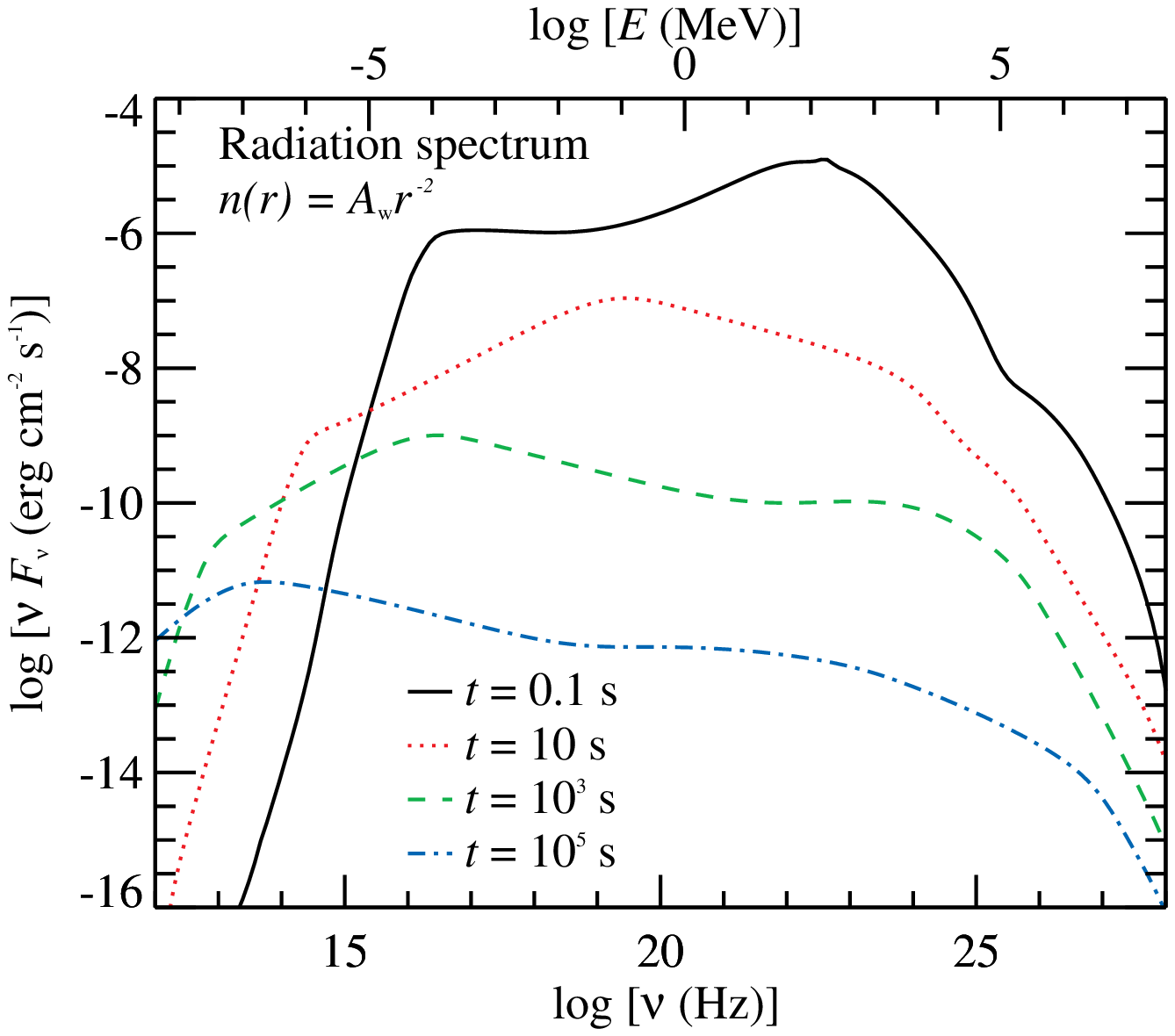}
\includegraphics[width= \columnwidth]{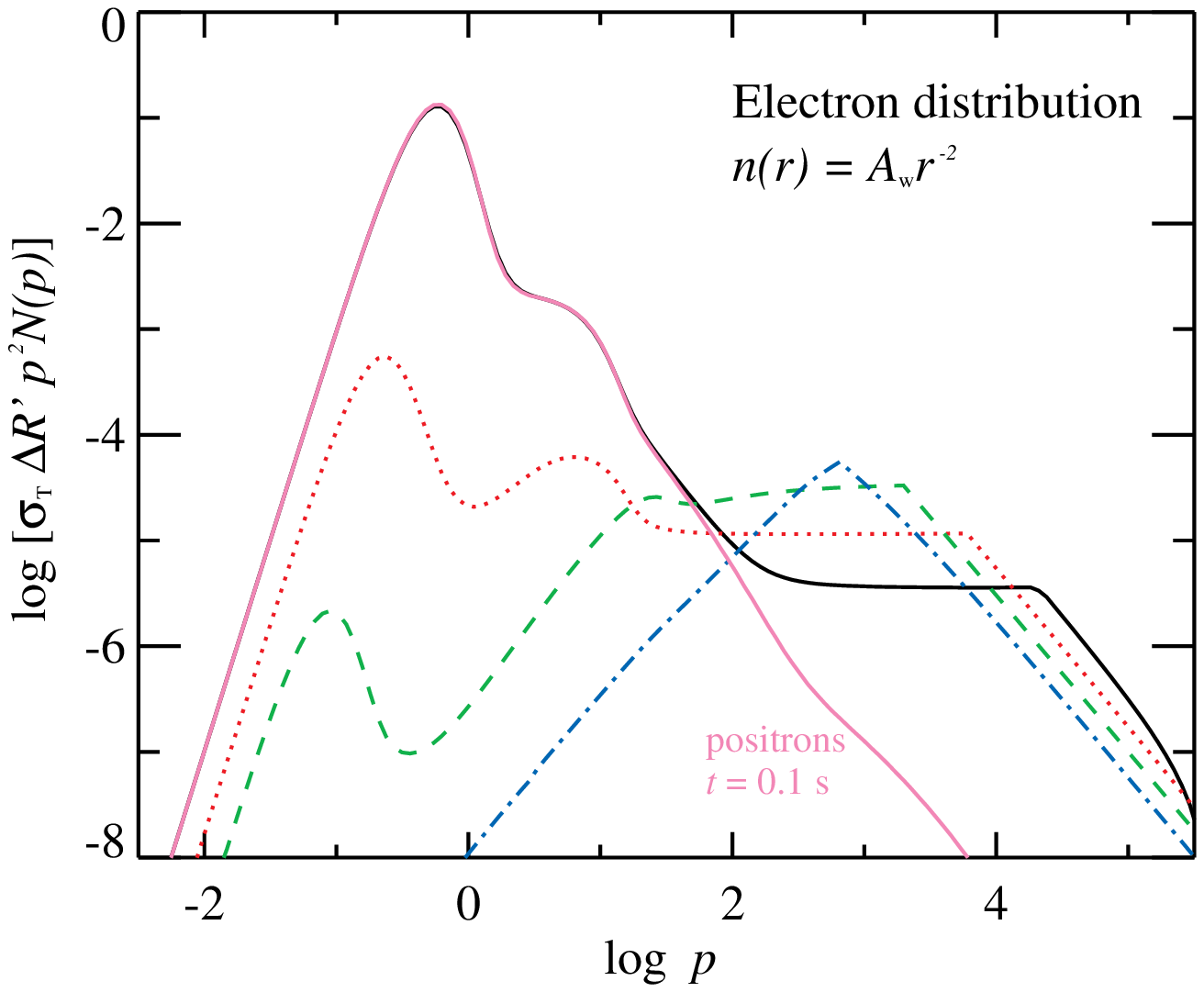}
\caption{\label{fig_wind_full}
Time evolution of the observer frame photon spectrum ({\it top panel}) and electron
distribution ({\it bottom panel}) in a wind medium when all the relevant radiative processes
(synchrotron, Compton and pair production) are accounted for in the simulation.
The positron distribution at $t = 0.1\:\textrm{s}$ is also shown as a solid magenta line in the
{\it bottom panel}.
The simulation parameters are the same as in Fig. \ref{fig_wind}.
} 
\end{center} 
\end{figure}

The density structure of the ambient medium has an impact
on the evolution of the bulk Lorentz factor $\Gamma$,
the injection rate of the electrons and the strength of the
shock-compressed magnetic field. In a
wind-type medium with $n(r)=\Aw r^{-2}$ all the 
synchrotron-emitting electrons are likely to be fast-cooling
at early times because of the high magnetic field,
provided that the magnetization parameter $\epsB$ is not
too low.
Synchrotron cooling is counteracted by self-absorption
at the low-energy end of the distribution. As long as the synchrotron
cooling time of the electrons is shorter than the lifetime of the flow
for all electron energies, the low-energy electrons are able to
thermalize due to self-absorption.

The difference between the pure synchrotron emission from the forward
shock in a wind-type and a constant-density medium is illustrated
in Fig. \ref{fig_wind}, which shows the simulated photon and electron
distributions for a typical set of parameters. Compton scattering
and pair production are left out of this simulation in order to
study the effect of synchrotron self-absorption heating on the
particle distributions. Again, we assume that the distribution of the shock-accelerated
electrons is a pure power law.

The deceleration radius (Eqs. (\ref{eq:Rdec}) and (\ref{eq:Rdec_wind}))
is reached earlier in a wind medium than a constant-density ISM. In
the simulations presented in Fig. \ref{fig_wind},
$\Rd = 7.1 \times 10^{14}\:\textrm{cm}$ in the wind case and
$\Rd = 8.6 \times 10^{16}\:\textrm{cm}$ in the ISM case. Because most of the blast
energy is dissipated at this radius, the peak luminosity is higher
in the wind case because the same amount of energy is emitted in a
shorter time compared to the ISM environment.

At the observed time $t = 0.1\:\textrm{s}$, the ambient particle density
in the wind case is $n = 1.1 \times 10^6\:\textrm{cm}^{-3}$, explaining the
large difference between the number of shocked electrons in the two
simulations. At this moment, all the electrons are fast-cooling in both simulations
due to the large fraction of energy given to the magnetic field.
However, the magnetic field in the wind case is much higher
because of the high density, even though the shock has already
entered the deceleration phase before $t = 0.1\:\textrm{s}$.
The electrons have now been able to cool
below $\gamma_{\min}$ and form a clear Maxwellian component
centered at $\gamma \sim 3$. The Maxwellian electrons have had plenty of time
to thermalize because their synchrotron cooling time in the
fluid comoving frame, $\tco' \sim 0.2\:\textrm{s}$ (Eqs. (\ref{Bfield}) and (\ref{eq:t_cool}))
is clearly shorter than the comoving lifetime of the flow, $t' = 40\:\textrm{s}$.
The electrons at higher energies are distributed according to
$N(\gamma) \propto \gamma^{-2}$ below $\gamma_{\min}$ and
$N(\gamma) \propto \gamma^{-s-1}$ between $\gamma_{\min}$ and
$\gamma_{\max}$, as expected. It is notable that the electrons would
be able to cool to much lower energies in a simulation without
the self-absorption heating term. Thus, the inclusion of this term
is necessary for the modeling of a wind-type medium with a
high magnetic compactness.
The thermalized electrons, however, only produce a very small
bump in the corresponding synchrotron spectrum.

At $t= 10\:\textrm{s}$, the Maxwellian component in the electron
distribution in a wind medium is already smaller because of the
decreasing magnetic field. The bump gradually disappears, and
the slope of the electron distribution at $t = 10^5\:\textrm{s}$
between $\gc \sim 600$ and
$\gamma_{\min} \sim 3 \times 10^3$ is becoming slightly shallower
because these electrons are no longer able to cool. At this
moment, the electron and photon distributions in the ISM case
look very similar to those in the wind case.

\begin{figure*}
\begin{center}
\includegraphics[width= 0.73\columnwidth]{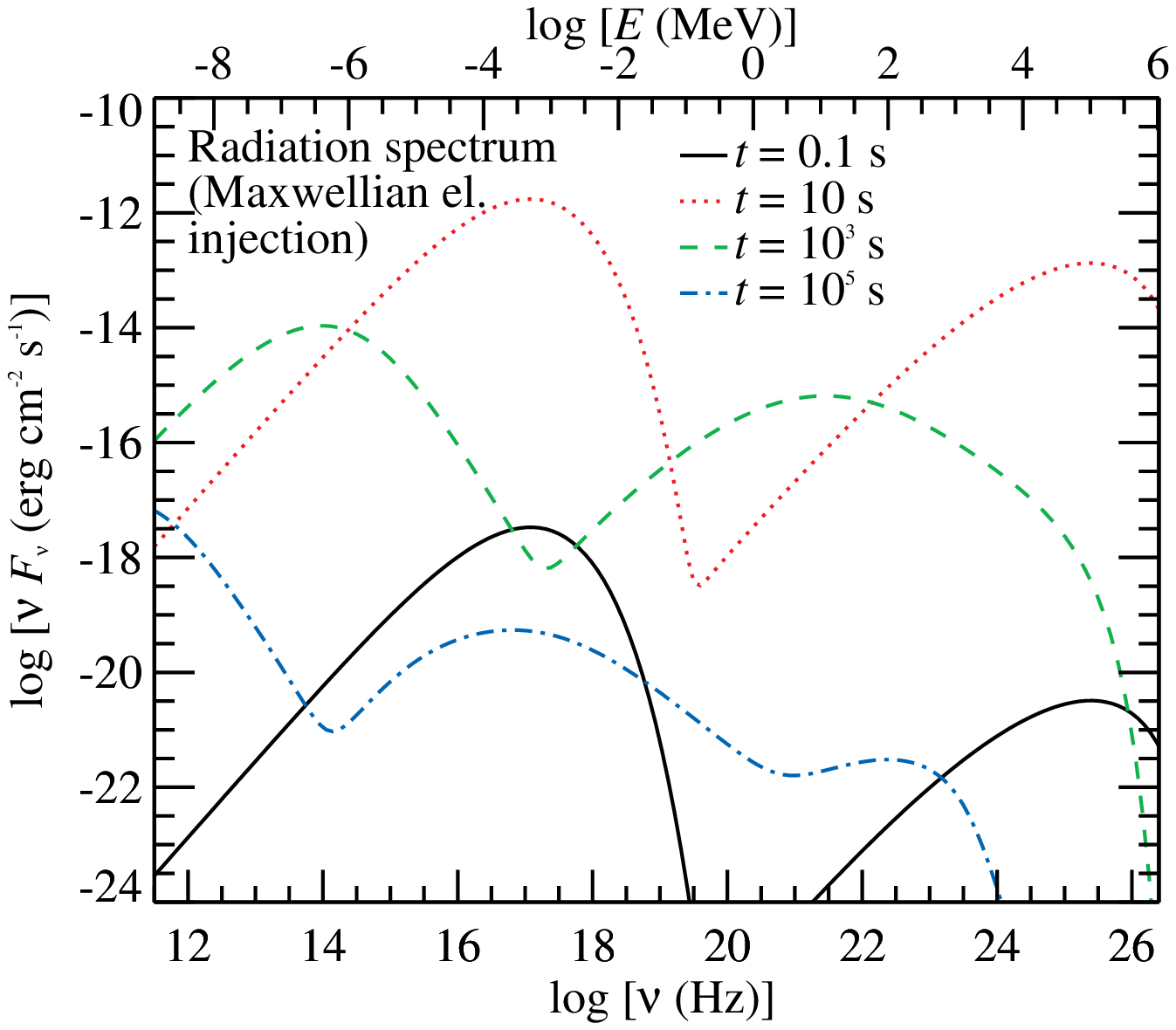}
\includegraphics[width= 0.73\columnwidth]{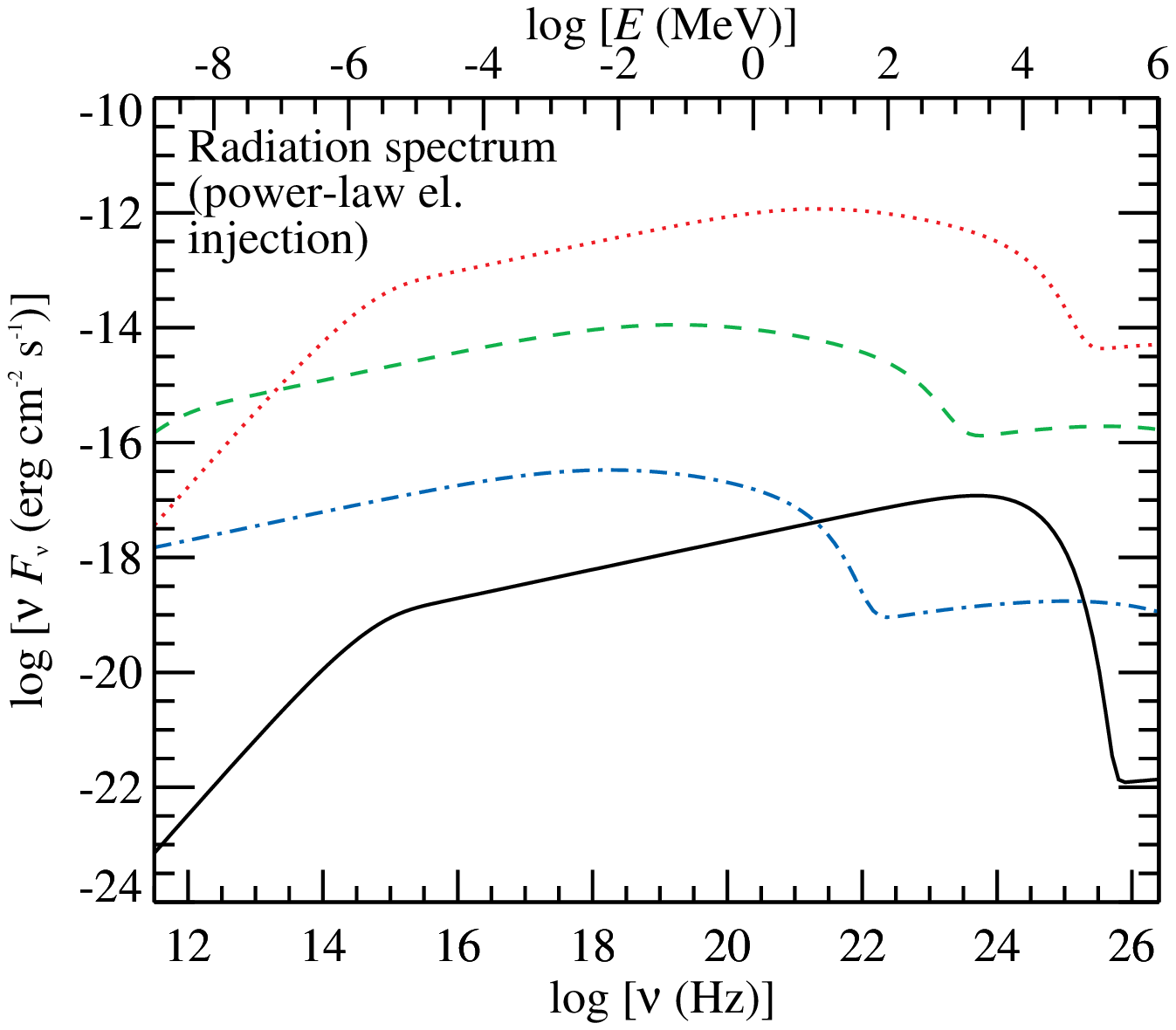}
\includegraphics[width= 0.73\columnwidth]{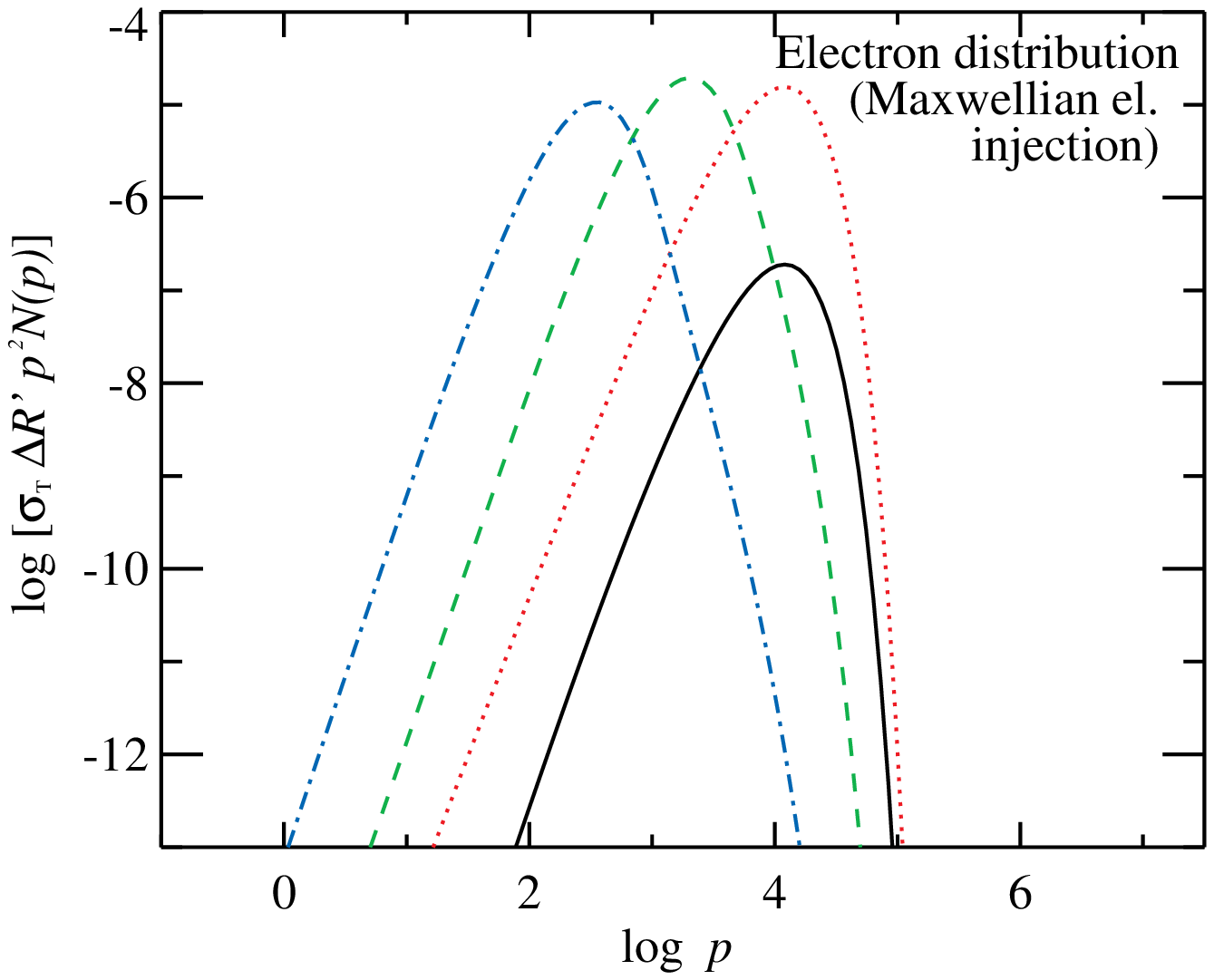}
\includegraphics[width= 0.73\columnwidth]{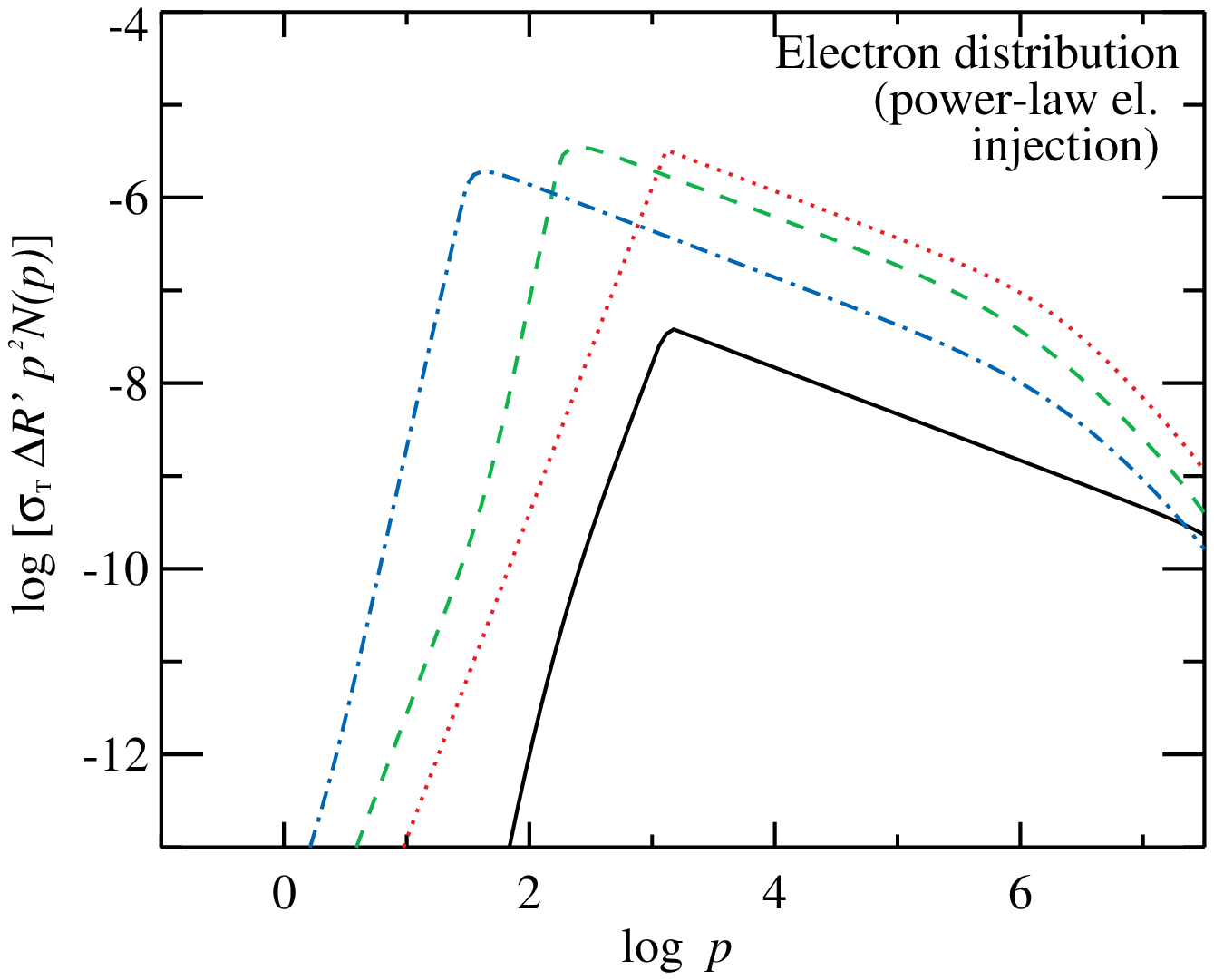}
\caption{\label{fig_max}
{\it Left panels:} time evolution of the observer frame photon spectrum ({\it top panel})
and electron distribution ({\it bottom panel}) resulting
from relativistic Maxwellian electron injection in a constant-density ISM.
The radiation processes are synchrotron emission/absorption, Compton scattering and
pair production/annihilation.
The simulation parameters are $E_0 = 10^{52}\:\textrm{erg}$,
$n_0 = 1\:\textrm{cm}^{-3}$, $\epse = 10^{-2}$, $\epsB = 10^{-4}$ and $\Gamma_0 = 200$,
and the redshift is $z = 1$.
The Maxwellian injection function
has the temperature $T' \sim \epse \Gamma \mpr c^2/k$.
{\it Right panels:} the evolving particle distributions in the case of a power-law
electron injection with $s = 2.5$. The other parameters are the same as
in the simulation with Maxwellian injection.
} 
\end{center} 
\end{figure*}

The simulations presented here show that the synchrotron
self-absorption heating of electrons can produce a prominent
Maxwellian component in the electron distribution for certain
parameter sets. Fig. \ref{fig_wind_var} shows how the size of
the Maxwellian bump at $t = 0.1\:\textrm{s}$ decreases if the parameters $\epsB$ and $\Aw$ are changed
from the fiducial values used to obtain Fig. \ref{fig_wind}
and how this affects the corresponding synchrotron spectrum.
When the magnetic energy fraction $\epsB$ is decreased, the synchrotron
cooling time is longer and the electrons are unable to cool to very low energies
or thermalize efficiently.

The magnetic field and the size of the thermalized bump
also decrease if the density of the wind goes down: when the
coefficient $\Awn \equiv \Aw/(10^{35}$ cm$^{-1})$ is decreased from
$\Awn = 3$ by a factor of 100, the low-energy
bump in both the electron and photon distributions has nearly
disappeared.
It should be noted that the bulk Lorentz factor $\Gamma$ is
lower in the simulation with $\Awn = 3$ compared to
the two cases with different $\Awn$, because the shock is already in its
deceleration phase at $t = 0.1\:\textrm{s}$
with $\Gamma \sim 300$ in the case with the highest density.
As a result, the shock radius is slightly smaller in the case of a decreasing
$\Gamma$, but only by a factor of $\sim$ a few.

In general, the magnetic field (see Eq. (\ref{Bfield}))
decreases at radii $r < \Rd$ (Eq. (\ref{eq:Rdec_wind})) due to the
radial dependence of the wind density: $B' \propto \Gamma n^{1/2} 
\propto \Gamma_0 r^{-1} \propto \Gamma_0^{-1} t^{-1}$, so a larger
$\Gamma_0$ indicates a lower magnetic field $B'$ at a given
observer time $t$ as long as the blast is its coasting phase.
The synchrotron cooling time of the electrons thus becomes longer,
working against the formation of a Maxwellian component.

The deceleration radius $\Rd$ depends on the parameters
$E_0, \Aw$ and $\Gamma_0$ but the hydrodynamic evolution at $r > \Rd$
is only affected by $E_0$ and $\Aw$ according to Eqs. (\ref{eq:r_time})
and (\ref{eq:g_time}). The magnetic field then depends on these
parameters and time as
$B' \propto \Gamma n^{1/2}
\propto (\Aw^3 E_0^{-1}t^{-3})^{1/4}$ (naturally, $B'$ also depends
on $\epsB$). Together with the width of the emission region, $\Delta R' \propto \Gamma t$
(Eq. (\ref{eq:dR})),
the magnetic field determines the magnetic
compactness $\lB \equiv \Delta R' \UB \sT/(\mel c^2)$, where $\UB = (B')^2/(8\pi)$
is the magnetic energy density. An electron with a random Lorentz factor $\gamma$
is able to cool due to synchrotron emission
if the ratio $t/\tco \sim \gamma \lB$ (see Eq. (\ref{eq:t_cool})) is large.
Because $\lB \propto \Delta R' (B')^2 \propto \Aw^{5/4}E_0^{-1/4}t^{-3/4}$,
one can obtain lower values of $t/\tco$ at a given time by
increasing $E_0$ or decreasing $\Aw$, which may
prevent the thermalization at low energies.

In reality, pair production has a significant
effect on the external shock emission as long as the wind density is high.
Fig. \ref{fig_wind_full} presents the time-evolving radiation spectrum and
electron distribution from a simulation where synchrotron, Compton and
pair production processes are all accounted for. At $t = 0.1\:\textrm{s}$, the interplay of these
processes results in a complex shape of the electron and positron distributions for
$\gamma < 100$. Numerous leptons appear at these energies because of pair production, and
are able to thermalize at low energies due to the high magnetic compactness.
However, the Maxwellian leptons are now
non-relativistic and do not result in an observable signature in the
corresponding photon spectrum. The spectrum at $t = 0.1\:\textrm{s}$ is clearly
different from the pure synchrotron case in Fig. \ref{fig_wind}.
In Fig. \ref{fig_wind_full}, the electrons
at $\gamma < 100$ produce a distinct spectral segment between $\sim 100\:\textrm{eV}$ and
$100\:\textrm{keV}$, and pair production shapes
the synchrotron and Compton components at high energies. The pair production
opacity decreases at later times and the spectrum becomes similar to the pure
synchrotron case, with the obvious exception that it extends to higher energies
because of Compton scattering.

\subsection{Emission from a Maxwellian distribution of injected electrons}\label{ex3}

Because a Maxwellian component can contain as much as $\sim 90 \%$
of the energy of the shocked electrons \citep{2008ApJ...682L...5S},
we have also simulated the forward shock emission resulting from
a Maxwellian electron injection function.
The relativistic
Maxwellian has the general form
\be
N(\gamma) \propto \gamma\sqrt{\gamma^2-1}\exp{\left[-\frac{\gamma}{kT'/\left(\mel c^2\right)}\right]},
\ee
where $T'$ is the temperature of the electrons. The temperature
can be determined from
\be \label{eq:max_pk}
3kT' \sim \epse \Gamma \mpr c^2,
\ee
where $\epse \Gamma \mpr c^2$ is the average energy transferred to a
shocked electron.

A comparison between the simulated photon and electron distributions resulting from
Maxwellian and power-law electron injection in a constant-density ISM is presented
in Fig. \ref{fig_max}. Synchrotron, Compton and pair production processes are
all included in these examples. The main difference between the two cases
is the width of the synchrotron component, which extends to higher energies
when a large fraction of the electron energy is in the power-law electrons
above the peak Lorentz factor.

In both simulations, the low-energy part of the synchrotron spectrum at $t = 0.1\:\textrm{s}$
and $t = 10\:\textrm{s}$ consists of
the tail emission from the peak electrons, having the spectral shape
$\nu F_{\nu} \propto \nu^{4/3}$. The synchrotron spectral component is naturally
broader in the case of power-law electron injection, since the radiating electrons
in the case of Maxwellian injection are nearly monoenergetic.
The $\sim 10\:\textrm{TeV}$ inverse Compton peak is due to the
scattering events by the electrons near the K-N limit.
At $t = 10^5\:\textrm{s}$, a second Compton scattering order appears as the first scattering
order moves to $\sim 100\:\textrm{eV}$. At all times,
a Maxwellian electron injection results in
a spectrum consisting of distinctly separate components,
which is very different from the case of power-law
electron injection.

\begin{figure}
\begin{center}
\includegraphics[width= \columnwidth]{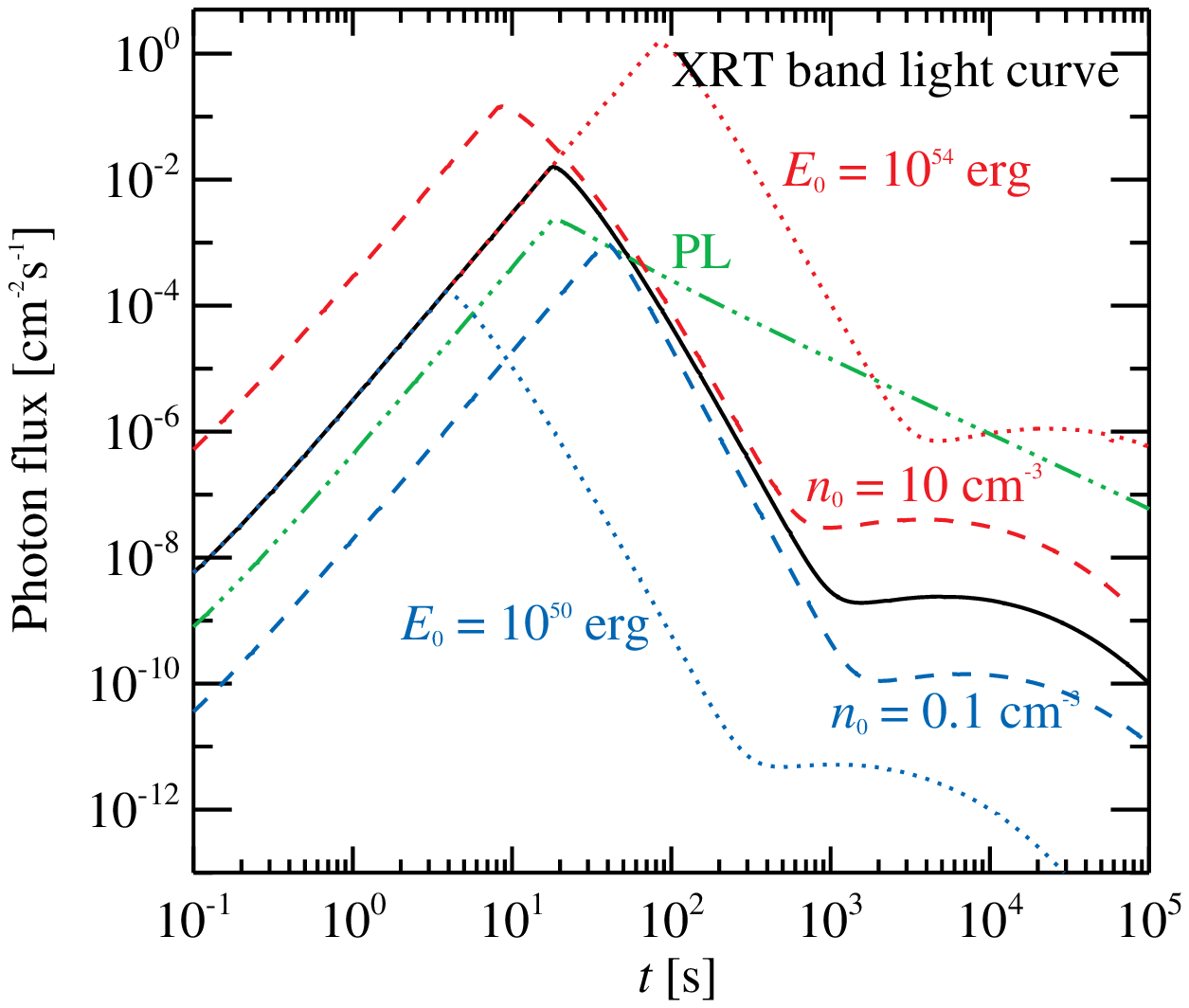}
\includegraphics[width= \columnwidth]{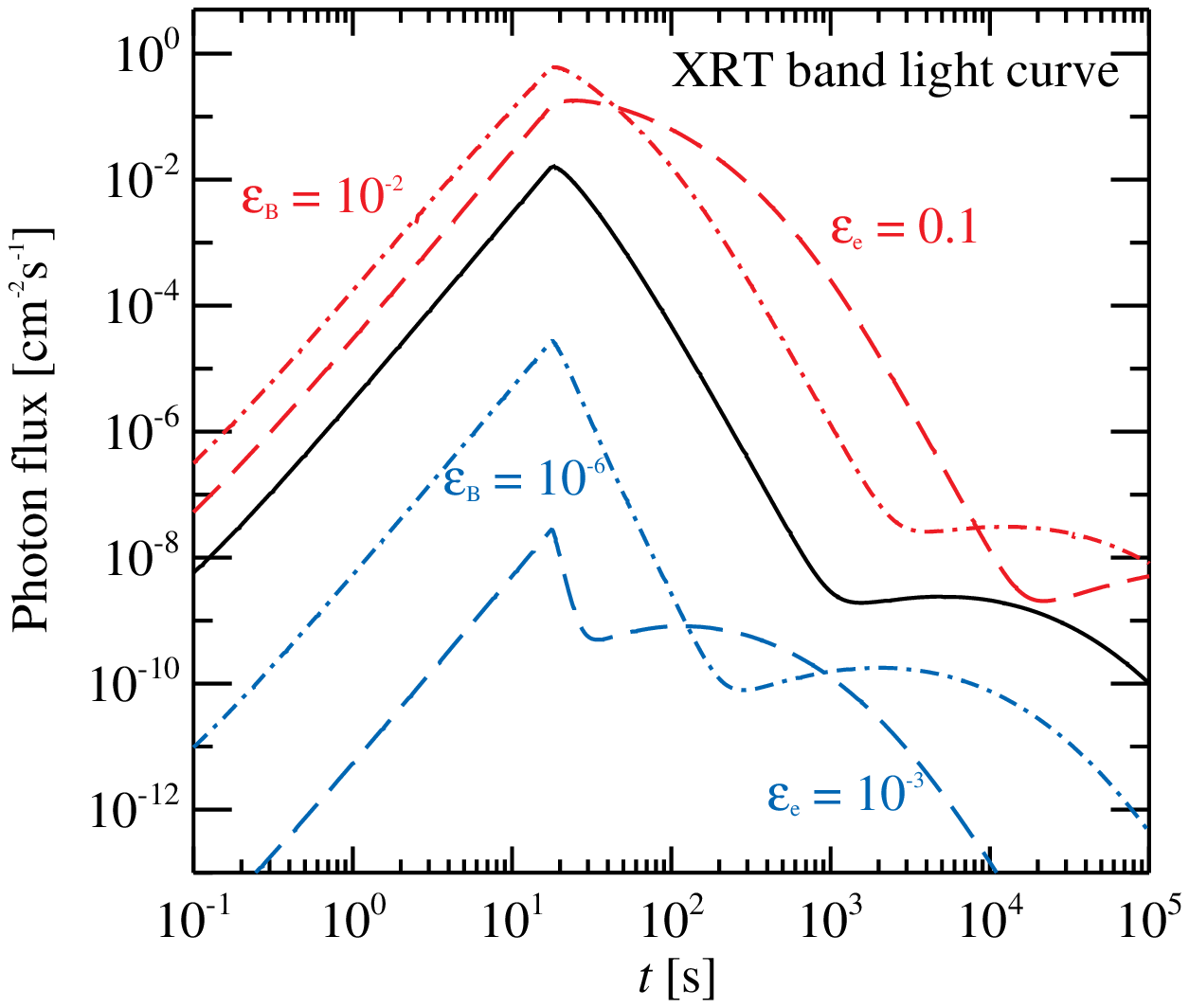}
\caption{\label{fig:lc_max}
Observer frame light curves in the {\it Swift}/XRT band ($0.3-10\:\textrm{keV}$)
resulting from different parameter sets. The black solid line in both panels shows
the fiducial light curve resulting from Maxwellian electron injection obtained with the
same parameters as the spectra in Fig. \ref{fig_max}.
The green dash-dot-dot-dot line in the
{\it top panel} represents the emission due to a power-law electron injection function.
The other curves represent simulations with a Maxwellian injection term where one
parameter at a time is varied from the fiducial value.
In the {\it top panel}, the dotted (dashed) lines show the light curves when the blast energy
$E_0$ (external density $n_0$) is increased/decreased by a factor of 100 (10).
In the {\it bottom panel}, the long-dashed (dash-dot) lines correspond to the emission
when the electron energy fraction $\epse$ (magnetic energy fraction $\epsB$) is
increased/decreased by a factor of 10 (100).
} 
\end{center} 
\end{figure}

\begin{figure}[!]
\begin{center}
\includegraphics[width= \columnwidth]{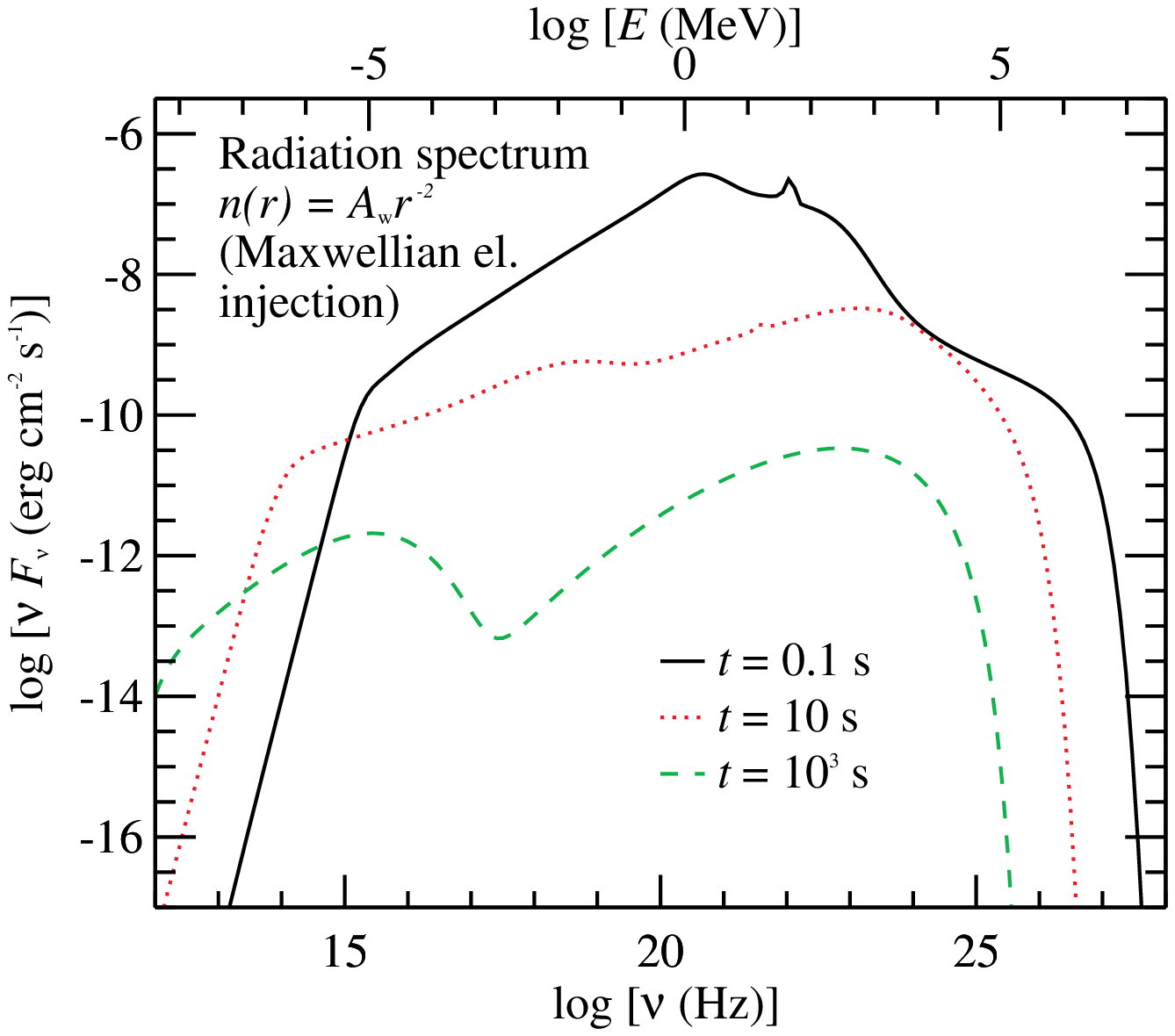}
\includegraphics[width= \columnwidth]{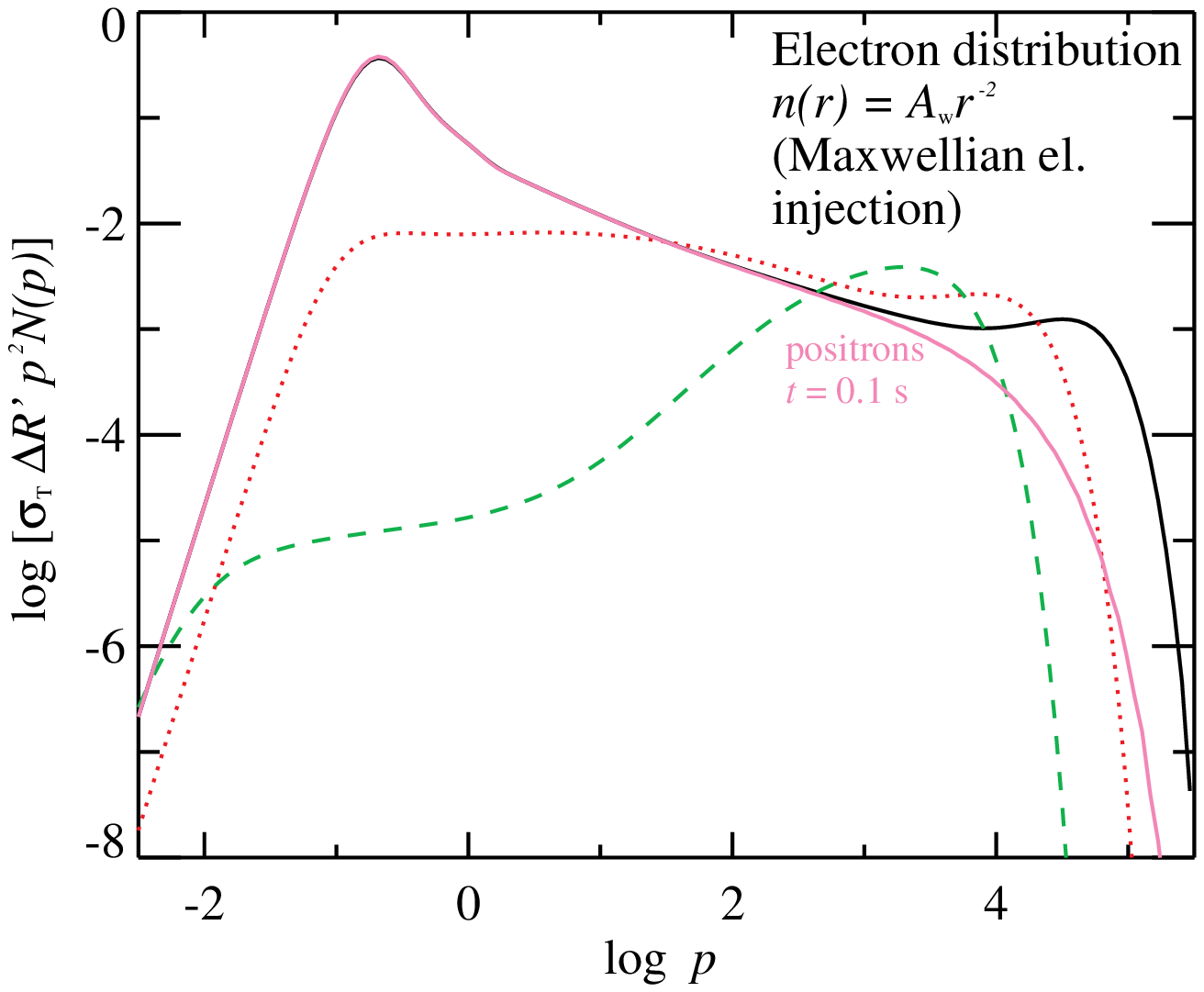}
\caption{\label{fig_max_wind}
Evolving photon ({\it top panel}) and electron ({\it bottom panel})
distributions from a simulation with
Maxwellian electron injection in a wind-type medium. The
magenta line in the {\it bottom panel} shows the positron
distribution at $t = 0.1\:\textrm{s}$. The simulation
parameters are the same as in Fig. \ref{fig_max}, except for
$\epse = 0.1$ and $\Awn = 3$.
} 
\end{center} 
\end{figure}

\begin{figure}[!]
\begin{center}
\includegraphics[width= \columnwidth]{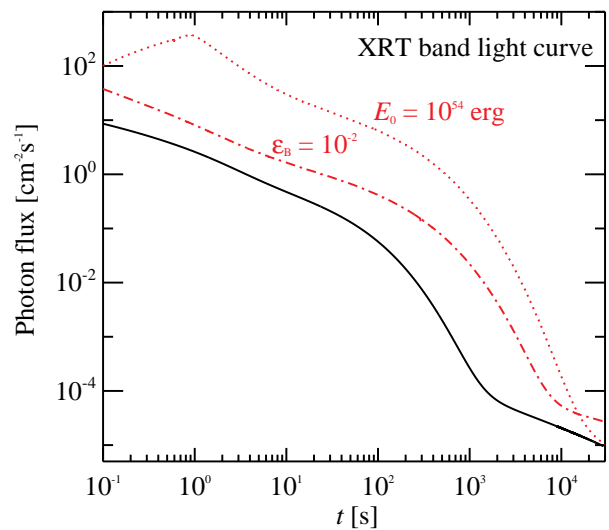}
\caption{\label{fig_lc_max_wind}
Observer frame {\it Swift}/XRT band light curves in the case
of Maxwellian electron injection in a wind medium. The solid line
is the fiducial light curve corresponding to Fig. \ref{fig_max_wind}.
The dotted (dash-dot) curve is obtained by increasing $E_0$
($\epsB$) by a factor of 100 from the fiducial value.
} 
\end{center} 
\end{figure}

Fig. \ref{fig:lc_max} shows the light curves in the {\it Swift}/XRT band
resulting from both the simulations in Fig. \ref{fig_max} compared to several cases
where one parameter at a time is changed. It is notable that a flattening or a
slight rebrightening appears in all the cases with a Maxwellian electron injection.
The flat segment is produced when the synchrotron component moves out of the observed energy range and
is gradually replaced by the inverse Compton emission, similarly as described
by \citet{2011A&A...531A..76P}, who simulated the external shock emission from power-law electrons
with a narrow energy distribution. A similar effect is also seen in the simulations
of \citet{2004MNRAS.352L..35S}. They considered prompt GRB emission due to a narrow
electron distribution that has developed due to the balance between heating and synchrotron/SSC
cooling. In this case, a flat section in the BATSE band light curve develops when the first Compton scattering order
moves out of the observed band and is replaced by the second scattering order.
A spectrum due to power-law electrons does not
contain such dramatic changes of the spectral slope, and consequently there
is no shallow decay segment in the light curve.

In all the cases shown in Fig. \ref{fig:lc_max}, the flux during the flattening
or rebrightening is very low. The flux is higher when $E_0, n_0, \epse$ or $\epsB$
is increased, but higher values of $E_0, \epse$ and $\epsB$ also lead to a delayed
start of the flat segment.
One can estimate that the flattening begins when the synchrotron
frequency of the peak Maxwellian electrons decreases below the observed waveband.
The peak electrons have $\gamma \sim \Gamma \epse \mpr/\mel$ (Eq. (\ref{eq:max_pk}))
and the characteristic synchrotron energy of the emitted photons is
$\xs \propto \Gamma \gamma^2 B' \propto E_0^{1/2} \epse^2 \epsB^{1/2} t^{-3/2}$
(see Eqs. (\ref{eq:syn_freq}), (\ref{eq:g_time}) and (\ref{Bfield})).
Increased values of $E_0, \epse$ and $\epsB$ thus correspond to a higher synchrotron
energy at a given time $t$, delaying the moment when the synchrotron bump crosses
the observing window. However, a higher external density $n_0$ does not significantly
delay the appearance of the shallow decay.

A more luminous shallow decay phase may occur in a wind-type medium.
The particle distributions resulting from a Maxwellian electron injection
function in a wind-type medium are presented in Fig. \ref{fig_max_wind}.
Similarly as in the case of power-law injection in a wind environment
(Fig. \ref{fig_wind_full}),
pair production shapes the electron distribution for $\gamma < 10^3$ at $t = 0.1\:\textrm{s}$,
where it is practically identical to the positron distribution. However, we are
now interested in the spectral evolution at later times when the shallow decay
may appear as the observing window becomes dominated by inverse Compton photons.
Fig. \ref{fig_lc_max_wind} shows that the flux during the shallow decay can
be high enough to be detected in the wind case. In the simulation presented
here, the flux decay is steeper
during this phase than in the ISM case, but still consistent with many
of the observed XRT afterglows.
The flux increases with higher values
of, say, $E_0$ and $\epsB$, but simultaneously the start of the shallow decay is
delayed, as described above.

The simulations with different parameter sets show that a flattening
in the X-ray light curve is likely to take place if the electron injection function
is Maxwellian. In the ISM case, the flux during the simulated flattening appears to be
much lower than a typical shallow decay flux observed by XRT. A wind environment,
on the other hand, can lead to a detectable photon flux during the shallow decay.

%______________________________________________________________

\section{Summary and conclusions}

In this paper, we have studied the modeling of GRB forward shock
emission for the first time with a code that treats all the radiation processes
self-consistently. The main advantages
of our code are the inclusion of synchrotron self-absorption
heating and Compton scattering
as mechanisms for electron thermalization,
and the exact treatment of Compton scattering
and pair production.
We have shown
that our results are in very good agreement with both analytic
estimations and numerical simulations that have been presented in the
literature. Inclusion of an accurate calculation of radiative
transport,
a multi-zone treatment of the emitting region,
and the emission from the reverse shock will be the
topics of future research.

We presented the results of test simulations where external
MeV photons, distributed similarly to prompt GRB emission,
interact with the shock-accelerated electrons
and are Compton scattered to TeV energies.
Such VHE emission is likely to dominate the synchrotron
self-Compton emission at times up to $t \gtrsim 10 \:\textrm{s}$
in GRBs with similar parameter sets
and might be an interesting target for observations
of low-redshift bursts
with future instruments such as the Cherenkov Telescope
Array \citep{2013APh....43..252I}.
We also showed that the
inverse Compton flux is not always quenched by
electron-positron pair production, owing to the
decrease in the pair production opacity before
the deceleration time.
A very prominent inverse Compton peak appears during the prompt emission,
the location of which is determined by the
maximum energy given to the photons by the peak electrons.

Even though the test simulations show that the inverse Compton
scattering of photons distributed according to a Band
function is likely to produce a high-energy
spectral component, the
current model does not explain the typical long duration
of the observed LAT emission in the case of GRB 090902B.
A possible explanation for the LAT data is that
the unscattered prompt photons are traveling within
a smaller solid angle than the scattered photons,
and the arrival time of the upscattered
radiation is delayed as a result \citep{2005ApJ...618L..13B,2013arXiv1307.2663B}.
This effect is not accounted for in the current version
of the code. However, the large-angle effect cannot
explain the LAT emission if it lasts much longer than the
prompt GRB.

As another example, we compared
the synchrotron emission in a wind-type and a constant-density ISM
medium for a set of typical parameters and studied the effect
of different simulation parameters on the thermalization of low-energy
electrons.
As long as the synchrotron cooling time is
shorter than the lifetime of the flow,
the low-energy end of a power-law electron
distribution is strongly thermalized
due to synchrotron self-absorption heating, which
is usually neglected in similar numerical codes.
The shape of the emerging radiation spectrum is still nearly identical
to the case with pure power-law electrons, although a small
bump is seen in the synchrotron spectrum thanks to the Maxwellian shape
of the low-energy electrons. However, the simulation shows
that it is necessary to
include the self-absorption heating term in the electron kinetic
equation to prevent the electrons from
cooling down to the low-energy edge of the grid
as long as thermalization is important.

Additionally, we have demonstrated that a flattening
or rebrightening segment
in the forward shock light curve is expected
in the case of purely Maxwellian electron injection
in a waveband where the observed flux gradually becomes
dominated by the inverse Compton component,
as described by \citet{2011A&A...531A..76P}.
Such flattenings are observed in a large fraction of X-ray light curves, but
their physical origin is still under debate.
In the ISM simulations, the flux during the flat
segments is much lower than in the case of a typical shallow decay phase
observed by {\it Swift}/XRT. However, the shallow decay flux in the wind case
could be detectable.

\begin{acknowledgements}
TP received the funding for this work from the
Finnish Doctoral Program in Astronomy and Space Physics.
\end{acknowledgements}

%\bibliographystyle{aa}
%\bibliography{GRBref}

\begin{thebibliography}{84}
\expandafter\ifx\csname natexlab\endcsname\relax\def\natexlab#1{#1}\fi

\bibitem[{{Abdo} {et~al.}(2009){Abdo}, {Ackermann}, {Ajello}, {Asano},
  {Atwood}, {Axelsson}, {Baldini}, {Ballet}, {Barbiellini}, {Baring},
  {Bastieri}, {Bechtol}, {Bellazzini}, {Berenji}, {Bhat}, {Bissaldi},
  {Blandford}, {Bloom}, {Bonamente}, {Borgland}, {Bouvier}, {Bregeon}, {Brez},
  {Briggs}, {Brigida}, {Bruel}, {Burgess}, {Burrows}, {Buson}, {Caliandro},
  {Cameron}, {Caraveo}, {Casandjian}, {Cecchi}, {{\c C}elik}, {Chekhtman},
  {Cheung}, {Chiang}, {Ciprini}, {Claus}, {Cohen-Tanugi}, {Cominsky},
  {Connaughton}, {Conrad}, {Cutini}, {d'Elia}, {Dermer}, {de Angelis}, {de
  Palma}, {Digel}, {Dingus}, {Silva}, {Drell}, {Dubois}, {Dumora}, {Farnier},
  {Favuzzi}, {Fegan}, {Finke}, {Fishman}, {Focke}, {Fortin}, {Frailis},
  {Fukazawa}, {Funk}, {Fusco}, {Gargano}, {Gehrels}, {Germani}, {Giavitto},
  {Giebels}, {Giglietto}, {Giordano}, {Glanzman}, {Godfrey}, {Goldstein},
  {Granot}, {Greiner}, {Grenier}, {Grove}, {Guillemot}, {Guiriec}, {Hanabata},
  {Harding}, {Hayashida}, {Hays}, {Horan}, {Hughes}, {Jackson},
  {J{\'o}hannesson}, {Johnson}, {Johnson}, {Johnson}, {Kamae}, {Katagiri},
  {Kataoka}, {Kawai}, {Kerr}, {Kippen}, {Kn{\"o}dlseder}, {Kocevski}, {Komin},
  {Kouveliotou}, {Kuss}, {Lande}, {Latronico}, {Lemoine-Goumard}, {Longo},
  {Loparco}, {Lott}, {Lovellette}, {Lubrano}, {Madejski}, {Makeev},
  {Mazziotta}, {McBreen}, {McEnery}, {McGlynn}, {Meegan}, {M{\'e}sz{\'a}ros},
  {Meurer}, {Michelson}, {Mitthumsiri}, {Mizuno}, {Moiseev}, {Monte},
  {Monzani}, {Moretti}, {Morselli}, {Moskalenko}, {Murgia}, {Nakamori},
  {Nolan}, {Norris}, {Nuss}, {Ohno}, {Ohsugi}, {Omodei}, {Orlando}, {Ormes},
  {Paciesas}, {Paneque}, {Panetta}, {Pelassa}, {Pepe}, {Pesce-Rollins},
  {Petrosian}, {Piron}, {Porter}, {Preece}, {Rain{\`o}}, {Rando}, {Rau},
  {Razzano}, {Razzaque}, {Reimer}, {Reimer}, {Reposeur}, {Ritz}, {Rochester},
  {Rodriguez}, {Roming}, {Roth}, {Ryde}, {Sadrozinski}, {Sanchez}, {Sander},
  {Saz Parkinson}, {Scargle}, {Schalk}, {Sgr{\`o}}, {Siskind}, {Smith},
  {Spinelli}, {Stamatikos}, {Stecker}, {Stratta}, {Strickman}, {Suson},
  {Swenson}, {Tajima}, {Takahashi}, {Tanaka}, {Thayer}, {Thayer}, {Thompson},
  {Tibaldo}, {Torres}, {Tosti}, {Tramacere}, {Uchiyama}, {Uehara}, {Usher},
  {van der Horst}, {Vasileiou}, {Vilchez}, {Vitale}, {von Kienlin}, {Waite},
  {Wang}, {Wilson-Hodge}, {Winer}, {Wood}, {Yamazaki}, {Ylinen}, \&
  {Ziegler}}]{2009ApJ...706L.138A}
{Abdo}, A.~A., {Ackermann}, M., {Ajello}, M., {et~al.} 2009, \apjl, 706, L138

\bibitem[{{Achterberg} {et~al.}(2001){Achterberg}, {Gallant}, {Kirk}, \&
  {Guthmann}}]{2001MNRAS.328..393A}
{Achterberg}, A., {Gallant}, Y.~A., {Kirk}, J.~G., \& {Guthmann}, A.~W. 2001,
  \mnras, 328, 393

\bibitem[{{Albert} {et~al.}(2007){Albert}, {Aliu}, {Anderhub}, {Antoranz},
  {Armada}, {Baixeras}, {Barrio}, {Bartko}, {Bastieri}, {Becker}, {Bednarek},
  {Berger}, {Bigongiari}, {Biland}, {Bock}, {Bordas}, {Bosch-Ramon}, {Bretz},
  {Britvitch}, {Camara}, {Carmona}, {Chilingarian}, {Ciprini}, {Coarasa},
  {Commichau}, {Contreras}, {Cortina}, {Costado}, {Curtef}, {Danielyan},
  {Dazzi}, {De Angelis}, {Delgado}, {de los Reyes}, {De Lotto},
  {Domingo-Santamar{\'{\i}}a}, {Dorner}, {Doro}, {Errando}, {Fagiolini},
  {Ferenc}, {Fern{\'a}ndez}, {Firpo}, {Flix}, {Fonseca}, {Font}, {Fuchs},
  {Galante}, {Garc{\'{\i}}a-L{\'o}pez}, {Garczarczyk}, {Gaug}, {Giller},
  {Goebel}, {Hakobyan}, {Hayashida}, {Hengstebeck}, {Herrero}, {H{\"o}hne},
  {Hose}, {Hsu}, {Jacon}, {Jogler}, {Kalekin}, {Kosyra}, {Kranich}, {Kritzer},
  {Laille}, {Lenisa}, {Liebing}, {Lindfors}, {Lombardi}, {Longo}, {L{\'o}pez},
  {L{\'o}pez}, {Lorenz}, {Majumdar}, {Maneva}, {Mannheim}, {Mansutti},
  {Mariotti}, {Mart{\'{\i}}nez}, {Mazin}, {Merck}, {Meucci}, {Meyer},
  {Miranda}, {Mirzoyan}, {Mizobuchi}, {Moralejo}, {Nilsson}, {Ninkovic},
  {O{\~n}a-Wilhelmi}, {Otte}, {Oya}, {Paneque}, {Panniello}, {Paoletti},
  {Paredes}, {Pasanen}, {Pascoli}, {Pauss}, {Pegna}, {Persic}, {Peruzzo},
  {Piccioli}, {Poller}, {Prandini}, {Puchades}, {Raymers}, {Rhode}, {Rib{\'o}},
  {Rico}, {Rissi}, {Robert}, {R{\"u}gamer}, {Saggion}, {S{\'a}nchez},
  {Sartori}, {Scalzotto}, {Scapin}, {Schmitt}, {Schweizer}, {Shayduk},
  {Shinozaki}, {Shore}, {Sidro}, {Sillanp{\"a}{\"a}}, {Sobczynska}, {Stamerra},
  {Stark}, {Takalo}, {Temnikov}, {Tescaro}, {Teshima}, {Tonello}, {Torres},
  {Turini}, {Vankov}, {Vitale}, {Wagner}, {Wibig}, {Wittek}, {Zanin}, \&
  {Zapatero}}]{2007ApJ...667..358A}
{Albert}, J., {Aliu}, E., {Anderhub}, H., {et~al.} 2007, \apj, 667, 358

\bibitem[{{Aleksi{\'c}} {et~al.}(2010){Aleksi{\'c}}, {Anderhub}, {Antonelli},
  {Antoranz}, {Backes}, {Baixeras}, {Balestra}, {Barrio}, {Bastieri}, {Becerra
  Gonz{\'a}lez}, {Becker}, {Bednarek}, {Berdyugin}, {Berger}, {Bernardini},
  {Biland}, {Bock}, {Bonnoli}, {Bordas}, {Borla Tridon}, {Bosch-Ramon}, {Bose},
  {Braun}, {Bretz}, {Britzger}, {Camara}, {Carmona}, {Carosi}, {Colin},
  {Commichau}, {Contreras}, {Cortina}, {Costado}, {Covino}, {Dazzi}, {de
  Angelis}, {de Cea Del Pozo}, {de Los Reyes}, {de Lotto}, {de Maria}, {de
  Sabata}, {Delgado Mendez}, {Doert}, {Dom{\'{\i}}nguez}, {Dominis Prester},
  {Dorner}, {Doro}, {Elsaesser}, {Errando}, {Ferenc}, {Fern{\'a}ndez}, {Firpo},
  {Fonseca}, {Font}, {Galante}, {Garc{\'{\i}}a L{\'o}pez}, {Garczarczyk},
  {Gaug}, {Godinovic}, {Goebel}, {Hadasch}, {Herrero}, {Hildebrand},
  {H{\"o}hne-M{\"o}nch}, {Hose}, {Hrupec}, {Hsu}, {Jogler}, {Klepser},
  {Kr{\"a}henb{\"u}hl}, {Kranich}, {La Barbera}, {Laille}, {Leonardo},
  {Lindfors}, {Lombardi}, {Longo}, {L{\'o}pez}, {Lorenz}, {Majumdar}, {Maneva},
  {Mankuzhiyil}, {Mannheim}, {Maraschi}, {Mariotti}, {Mart{\'{\i}}nez},
  {Mazin}, {Meucci}, {Miranda}, {Mirzoyan}, {Miyamoto}, {Mold{\'o}n}, {Moles},
  {Moralejo}, {Nieto}, {Nilsson}, {Ninkovic}, {Orito}, {Oya}, {Paoletti},
  {Paredes}, {Pasanen}, {Pascoli}, {Pauss}, {Pegna}, {Perez-Torres}, {Persic},
  {Peruzzo}, {Prada}, {Prandini}, {Puchades}, {Puljak}, {Reichardt}, {Rhode},
  {Rib{\'o}}, {Rico}, {Rissi}, {R{\"u}gamer}, {Saggion}, {Saito}, {Salvati},
  {S{\'a}nchez-Conde}, {Satalecka}, {Scalzotto}, {Scapin}, {Schweizer},
  {Shayduk}, {Shore}, {Sierpowska-Bartosik}, {Sillanp{\"a}{\"a}}, {Sitarek},
  {Sobczynska}, {Spanier}, {Spiro}, {Stamerra}, {Steinke}, {Strah}, {Struebig},
  {Suric}, {Takalo}, {Tavecchio}, {Temnikov}, {Tescaro}, {Teshima}, {Torres},
  {Turini}, {Vankov}, {Wagner}, {Zabalza}, {Zandanel}, {Zanin}, {Zapatero}, {de
  Ugarte-Postigo}, \& {MAGIC Collaboration}}]{2010A&A...517A...5A}
{Aleksi{\'c}}, J., {Anderhub}, H., {Antonelli}, L.~A., {et~al.} 2010, \aap,
  517, A5

\bibitem[{{Aleksi{\'c}} {et~al.}(2014){Aleksi{\'c}}, {Ansoldi}, {Antonelli},
  {Antoranz}, {Babic}, {de Almeida}, {Barrio}, {Gonz{\'a}lez}, {Bednarek},
  {Berger}, {Bernardini}, {Biland}, {Blanch}, {Bock}, {Boller}, {Bonnefoy},
  {Bonnoli}, {Borracci}, {Bretz}, {Carmona}, {Carosi}, {Fidalgo}, {Colin},
  {Colombo}, {Contreras}, {Cortina}, {Cossio}, {Covino}, {Da Vela}, {Dazzi},
  {De Angelis}, {De Caneva}, {De Lotto}, {Mendez}, {Doert}, {Dom{\'{\i}}nguez},
  {Prester}, {Dorner}, {Doro}, {Eisenacher}, {Elsaesser}, {Farina}, {Ferenc},
  {Fonseca}, {Font}, {Frantzen}, {Fruck}, {L{\'o}pez}, {Garczarczyk},
  {Terrats}, {Gaug}, {Giavitto}, {Godinovi{\'c}}, {Munoz}, {Gozzini},
  {Hadamek}, {Hadasch}, {Herrero}, {Hose}, {Hrupec}, {Idec}, {Kadenius},
  {Knoetig}, {Kr{\"a}henb{\"u}hl}, {Krause}, {Kushida}, {Barbera}, {Lelas},
  {Lewandowska}, {Lindfors}, {Lombardi}, {L{\'o}pez-Coto}, {L{\'o}pez},
  {L{\'o}pez-Oramas}, {Lorenz}, {Lozano}, {Makariev}, {Mallot}, {Maneva},
  {Mankuzhiyil}, {Mannheim}, {Maraschi}, {Marcote}, {Mariotti},
  {Mart{\'{\i}}nez}, {Masbou}, {Mazin}, {Menzel}, {Meucci}, {Miranda},
  {Mirzoyan}, {Mold{\'o}n}, {Moralejo}, {Munar-Adrover}, {Nakajima},
  {Niedzwiecki}, {Nilsson}, {Nowak}, {Orito}, {Overkemping}, {Paiano},
  {Palatiello}, {Paneque}, {Paoletti}, {Paredes}, {Partini}, {Persic}, {Prada},
  {Moroni}, {Prandini}, {Preziuso}, {Puljak}, {Reichardt}, {Reinthal}, {Rhode},
  {Rib{\'o}}, {Rico}, {Garcia}, {R{\"u}gamer}, {Saggion}, {Saito}, {Saito},
  {Salvati}, {Satalecka}, {Scalzotto}, {Scapin}, {Schultz}, {Schweizer},
  {Shore}, {Sillanp{\"a}{\"a}}, {Sitarek}, {Snidaric}, {Sobczynska}, {Spanier},
  {Stamatescu}, {Stamerra}, {Storz}, {Sun}, {Suri{\'c}}, {Takalo}, {Tavecchio},
  {Temnikov}, {Terzi{\'c}}, {Tescaro}, {Teshima}, {Thaele}, {Tibolla},
  {Torres}, {Toyama}, {Treves}, {Uellenbeck}, {Vogler}, {Wagner}, {Weitzel},
  {Zandanel}, {Zanin}, {Bouvier}, {Hayashida}, {Tajima}, \&
  {Longo}}]{2014MNRAS.437.3103A}
{Aleksi{\'c}}, J., {Ansoldi}, S., {Antonelli}, L.~A., {et~al.} 2014, \mnras,
  437, 3103

\bibitem[{{Band} {et~al.}(1993){Band}, {Matteson}, {Ford}, {Schaefer},
  {Palmer}, {Teegarden}, {Cline}, {Briggs}, {Paciesas}, {Pendleton}, {Fishman},
  {Kouveliotou}, {Meegan}, {Wilson}, \& {Lestrade}}]{1993ApJ...413..281B}
{Band}, D., {Matteson}, J., {Ford}, L., {et~al.} 1993, \apj, 413, 281

\bibitem[{{Beloborodov}(2005{\natexlab{a}})}]{2005ApJ...627..346B}
{Beloborodov}, A.~M. 2005{\natexlab{a}}, \apj, 627, 346

\bibitem[{{Beloborodov}(2005{\natexlab{b}})}]{2005ApJ...618L..13B}
{Beloborodov}, A.~M. 2005{\natexlab{b}}, \apjl, 618, L13

\bibitem[{{Beloborodov} {et~al.}(2013){Beloborodov}, {Hascoet}, \&
  {Vurm}}]{2013arXiv1307.2663B}
{Beloborodov}, A.~M., {Hascoet}, R., \& {Vurm}, I. 2013, arXiv: 1307.2663

\bibitem[{{Beloborodov} \& {Uhm}(2006)}]{2006ApJ...651L...1B}
{Beloborodov}, A.~M. \& {Uhm}, Z.~L. 2006, \apjl, 651, L1

\bibitem[{{Birnbaum} {et~al.}(2012){Birnbaum}, {Zhang}, {Zhang}, \&
  {Liang}}]{2012MNRAS.422..393B}
{Birnbaum}, T., {Zhang}, B., {Zhang}, B.-B., \& {Liang}, E.-W. 2012, \mnras,
  422, 393

\bibitem[{{Blandford} \& {McKee}(1976)}]{1976PhFl...19.1130B}
{Blandford}, R.~D. \& {McKee}, C.~F. 1976, Physics of Fluids, 19, 1130

\bibitem[{{Blumenthal} \& {Gould}(1970)}]{1970RvMP...42..237B}
{Blumenthal}, G.~R. \& {Gould}, R.~J. 1970, Reviews of Modern Physics, 42, 237

\bibitem[{{Chevalier} \& {Li}(2000)}]{2000ApJ...536..195C}
{Chevalier}, R.~A. \& {Li}, Z.-Y. 2000, \apj, 536, 195

\bibitem[{{Chiang} \& {Dermer}(1999)}]{1999ApJ...512..699C}
{Chiang}, J. \& {Dermer}, C.~D. 1999, \apj, 512, 699

\bibitem[{{Dermer}(2007)}]{2007ApJ...664..384D}
{Dermer}, C.~D. 2007, \apj, 664, 384

\bibitem[{{Downes} {et~al.}(2002){Downes}, {Duffy}, \&
  {Komissarov}}]{2002MNRAS.332..144D}
{Downes}, T.~P., {Duffy}, P., \& {Komissarov}, S.~S. 2002, \mnras, 332, 144

\bibitem[{{Eichler} \& {Granot}(2006)}]{2006ApJ...641L...5E}
{Eichler}, D. \& {Granot}, J. 2006, \apjl, 641, L5

\bibitem[{{Fan} \& {Piran}(2006)}]{2006MNRAS.369..197F}
{Fan}, Y. \& {Piran}, T. 2006, \mnras, 369, 197

\bibitem[{{Fan} {et~al.}(2013){Fan}, {Tam}, {Zhang}, {Liang}, {He}, {Zhou},
  {Yang}, {Jin}, \& {Wei}}]{2013ApJ...776...95F}
{Fan}, Y.-Z., {Tam}, P.~H.~T., {Zhang}, F.-W., {et~al.} 2013, \apj, 776, 95

\bibitem[{{Fan} {et~al.}(2005){Fan}, {Zhang}, \& {Wei}}]{2005ApJ...629..334F}
{Fan}, Y.~Z., {Zhang}, B., \& {Wei}, D.~M. 2005, \apj, 629, 334

\bibitem[{{Gao} {et~al.}(2009){Gao}, {Mao}, {Xu}, \&
  {Fan}}]{2009ApJ...706L..33G}
{Gao}, W.-H., {Mao}, J., {Xu}, D., \& {Fan}, Y.-Z. 2009, \apjl, 706, L33

\bibitem[{{Genet} {et~al.}(2007){Genet}, {Daigne}, \&
  {Mochkovitch}}]{2007MNRAS.381..732G}
{Genet}, F., {Daigne}, F., \& {Mochkovitch}, R. 2007, \mnras, 381, 732

\bibitem[{{Ghisellini} {et~al.}(2010){Ghisellini}, {Ghirlanda}, {Nava}, \&
  {Celotti}}]{2010MNRAS.403..926G}
{Ghisellini}, G., {Ghirlanda}, G., {Nava}, L., \& {Celotti}, A. 2010, \mnras,
  403, 926

\bibitem[{{Ghisellini} {et~al.}(2007){Ghisellini}, {Ghirlanda}, {Nava}, \&
  {Firmani}}]{2007ApJ...658L..75G}
{Ghisellini}, G., {Ghirlanda}, G., {Nava}, L., \& {Firmani}, C. 2007, \apjl,
  658, L75

\bibitem[{{Ghisellini} {et~al.}(1988){Ghisellini}, {Guilbert}, \&
  {Svensson}}]{1988ApJ...334L...5G}
{Ghisellini}, G., {Guilbert}, P.~W., \& {Svensson}, R. 1988, \apjl, 334, L5

\bibitem[{{Giannios} \& {Spitkovsky}(2009)}]{2009MNRAS.400..330G}
{Giannios}, D. \& {Spitkovsky}, A. 2009, \mnras, 400, 330

\bibitem[{{Granot} {et~al.}(2006){Granot}, {K{\"o}nigl}, \&
  {Piran}}]{2006MNRAS.370.1946G}
{Granot}, J., {K{\"o}nigl}, A., \& {Piran}, T. 2006, \mnras, 370, 1946

\bibitem[{{Granot} \& {Kumar}(2006)}]{2006MNRAS.366L..13G}
{Granot}, J. \& {Kumar}, P. 2006, \mnras, 366, L13

\bibitem[{{Granot} \& {Sari}(2002)}]{2002ApJ...568..820G}
{Granot}, J. \& {Sari}, R. 2002, \apj, 568, 820

\bibitem[{{Guilbert} {et~al.}(1983){Guilbert}, {Fabian}, \&
  {Rees}}]{1983MNRAS.205..593G}
{Guilbert}, P.~W., {Fabian}, A.~C., \& {Rees}, M.~J. 1983, \mnras, 205, 593

\bibitem[{{He} {et~al.}(2012){He}, {Zhang}, {Wang}, {Li}, \&
  {M{\'e}sz{\'a}ros}}]{2012ApJ...753..178H}
{He}, H.-N., {Zhang}, B.-B., {Wang}, X.-Y., {Li}, Z., \& {M{\'e}sz{\'a}ros}, P.
  2012, \apj, 753, 178

\bibitem[{{Inoue} {et~al.}(2013){Inoue}, {Granot}, {O'Brien}, {Asano},
  {Bouvier}, {Carosi}, {Connaughton}, {Garczarczyk}, {Gilmore}, {Hinton},
  {Inoue}, {Ioka}, {Kakuwa}, {Markoff}, {Murase}, {Osborne}, {Otte},
  {Starling}, {Tajima}, {Teshima}, {Toma}, {Wagner}, {Wijers}, {Williams},
  {Yamamoto}, {Yamazaki}, \& {for the CTA Consortium}}]{2013APh....43..252I}
{Inoue}, S., {Granot}, J., {O'Brien}, P.~T., {et~al.} 2013, Astroparticle
  Physics, 43, 252

\bibitem[{{Kardashev}(1962)}]{1962SvA.....6..317K}
{Kardashev}, N.~S. 1962, \sovast, 6, 317

\bibitem[{{Kobayashi} {et~al.}(1999){Kobayashi}, {Piran}, \&
  {Sari}}]{1999ApJ...513..669K}
{Kobayashi}, S., {Piran}, T., \& {Sari}, R. 1999, \apj, 513, 669

\bibitem[{{Kumar} \& {Barniol Duran}(2010)}]{2010MNRAS.409..226K}
{Kumar}, P. \& {Barniol Duran}, R. 2010, \mnras, 409, 226

\bibitem[{{Kumar} {et~al.}(2008){Kumar}, {Narayan}, \&
  {Johnson}}]{2008Sci...321..376K}
{Kumar}, P., {Narayan}, R., \& {Johnson}, J.~L. 2008, Science, 321, 376

\bibitem[{{Li} {et~al.}(2012){Li}, {Liang}, {Tang}, {Chen}, {Xi}, {L{\"u}},
  {Gao}, {Zhang}, {Zhang}, {Yi}, {Lu}, {L{\"u}}, \&
  {Wei}}]{2012ApJ...758...27L}
{Li}, L., {Liang}, E.-W., {Tang}, Q.-W., {et~al.} 2012, \apj, 758, 27

\bibitem[{{Liu} {et~al.}(2013){Liu}, {Wang}, \& {Wu}}]{2013ApJ...773L..20L}
{Liu}, R.-Y., {Wang}, X.-Y., \& {Wu}, X.-F. 2013, \apjl, 773, L20

\bibitem[{{Martins} {et~al.}(2009){Martins}, {Fonseca}, {Silva}, \&
  {Mori}}]{2009ApJ...695L.189M}
{Martins}, S.~F., {Fonseca}, R.~A., {Silva}, L.~O., \& {Mori}, W.~B. 2009,
  \apjl, 695, L189

\bibitem[{{Maxham} {et~al.}(2011){Maxham}, {Zhang}, \&
  {Zhang}}]{2011MNRAS.415...77M}
{Maxham}, A., {Zhang}, B.-B., \& {Zhang}, B. 2011, \mnras, 415, 77

\bibitem[{{Meliani} {et~al.}(2007){Meliani}, {Keppens}, {Casse}, \&
  {Giannios}}]{2007MNRAS.376.1189M}
{Meliani}, Z., {Keppens}, R., {Casse}, F., \& {Giannios}, D. 2007, \mnras, 376,
  1189

\bibitem[{{M{\'e}sz{\'a}ros}(2006)}]{2006RPPh...69.2259M}
{M{\'e}sz{\'a}ros}, P. 2006, Reports on Progress in Physics, 69, 2259

\bibitem[{{Mimica} {et~al.}(2009){Mimica}, {Giannios}, \&
  {Aloy}}]{2009A&A...494..879M}
{Mimica}, P., {Giannios}, D., \& {Aloy}, M.~A. 2009, \aap, 494, 879

\bibitem[{{Murase} {et~al.}(2011){Murase}, {Toma}, {Yamazaki}, \&
  {M{\'e}sz{\'a}ros}}]{2011ApJ...732...77M}
{Murase}, K., {Toma}, K., {Yamazaki}, R., \& {M{\'e}sz{\'a}ros}, P. 2011, \apj,
  732, 77

\bibitem[{{Nousek} {et~al.}(2006){Nousek}, {Kouveliotou}, {Grupe}, {Page},
  {Granot}, {Ramirez-Ruiz}, {Patel}, {Burrows}, {Mangano}, {Barthelmy},
  {Beardmore}, {Campana}, {Capalbi}, {Chincarini}, {Cusumano}, {Falcone},
  {Gehrels}, {Giommi}, {Goad}, {Godet}, {Hurkett}, {Kennea}, {Moretti},
  {O'Brien}, {Osborne}, {Romano}, {Tagliaferri}, \&
  {Wells}}]{2006ApJ...642..389N}
{Nousek}, J.~A., {Kouveliotou}, C., {Grupe}, D., {et~al.} 2006, \apj, 642, 389

\bibitem[{{Panaitescu} \& {Kumar}(2000)}]{2000ApJ...543...66P}
{Panaitescu}, A. \& {Kumar}, P. 2000, \apj, 543, 66

\bibitem[{{Panaitescu} \& {Meszaros}(1998)}]{1998ApJ...501..772P}
{Panaitescu}, A. \& {Meszaros}, P. 1998, \apj, 501, 772

\bibitem[{{Panaitescu} {et~al.}(2006){Panaitescu}, {M{\'e}sz{\'a}ros},
  {Burrows}, {Nousek}, {Gehrels}, {O'Brien}, \&
  {Willingale}}]{2006MNRAS.369.2059P}
{Panaitescu}, A., {M{\'e}sz{\'a}ros}, P., {Burrows}, D., {et~al.} 2006, \mnras,
  369, 2059

\bibitem[{{Panaitescu} \& {Vestrand}(2011)}]{2011MNRAS.414.3537P}
{Panaitescu}, A. \& {Vestrand}, W.~T. 2011, \mnras, 414, 3537

\bibitem[{{Pandey} {et~al.}(2010){Pandey}, {Swenson}, {Perley}, {Guidorzi},
  {Wiersema}, {Malesani}, {Akerlof}, {Ashley}, {Bersier}, {Cano}, {Gomboc},
  {Ilyin}, {Jakobsson}, {Kleiser}, {Kobayashi}, {Kouveliotou}, {Levan},
  {McKay}, {Melandri}, {Mottram}, {Mundell}, {O'Brien}, {Phillips}, {Rex},
  {Siegel}, {Smith}, {Steele}, {Stratta}, {Tanvir}, {Weights}, {Yost}, {Yuan},
  \& {Zheng}}]{2010ApJ...714..799P}
{Pandey}, S.~B., {Swenson}, C.~A., {Perley}, D.~A., {et~al.} 2010, \apj, 714,
  799

\bibitem[{{Petropoulou} \& {Mastichiadis}(2009)}]{2009A&A...507..599P}
{Petropoulou}, M. \& {Mastichiadis}, A. 2009, \aap, 507, 599

\bibitem[{{Petropoulou} {et~al.}(2011){Petropoulou}, {Mastichiadis}, \&
  {Piran}}]{2011A&A...531A..76P}
{Petropoulou}, M., {Mastichiadis}, A., \& {Piran}, T. 2011, \aap, 531, A76

\bibitem[{{Piran}(2004)}]{2004RvMP...76.1143P}
{Piran}, T. 2004, Reviews of Modern Physics, 76, 1143

\bibitem[{{Qin}(2008)}]{2008ApJ...683..900Q}
{Qin}, Y.-P. 2008, \apj, 683, 900

\bibitem[{{Ramirez-Ruiz} \& {MacFadyen}(2010)}]{2010ApJ...716.1028R}
{Ramirez-Ruiz}, E. \& {MacFadyen}, A.~I. 2010, \apj, 716, 1028

\bibitem[{{Rees} \& {Meszaros}(1992)}]{1992MNRAS.258P..41R}
{Rees}, M.~J. \& {Meszaros}, P. 1992, \mnras, 258, 41P

\bibitem[{{Rhoads}(1999)}]{1999ApJ...525..737R}
{Rhoads}, J.~E. 1999, \apj, 525, 737

\bibitem[{{Sari} \& {Esin}(2001)}]{2001ApJ...548..787S}
{Sari}, R. \& {Esin}, A.~A. 2001, \apj, 548, 787

\bibitem[{{Sari} {et~al.}(1996){Sari}, {Narayan}, \&
  {Piran}}]{1996ApJ...473..204S}
{Sari}, R., {Narayan}, R., \& {Piran}, T. 1996, \apj, 473, 204

\bibitem[{{Sari} {et~al.}(1998){Sari}, {Piran}, \&
  {Narayan}}]{1998ApJ...497L..17S}
{Sari}, R., {Piran}, T., \& {Narayan}, R. 1998, \apjl, 497, L17

\bibitem[{{Sedov}(1959)}]{1959sdmm.book.....S}
{Sedov}, L.~I. 1959, {Similarity and Dimensional Methods in Mechanics} (New York: Academic Press)

\bibitem[{{Shao} \& {Dai}(2007)}]{2007ApJ...660.1319S}
{Shao}, L. \& {Dai}, Z.~G. 2007, \apj, 660, 1319

\bibitem[{{Sironi} {et~al.}(2013){Sironi}, {Spitkovsky}, \&
  {Arons}}]{2013ApJ...771...54S}
{Sironi}, L., {Spitkovsky}, A., \& {Arons}, J. 2013, \apj, 771, 54

\bibitem[{{Spitkovsky}(2008)}]{2008ApJ...682L...5S}
{Spitkovsky}, A. 2008, \apjl, 682, L5

\bibitem[{{Stern} \& {Poutanen}(2004)}]{2004MNRAS.352L..35S}
{Stern}, B.~E. \& {Poutanen}, J. 2004, \mnras, 352, L35

\bibitem[{{Tam} {et~al.}(2013){Tam}, {Tang}, {Hou}, {Liu}, \&
  {Wang}}]{2013ApJ...771L..13T}
{Tam}, P.-H.~T., {Tang}, Q.-W., {Hou}, S.-J., {Liu}, R.-Y., \& {Wang}, X.-Y.
  2013, \apjl, 771, L13

\bibitem[{{Taylor}(1950)}]{1950RSPSA.201..159T}
{Taylor}, G. 1950, Royal Society of London Proceedings Series A, 201, 159

\bibitem[{{Uhm} \& {Beloborodov}(2007)}]{2007ApJ...665L..93U}
{Uhm}, Z.~L. \& {Beloborodov}, A.~M. 2007, \apjl, 665, L93

\bibitem[{{Uhm} \& {Zhang}(2014)}]{2014ApJ...780...82U}
{Uhm}, Z.~L. \& {Zhang}, B. 2014, \apj, 780, 82

\bibitem[{{Uhm} {et~al.}(2012){Uhm}, {Zhang}, {Hasco{\"e}t}, {Daigne},
  {Mochkovitch}, \& {Park}}]{2012ApJ...761..147U}
{Uhm}, Z.~L., {Zhang}, B., {Hasco{\"e}t}, R., {et~al.} 2012, \apj, 761, 147

\bibitem[{{van Eerten} {et~al.}(2010{\natexlab{a}}){van Eerten}, {Zhang}, \&
  {MacFadyen}}]{2010ApJ...722..235V}
{van Eerten}, H., {Zhang}, W., \& {MacFadyen}, A. 2010{\natexlab{a}}, \apj,
  722, 235

\bibitem[{{van Eerten} {et~al.}(2010{\natexlab{b}}){van Eerten}, {Leventis},
  {Meliani}, {Wijers}, \& {Keppens}}]{2010MNRAS.403..300V}
{van Eerten}, H.~J., {Leventis}, K., {Meliani}, Z., {Wijers}, R.~A.~M.~J., \&
  {Keppens}, R. 2010{\natexlab{b}}, \mnras, 403, 300

\bibitem[{{van Eerten} {et~al.}(2011){van Eerten}, {Meliani}, {Wijers}, \&
  {Keppens}}]{2011MNRAS.410.2016V}
{van Eerten}, H.~J., {Meliani}, Z., {Wijers}, R.~A.~M.~J., \& {Keppens}, R.
  2011, \mnras, 410, 2016

\bibitem[{{Veledina} {et~al.}(2013){Veledina}, {Poutanen}, \&
  {Vurm}}]{2013MNRAS.430.3196V}
{Veledina}, A., {Poutanen}, J., \& {Vurm}, I. 2013, \mnras, 430, 3196

\bibitem[{{Veledina} {et~al.}(2011){Veledina}, {Vurm}, \&
  {Poutanen}}]{2011MNRAS.414.3330V}
{Veledina}, A., {Vurm}, I., \& {Poutanen}, J. 2011, \mnras, 414, 3330

\bibitem[{{Vurm} {et~al.}(2011){Vurm}, {Beloborodov}, \&
  {Poutanen}}]{2011ApJ...738...77V}
{Vurm}, I., {Beloborodov}, A.~M., \& {Poutanen}, J. 2011, \apj, 738, 77

\bibitem[{{Vurm} \& {Poutanen}(2009)}]{2009ApJ...698..293V}
{Vurm}, I. \& {Poutanen}, J. 2009, \apj, 698, 293

\bibitem[{{Wang} {et~al.}(2013){Wang}, {Liu}, \&
  {Lemoine}}]{2013ApJ...771L..33W}
{Wang}, X.-Y., {Liu}, R.-Y., \& {Lemoine}, M. 2013, \apjl, 771, L33

\bibitem[{{Wygoda} {et~al.}(2011){Wygoda}, {Waxman}, \&
  {Frail}}]{2011ApJ...738L..23W}
{Wygoda}, N., {Waxman}, E., \& {Frail}, D.~A. 2011, \apjl, 738, L23

\bibitem[{{Yamazaki}(2009)}]{2009ApJ...690L.118Y}
{Yamazaki}, R. 2009, \apjl, 690, L118

\bibitem[{{Zdziarski}(1988)}]{1988ApJ...335..786Z}
{Zdziarski}, A.~A. 1988, \apj, 335, 786

\bibitem[{{Zhang} {et~al.}(2006){Zhang}, {Fan}, {Dyks}, {Kobayashi},
  {M{\'e}sz{\'a}ros}, {Burrows}, {Nousek}, \& {Gehrels}}]{2006ApJ...642..354Z}
{Zhang}, B., {Fan}, Y.~Z., {Dyks}, J., {et~al.} 2006, \apj, 642, 354

\bibitem[{{Zhang} \& {MacFadyen}(2009)}]{2009ApJ...698.1261Z}
{Zhang}, W. \& {MacFadyen}, A. 2009, \apj, 698, 1261

\end{thebibliography}

\end{document}